\documentclass[12pt]{article}
\pdfoutput=1

\usepackage{draft} 
\usepackage{hyperref}
\usepackage{graphicx,color,subfigure}
\usepackage{cite}
\usepackage{mciteplus}
\usepackage{skak}
\DeclareFontFamily{OT1}{pzc}{}
\DeclareFontShape{OT1}{pzc}{m}{it}{<-> s * [1.10] pzcmi7t}{}
\DeclareMathAlphabet{\mathpzc}{OT1}{pzc}{m}{it}

\def\be#1\ee{\begin{align}#1\end{align}}
\newcommand\nn{\nonumber}

\begin{document}

\unitlength = .8mm

\begin{titlepage}
\rightline{MIT-CTP/4728}

\begin{center}

\hfill \\
\hfill \\
\vskip 1cm

\title{${\cal N}=4$ Superconformal Bootstrap of the K3 CFT}

\author{Ying-Hsuan Lin,$^\textsymrook$ Shu-Heng Shao,$^\textsymrook$ David Simmons-Duffin,$^\textsymbishop$ Yifan Wang,$^\textsymknight$ Xi Yin$^\textsymrook$}

\address{
$^\textsymrook$Jefferson Physical Laboratory, Harvard University, \\
Cambridge, MA 02138 USA
\\
$^\textsymbishop$School of Natural Sciences, Institute for Advanced Study, \\
Princeton, NJ 08540 USA
\\
$^\textsymknight$Center for Theoretical Physics, Massachusetts Institute of Technology, \\
Cambridge, MA 02139 USA
}

\email{yhlin@physics.harvard.edu, shshao@physics.harvard.edu, dsd@ias.edu, yifanw@mit.edu,
xiyin@fas.harvard.edu}

\end{center}

\abstract{ We study two-dimensional $(4,4)$ superconformal field theories of central charge $c=6$, corresponding to nonlinear sigma models on K3 surfaces, using the superconformal bootstrap.  This is made possible through a surprising relation between the BPS ${\cal N}=4$ superconformal blocks with $c=6$ and bosonic Virasoro conformal blocks with $c=28$, and an exact result on the moduli dependence of a certain integrated BPS 4-point function. Nontrivial bounds on the non-BPS spectrum in the K3 CFT are obtained as functions of the CFT moduli, that interpolate between the free orbifold points and singular CFT points. We observe directly from the CFT perspective the signature of a continuous spectrum above a gap at the singular moduli, and find numerically an upper bound on this gap that is saturated by the $A_1$ ${\cal N}=4$ cigar CFT. We also derive an analytic upper bound on the first nonzero eigenvalue of the scalar Laplacian on K3 in the large volume regime, that depends on the K3 moduli data. As two byproducts, we find an exact equivalence between a class of BPS ${\cal N}=2$ superconformal blocks and Virasoro conformal blocks in two dimensions, and an upper bound on the four-point functions of operators of sufficiently low scaling dimension in three and four dimensional CFTs. }

\vfill

\end{titlepage}

\eject

\tableofcontents

\section{Introduction}

The conformal bootstrap \cite{polyakov1974nonhamiltonian,Ferrara:1973yt,Mack:1975jr}, the idea that a conformal field theory can be determined entirely based on (possibly extended) conformal symmetry, unitarity, and simple assumptions about the spectrum, has proven to be remarkably powerful. Such methods have been implemented analytically to solve two-dimensional rational CFTs \cite{Belavin:1984vu,Knizhnik:1984nr,Gepner:1986wi,Bouwknegt:1992wg}, and later extended to certain irrational CFTs \cite{Zamolodchikov:1995aa,Teschner:1997ft,Teschner:1999ug, Teschner:2001rv}. The numerical approach to the conformal bootstrap has been applied successfully to higher dimensional theories \cite{Rattazzi:2008pe, Rychkov:2009ij, Poland:2010wg, Poland:2011ey, ElShowk:2012ht,Kos:2013tga, Beem:2013qxa, Beem:2014zpa, El-Showk:2014dwa,Chester:2014fya,Chester:2014mea,Chester:2014gqa,Bae:2014hia,Chester:2015qca,Iliesiu:2015qra,Kos:2015mba,Beem:2015aoa,Lemos:2015awa}, as well as putting nontrivial constraints on the spectrum of two-dimensional theories that have been previously unattainable with analytic methods \cite{Hellerman:2009bu, Keller:2012mr,Fiset:2015pta}.
 
In this paper, we analyze $c=6$ $(4,4)$ superconformal field theories using the conformal bootstrap. Our primary example\footnote{For noncompact target spaces, there are other interesting $c=6$ (4,4) non-linear sigma models including the ALF CFT \cite{Harvey:2014nha},  for which our bootstrap method also applies.} is the supersymmetric nonlinear sigma model with the K3 surface as its target space. We refer to this theory as the K3 CFT. The conformal manifold and BPS spectrum of the K3 CFT has been well known \cite{Seiberg:1988pf, Eguchi:1988vra, Cecotti:1990kz, Cecotti:1991me, Aspinwall:1994rg,Nahm:1999ps}. Much less was known about the non-BPS spectrum of the theory, except at special solvable points in the moduli space \cite{Gepner:1987qi,Gaberdiel:2011fg,Gaberdiel:2013psa}, and in the vicinity of points where the CFT becomes singular \cite{Ooguri:1995wj,Eguchi:2004ik,Eguchi:2008ct}. To understand the non-BPS spectrum of the K3 CFT is the subject of this paper.

There are two essential technical ingredients that will enable us to bootstrap the K3 CFT. The first ingredient is an exact relation between the BPS ${\cal N}=4$ superconformal block at central charge $c=6$ and the bosonic Virasoro conformal block at central charge $c=28$ discussed in Section \ref{sec:block}. More precisely, we consider the sphere four-point block of the small ${\cal N}=4$ super-Virasoro algebra, with four external BPS operators of weight and spin $(h,j)=({1\over 2},{1\over 2})$ in the NS sector or $(h,j)=({1\over 4},0)$ in the R sector, and a generic non-BPS intermediate primary of weight $h$. This ${\cal N}=4$ block will be equal to, up to a simple factor, the sphere four-point bosonic Virasoro conformal block of central charge 28, with external weights 1 and internal primary weight $h+1$. This relation is observed by comparing the four-point function of normalizable BPS operators in the ${\cal N}=4$ $A_1$ cigar CFT to correlators in the bosonic Liouville theory, through the relation of Ribault and Teschner that expresses $SL(2)$ WZW model correlators in terms of Liouville correlators \cite{Ribault:2005wp, Chang:2014jta}.

We generalize the above argument to establish an exact equivalence between a class of BPS $\mathcal{N}=2$ superconformal blocks of $c=3(k+2)/k$ with bosonic Virasoro conformal blocks of $c=13+6k+6/k$ in Section \ref{sec:N=2block}.

The second ingredient is the exact moduli dependence of certain integrated four-point functions $A_{ijk\ell}$ of ${1\over 2}$-BPS operators (corresponding to marginal deformations) in the K3 CFT. They are obtained from the weak coupling limit of the non-perturbatively exact results on 4- and 6-derivative terms in the spacetime effective action of type IIB string theory compactified on the K3 surface \cite{Kiritsis:2000zi,Lin:2015dsa}. This allows us to encode the moduli of the K3 CFT directly in terms of CFT data applicable in the bootstrap method, namely the four-point function.

The numerical bootstrap then proceeds by analyzing the crossing equation, where the ${\cal N}=4$ blocks, re-expressed in terms of Virasoro conformal blocks, are evaluated using Zamolodchikov's recurrence relations \cite{Zamolodchikov:1985ie,Zamolodchikov:1995aa}. The reality condition on the OPE coefficients, which follows from unitarity, leads to two kinds of bounds on the scaling dimension of non-BPS operators, which we refer to as the gap dimension $\Delta_{gap}$ and a \textit{critical dimension} $\widehat\Delta_{crt}$. $\Delta_{gap}$ is the scaling dimension of the lowest non-BPS primary that appear in the OPE of a pair of ${1\over 2}$-BPS operators. $\widehat\Delta_{crt}$ is defined such that, roughly speaking, the OPE coefficients of (and contributions to the four-point function from) the non-BPS primaries at dimension $\Delta>\widehat\Delta_{crt}$ are bounded from above by those of the primaries of dimension $\Delta\leq \widehat\Delta_{crt}$. A consequence is that,  when the four-point function diverges at special points on the conformal manifold, the CFT either develops a continuum that contains $\widehat\Delta_{crt}$ or some of its OPE coefficients diverge. In the case  when the OPE coefficients are bounded (which is not always true as we will discuss in Section \ref{sec:mixed}), $\widehat\Delta_{crt}$ provides an upper bound on the gap below the continuum of the spectrum that is developed when the CFT becomes singular.

We will see that the numerical bounds on $\widehat\Delta_{crt}$ and $\Delta_{gap}$ are saturated by the free orbifold $T^4/\mathbb{Z}_2$ CFT, as well as the $A_1$ cigar CFT, and interpolate between the two as we move along the moduli space. The moduli dependence is encoded in the integrated four-point function of ${1\over 2}$-BPS operators $A_{ijk\ell}$, which has been determined as an exact function of the moduli. Our results provide  direct evidence for the emergence of a continuum in the CFT spectrum, at the points on the conformal manifold where the K3 surface develops ADE singularities, using purely CFT methods (as opposed to the knowledge of the spacetime BPS spectrum of string theory \cite{Aspinwall:1994rg, Ooguri:1995wj, Aspinwall:1995zi}). Our bounds are also consistent with, but not saturated by, the OPE of twist fields in the free orbifold CFT.

We further discuss analytic and numerical bounds on $\widehat \Delta_{crt}$ in general CFTs in 2,3, and 4 dimensions. Using crossing equations, we derive a crude analytic bound $\widehat \Delta_{crt}\le \sqrt{2} \Delta_\phi$,  where $\Delta_\phi$ is  the scaling dimension of the external scalar operator.   This bound on $\widehat\Delta_{crt}$ is then refined  numerically, and we observe that it  meets at the unitarity bound for  $\Delta_\phi\lesssim 1$ in 3 dimensions and $\Delta_\phi \lesssim 2$ in 4 dimensions, thus giving  universal upper bounds on the four-point functions for this range of external operator dimension.

In the large volume limit of the K3 target space, the spectrum of the CFT is captured by the eigenvalues of the Laplacian on the K3. Using a positivity condition on the $q$-expansion of conformal blocks and four-point functions \cite{Hartman:2015lfa,Maldacena:2015iua}, we will derive an upper bound on the gap in the spectrum, or equivalently on the first nonzero eigenvalue of the scalar Laplacian on the K3, that depends on the moduli and remains nontrivial in the large volume limit. Namely, it scales with the volume $V$ as $V^{-{1\over 2}}$ and thereby provides a bound on the first nonzero eigenvalue of the scalar Laplacian on the K3. 

We summarize our results and discuss possible extensions of the current work in the concluding section.  Various  technical details are presented in the appendices. 
 In Appendix \ref{app:T4}, we fix the normalization of the integrated four-point function by comparing with known results at the free orbifold point. In Appendix \ref{Sec:qmap}, we review the $q$-expansion of the Virasoro conformal blocks and Zamolodchikov's recurrence relations. In Appendix \ref{app:Aijkl}, we explain the subtle technical details on how to incorporate the integrated four-point function $A_{ijk\ell}$ into the bootstrap equations, and also derive a bound on the integrated four-point function by the four-point function evaluated at $z={1 \over 2}$. In Appendix \ref{app:critical}, we discuss how the critical dimension $\widehat\Delta_{crt}$ gives an upper bound on the gap below the continuum when the integrated four-point function diverges at some points on the moduli space.

\section{Review of $\cN=4$ Superconformal Representation Theory}
\label{sec:review}
The small $\cN=4$ superconformal algebra (SCA) with central charge $c=6k'$, current algebra $SU(2)_R$ and outer-automorphism $SU(2)_{out}$ is generated by a energy-momentum tensor $T$, super-currents $G^{\A A}$ transforming as $(\bf 2,2)$ under $SU(2)_R\times SU(2)_{out}$ and the $SU(2)_R$ current $J^i$. In terms of their Fourier components $L_n$, $G^{\A A}_r$ and $J^i$, the small $\cN=4$ SCA is captured by the commutation relations
\ie
& [L_m, L_n] = (m-n) L_{m+n} + {k'\over 2}(m^3-m)\delta_{m+n},
\\
& [L_m, G^{\A A}_r] = ({m\over 2}-r) G^{\A A}_{m+r},~~~~ [L_m, J_n^i] = -n J_{m+n}^i,
\\
& \{ G^{\A A}_r, G^{\B B}_s\} = 2\epsilon^{\A\B}\epsilon^{AB} L_{r+s} - 2(r-s) \epsilon^{AB}\sigma_i^{\A\B} J^{i}_{r+s} + {k'\over 2} (4r^2-1) \epsilon^{\A\B}\epsilon^{AB} \delta_{r+s},
\\
& [J_m^{i}, G^{\A A}_r] =-{1\over 2}(\sigma_i)^\A{}_\B G^{\B A}_{m+r}, ~~~~ [J^i_m, J^j_n] = i\epsilon^{ijk} J^k_{m+n} + m {k'\over 2}\delta^{ij}\delta_{m+n}
\fe
where $(\sigma_i)^\A_{~\B}$ are the Pauli matrices and $(\sigma_i)^{\A\B}=(\sigma_i)^\A{}_\C\epsilon^{\B\C}$ with $\epsilon_{+-} =\epsilon^{+-}=+1$. Here we are focusing on the left-moving (holomorphic) part. The subscripts $r,s$ take half-integer values for the NS sector and integer values for the R sector.

The $\cN=4$ SCA enjoys an inner automorphism known as spectral flow, which acts as \cite{Schwimmer:1986mf},
\ie
 J^3_n&\rightarrow J^3_n+ \eta {k'}\D_{n,0},~~J^\pm_n\rightarrow J^\pm_{n\pm 2\eta}
\\
 L_n&\rightarrow L_n+2\eta J_n^3+\eta^2{k'}\D_{n,0},~~
G_r^{\pm A} \rightarrow  G_{r\pm {\eta}}^{\pm A}
\fe
where $\eta\in \mathbb{Z}/2$. In particular, spectral flow with $\eta\in \mathbb{Z}+{1\over 2}$ connects the NS and R sectors. 

To obtain a unitary representation of the $\cN=4$ SCA, $k'$ must be a positive integer. Furthermore, if the highest weight state ($\cN=4$ superconformal primary) has weight $h$ and $SU(2)_R$ spin $\ell\in\mathbb{Z}/2$, unitarity imposes the constraints $h\geq \ell$ in the NS sector and  $h\geq {k'\over 4}$ in the R sector. There are two classes of unitary representations of $\cN=4$ SCA: the BPS (massless or short) representations and the non-BPS (massive or long) representations, which are summarized in Table~\ref{scareps}. In the full $\cN=(4,4)$ SCFT, operators which are BPS on both the left and right sides are called ${1\over 2}$-BPS; the operators which are BPS on one side and non-BPS on the other are ${1\over 4}$-BPS. We should emphasize that our terminology of BPS operators exclude the currents which will be lifted at generic moduli of the K3 CFT.
 
\begin{table}[htb]
\centering
\begin{tabular}{|c|c|c|}
\hline
 &   BPS  & non-BPS  \\\hline
NS & ~$h=\ell$,~~$0\leq \ell\leq {k'\over 2}$ ~&~ $h>\ell $,~~$0\leq \ell\leq {k'-1\over 2}$~
\\\hline
R  &  ~ $h={k'\over 4}$,~~$0\leq \ell\leq {k'\over 2}$ ~&~ $h>{k'\over 4}$,~~${1\over 2}\leq \ell\leq {k'\over 2}$~
\\\hline 
\end{tabular}
\caption{$\cN=4$ superconformal primaries in BPS and non-BPS representations.}
\label{scareps}
\end{table}

The character for the BPS representation in the NS sector is
\ie
&ch^{BPS}_{h=\ell}(q,z,y) = q^\ell \prod_{n=1}^\infty {(1+yz q^{n-{1\over 2}}) (1+y^{-1} z q^{n-{1\over 2}}) (1+yz^{-1} q^{n-{1\over 2}}) (1+y^{-1} z^{-1} q^{n-{1\over 2}}) \over (1-q^n)^2 (1-z^2 q^n)(1-z^{-2} q^{n})}
\\
&~~~\times \sum_{m=-\infty}^\infty {q^{(k'+1) m^2+(2\ell+1)m}\over 1-z^{-2}} \left[  { z^{2((k'+1)m+\ell)} \over (1+yzq^{m+{1\over 2}}) (1+y^{-1} z q^{m+{1\over 2}})} - { z^{-2((k'+1)m+\ell+1)} \over (1+yz^{-1} q^{m+{1\over 2}}) (1+y^{-1}z^{-1} q^{m+{1\over 2}}) } \right].
\fe
while the non-BPS NS sector character is
\ie
ch^{non-BPS}_{h,\ell}(q,z,y) &= q^{h} \prod_{n=1}^\infty {(1+yz q^{n-{1\over 2}}) (1+y^{-1} z q^{n-{1\over 2}}) (1+yz^{-1} q^{n-{1\over 2}}) (1+y^{-1} z^{-1} q^{n-{1\over 2}}) \over (1-q^n)^2 (1-z^2 q^n)(1-z^{-2} q^{n})}
\\
&~~~\times \sum_{m=-\infty}^\infty q^{(k'+1) m^2+(2\ell+1)m} { z^{2((k'+1)m+\ell)} - z^{-2((k'+1)m+\ell+1)} \over 1-z^{-2}},
\fe
where $z$ and $y$ are the fugacities for the third components of $SU(2)_R$ and $SU(2)_{out}$, respectively. 
The Ramond sector characters are related to the above by spectral flow. 

We will now specialize to the K3 CFT which admits a small $\cN=4$ SCA containing left and right moving $SU(2)_R$ R-current at level $k'=1$. In this case, the $  1\over 2 $-BPS primaries  in the (NS,NS) sector consist of the identity operator $(h=\ell=\bar h=\bar \ell=0)$ and 20 others labelled by ${\cal O}_{i}^{\pm\pm}$ with $h=\ell=\bar h=\bar\ell={1\over 2}$ which correspond to the 20 $(1,1)$-harmonic forms on K3 ($i=1,\cdots ,20$). In particular, the weight-${1\over 2}$ BPS primaries ${\cal O}_i^{\pm\pm}$ correspond to exactly marginal operators of the K3 CFT.
Under spectral flow, the identity operator is mapped to the unique  $h=\bar h={1\over 4},\ell=\bar\ell={1\over 2}$ ground state ${\cal O}_0^{\pm\pm}$ in the (R,R) sector, whereas ${\cal O}^{\pm\pm}_i$ give rise to 20 $h=\bar h={1\over 4},\ell=\bar\ell=0$ (R,R) sector ground states denoted by $\phi_i^{RR} $.  The K3 CFT also contains ${1\over 4}$ BPS primaries of weight $(s,{1\over 2})$ and $({1\over 2},s)$, for integer $s\geq 1$.\footnote{Note that the ${1\over 4}$-BPS primaries are fermionic with half integer spin, and are themselves projected out in the spectrum of the K3 SCFT. Rather, their integer spin $(4,4)$ SCA descendants comprise the true ${1\over 4}$ BPS operators of the K3 SCFT.} The weight $(s,{1\over 2})$ ${1\over 4}$-BPS primaries have left $SU(2)_R$ spin 0 and right $SU(2)_R$ spin ${1\over 2}$. They are captured by the K3 elliptic genus (NS sector) decomposed into $\cN=4$ characters \cite{Eguchi:1987sm, Eguchi:1987wf, Eguchi:1988vra},
\ie
Z_{K3}^{NS} = 20 ch^{BPS}_{1\over 2} + ch^{BPS}_{0} - ch^{non-BPS}_{0} (90q+462q^2+1540 q^3+\cdots).
\label{EG}
\fe
where the $(s,{1\over 2})$ BPS primaries are counted by the character
\ie
90q+462q^2+1540 q^3+\cdots.
\fe
We assume the absence of currents at generic moduli of the K3 CFT, which may be justified by conformal perturbation theory, so that the ${1\over 4}$ BPS primaries are the only contributions to the non-BPS character terms in the elliptic genus \eqref{EG}. While the currents (of general spin) may appear at special points in the moduli space, they can be viewed as limits of non-BPS operators and therefore do not affect our bootstrap analysis.

We are interested in the four-point function of ${\cal O}_i^{\pm\pm}$ (or $\phi^{RR}_i$ by spectral flow). Below we will make a general argument, based on $\cN=4$ superconformal algebra at general $c=6k'$, that the OPE of two BPS primaries $\phi_1^{\ell_1,m_1}$ and $\phi_2^{\ell_2,m_2}$ with  $SU(2)_R$  spin $\ell_1$ and $\ell_2$ respectively can only contain superconformal primaries ${\cal O}^{\ell,m}$ (and descendants of), with $SU(2)_R$ spin $\ell$  within the range  $|\ell_1-\ell_2|,|\ell_1-\ell_2|+1,\dots, \ell_1+\ell_2-1,\ell_1+\ell_2$ and $m$ labels its $J^3_R$ charge.\footnote{We will focus on the holomorphic part in this argument.}\footnote{Similar contour arguments have been used in \cite{deBoer:2008ss,Baggio:2012rr} to argue that the three point functions of BPS primaries are covariantly constant over the moduli space.} In particular, this will imply at $k'=1$ for K3 CFT, only (descendants of) the identity operator and non-BPS operators can appear. Consequently, only the identity block and non-BPS blocks contribute to the four-point function of ${1\over 2}$-BPS primaries ${\cal O}_i^{\pm\pm}$.

We start with the 3-point function 
\ie
\la\phi_1^{\ell_1,m_1}(x_1)\phi_2^{\ell_2,m_2}(x_2)[W^{-m_1-m_2-m},{\cal O}^{\ell,m}(x_3)]\ra
\label{3ptv}
\fe
where $W^{-m-m_1-m_2}$ is an arbitrary word with $J_0^3=-m-m_1-m_2$ under left $SU(2)_R$ and composed of raising operators $L_{-n}$, $J^i_{-n}$, $G_{-r}^{\A A}$ with $n>0,r>1/2$, and $J_0^+$, $G^{-A}_{-1/2}$.  We would like to argue by $\cN=4$ superconformal invariance that such a correlator vanishes identically. The main idea is to perform contour deformation a number of times to strip off $W^{-m-m_1-m_2}$ completely while either leaving behind a $G^{\A A}_{1/2}$ which annihilates ${\cal O}^{j,m}$ or just the correlator of the superconformal primaries themselves which vanish due to $SU(2)_R$ invariance. 

Let us suppose $\ell$ does not belong to $|\ell_1-\ell_2|,|\ell_1-\ell_2|+1,\dots, \ell_1+\ell_2-1,\ell_1+\ell_2$. By inserting an appropriate number of $J^{\pm}_0$ at $x_1$ and $x_2$ in \eqref{3ptv}, and redistributing them by contour deformations, we can reduce the correlator \eqref{3ptv} to
\ie
\la\phi_1^{\ell_1,\ell_1}(x_1)\phi_2^{\ell_2,-\ell_2}(x_2)[W^{\ell_2-\ell_1-m},{\cal O}^{\ell,m}(x_3)]\ra
\label{3ptve}.
\fe
We can immediately strip off all Virasoro generators $L_{-n}$ in $W^{\ell_2-\ell_1-m}$ by deforming the contour of $\oint {dz\over 2\pi i} (z-x_3)^{1-n} T(z)$. This will relate the original three-point correlator to the derivatives of those without $L_{-n}$. Similarly, we can deform the contour of $\oint {dz\over 2\pi i} (z-x_3)^{-n} J^3(z)$ to move $J^3_{-n}$ on $\phi_1^{\ell_1,\ell_1}$ to $J_0^3$ on $\phi_1^{\ell_1,\ell_1}$ and $\phi_2^{\ell_2,-\ell_2}$. As for $J^+_{-n}=\oint {dz\over 2\pi i} (z-x_3)^{-n} J^+(z)$, we can replace its insertion by 
\ie\label{repJ}
(x_3-x_2)J^+_{-n}=-J^+_{-n+1}+\oint {dz\over 2\pi i}\, J^+(z)(z-x_3)^{-n}(z-x_2) 
\fe
and deforming the contour. Note that the second term in \eqref{repJ} has a vanishing contribution when we deform the contour to encircle either $\phi_1^{\ell_1,\ell_1}$ or $\phi_2^{\ell_2,-\ell_2}$, hence the original three-point function with $J^+_{-n}$ in $W^{\ell_2-\ell_1-m}$ is related to another with the operator replaced by $J^+_{-n+1}$ in $W^{\ell_2-\ell_1-m}$. Repeating this procedure a number of times, we can be replace $J^+_{-n}$ by $J^+_0$.\footnote{Note that we do not have contributions when deforming the contour past infinity for $n\ge 0$.} Similarly we can substitute $J^-_{-n}$ by $J^-_0$. By commuting $J_0^i$ all the way to right, we obtain a bunch of three point correlators of the form \eqref{3ptve} with $W^{\ell_2-\ell_1-m}$ purely made of $G^{\A A}_{-r}$. Consider for example the case when $G^{+A}_{-n-1/2}=\oint {dz\over 2\pi i}\, G^{+A}(z)(z-x_3)^{-n}$ is the leftmost letter in $W^-$. As before for $J^+_{-n}$, we can replace this insertion in the three-point function by $G^{+A}_{-n+1/2}$ for $n\geq 0$ using 
\ie\label{repG}
 (x_3-x_2)G^{+A}_{-n-1/2}=-G^{+A}_{-n+1/2}+\oint {dz\over 2\pi i}\, G^{+A}(z)(z-x_3)^{-n}(z-x_2) .
\fe
 Iterating this a number of times, we can replace $G^{-A}_{-n-1/2}$ by $G^{-A}_{1/2}$.\footnote{One can apply a similar procedure if $G^{+A}_{-r}$ is the leftmost letter in $W^-$.} Now we can commute $G^{-A}_{1/2}$ all the way to the right which will produce $L_{-n}$ and $J^i_{-m}$ via anti-commutators and reduce the number of $G^{\A A}_{-r}$'s in $W^{\ell_2-\ell_1-m}$ by two.  Therefore we have reduced the correlator to that of the form \eqref{3ptve} with $W^{\ell_2-\ell_1-m}$ being either $G^{\A A}_{-r}$ or removed completely. In the former case, we can perform the replacement \eqref{repG} and contour deformation again and conclude the reduced three-point function vanishes. In the latter case, the resulting 3-point correlator also vanishes due to $SU(2)_R$ invariance. This completes the argument.

\section{$\cN=4$ Superconformal Blocks}\label{sec:block}

For the purpose of bootstrapping the K3 CFT, we will need the sphere four-point superconformal block of the small ${\cal N}=4$ superconformal algebra of central charge $c=6$, with the four external primaries being those of BPS representations with $(h,j)=({1\over 2},{1\over 2})$ in the NS sector, or equivalently by spectral flow, BPS representations with $(h,j)=({1\over 4},0)$ in the R sector. The intermediate representation will be taken to be that of a non-BPS primary of weight $h$ (and necessarily $SU(2)_R$ spin 0). Let us denote the NS BPS primary by ${\cal O}^\pm$ (exhibiting the left $SU(2)_R$ doublet index only), and the Ramond BPS primary by $\phi^R$. We shall denote the chiral-anti-chiral ${\cal N}=4$ superconformal block\footnote{By a contour argument similar to the one in Section~\ref{sec:review}, one can show there is only one independent OPE coefficient between two BPS superconformal primaries.
}
associated with an NS sector BPS correlator of the form $\langle {\cal O}^+(z){\cal O}^-(0){\cal O}^+(1){\cal O}^-(\infty)\rangle$ by ${\cal F}^{{\cal N}=4, NS}_h(z)$, and the corresponding block with R sector external primaries, associated with a correlator of the form $\langle\phi^R(z) \phi^R(0) \phi^R(1) \phi^R(\infty) \rangle$, by ${\cal F}^{{\cal N}=4, R}_h(z)$. The NS and R sector blocks are related by
\ie
{\cal F}^{{\cal N}=4, NS}_h(z) = z^{-{1\over 2}} (1-z)^{1\over 2} {\cal F}^{{\cal N}=4, R}_h(z).
\fe
Note that the $j={1\over 2}$ BPS representation does not appear in the superconformal block decomposition of the BPS four-point function in the K3 CFT, because neither the ${1\over 2}$-BPS nor the ${1\over 4}$-BPS operators appear in the OPE of a pair of ${1\over 2}$-BPS primaries, as demonstrated in the previous section. The identity representation superconformal block, on the other hand, can simply be obtained by taking the $h\to 0$ limit of ${\cal F}^{{\cal N}=4}_h(z)$.

\begin{figure}
\centering
\includegraphics[scale=1]{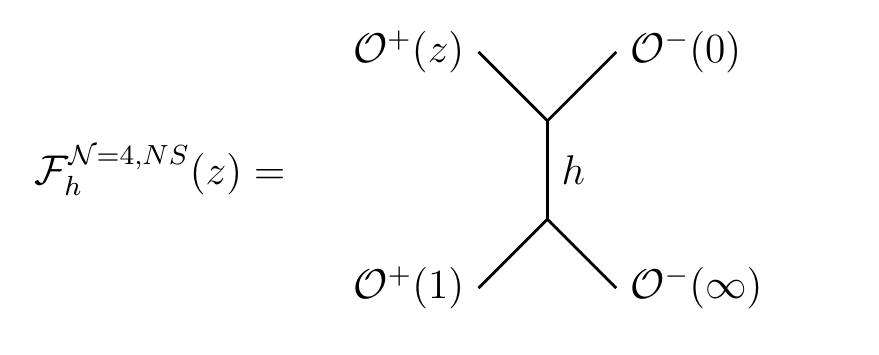}
\caption{The chiral-anti-chiral $c=6$ NS $\mathcal{N}=4$ superconformal block with external BPS primaries $\mathcal{O}^\pm$ and intermediate non-BPS primary of weight $h$.}
\end{figure}
\bigskip

\noindent{\bf Claim:} \textit{The chiral-anti-chiral $c=6$ ${\cal N}=4$ superconformal block with BPS external primaries and internal non-BPS primary of weight $h$ is identified with the {\it bosonic} Virasoro conformal block of central charge $c=28$, with external primaries of weight 1, and shifted weight $h+1$ for the internal primary, through the relation
\ie\label{idkey}
\boxed{
{\cal F}^{{\cal N}=4, R}_h(z) = z^{1\over 2} (1-z)^{1\over 2} F^{Vir}_{c=28} (1,1,1,1;h+1;z).}
\fe
Here $F^{Vir}_c(h_1,h_2,h_3,h_4;h';z)$ denotes the sphere four-point Virasoro conformal block with central charge $c$, external weights $h_i$, and internal weight $h'$.}\footnote{We will omit the $\mathcal{N}=4$ superscript for the $\mathcal{N}=4$ superconformal blocks from now on, but keep the superscript $Vir$ for the bosonic Virasoro conformal blocks.}\footnote{A similar relation between superconformal blocks and non-SUSY blocks with shifted weights was found in \cite{Fitzpatrick:2014oza,Khandker:2014mpa,Bobev:2015jxa,Lemos:2015awa} for SCFTs in $d>2$.}

We will discuss an explicit check of \eqref{idkey} on the $z$-expansion coefficients of the conformal block in Section \ref{sec:N=2block}.
\bigskip

\subsection{${\cal N}=4$ $A_{k-1}$ Cigar CFT}
We will justify the above claim by inspecting the ${\cal N}=4$ $A_1$ cigar CFT, which can be described as a $\mathbb{Z}_2$ orbifold of the ${\cal N}=2$ superconformal coset $SL(2)/U(1)$ at level $k=2$.\footnote{The $k$ of the $\mathcal{N}=4$ $A_{k-1}$ cigar CFT is  not to be confused with the level $k'$ of the $\mathcal{N}=4$ algebra. In particular, the $A_{k-1}$ cigar CFT has $c=6$ and hence $k'=1$ for its $\mathcal{N}=4$ algebra.} This is a special case of the ${\cal N}=4$ $A_{k-1}$ cigar CFT, constructed as a $\mathbb{Z}_k$ orbifold of the product of ${\cal N}=2$ coset SCFTs \cite{Ooguri:1995wj,Kutasov:1995te, Giveon:1999px},
\ie
SL(2)_k/U(1)\times SU(2)_k/U(1).
\fe
 The ${\cal N}=4$ $A_{k-1}$ cigar theory has $4(k-1)$ normalizable weight $({1\over 2},{1\over 2})$ BPS primaries, corresponding to $4(k-1)$ exactly marginal deformations,\footnote{In the 6d $A_{k-1}$ IIA little string theory, they parametrize the Coulomb branch moduli space ${\mathbb R}^{4(k-1)}/S_k$.} and a continuum of delta function normalizable non-BPS primaries above the gap
\ie\label{Ddivcigar}
\Delta_{cont} = {1\over 2k}
\fe
 in the scaling dimension. Later when we consider a sector of primaries with nonzero R-charges, the continuum develops above a gap of larger value and there may also be discrete, normalizable non-BPS primaries below the gap. The continuum states are in correspondence with those of the supersymmetric $SU(2)_k\times \mathbb{R}_\phi$ CFT, where $\mathbb{R}_\phi$ is a linear dilaton, with background charge $1/\sqrt{k}$, which describes the asymptotic region of the cigar.

\subsection{Four-Point Function and the Ribault-Teschner Relation}
\label{rite}

Let us recall the computation of the sphere four-point function of the BPS primaries in the $A_{k-1}$ cigar CFT, studied in \cite{Chang:2014jta}. The weight $({1\over4},{1\over 4})$ ${1\over 2}$-BPS RR sector primaries lie in the twisted sectors of the $\mathbb{Z}_k$ orbifold, labeled by an integer $\ell+1$, with $\ell=0,1,\cdots,k-2$. Note that $\ell+1$ is also the charge with respect to a $\widetilde{\mathbb{Z}_k}$ symmetry that acts on the twisted sectors, and is conserved modulo $k$. They can be constructed from $SL(2)$ and $SU(2)$ coset primaries as either
\ie\label{VR+}
& V_{R,\ell}^+ = V^{s\ell, (-{1\over 2},-{1\over 2})}_{{\ell\over 2},{\ell+2\over 2},{\ell+2\over 2}} V^{su, ({1\over 2},{1\over 2})}_{{\ell\over 2},{\ell\over 2},{\ell\over 2}},
\fe
or
\ie\label{VR-}
& V_{R,\ell}^- = V^{s\ell, ({1\over 2},{1\over 2})}_{{\ell\over 2},-{\ell+2\over 2},-{\ell+2\over 2}} V^{su, (-{1\over 2},-{1\over 2})}_{{\ell\over 2},-{\ell\over 2},-{\ell\over 2}}.
\fe
Here $V^{s\ell, (\eta,\bar\eta)}_{j,m,\bar m}(z,\bar z)$ and $V^{su,(\eta',\bar\eta')}_{j',m',\bar m'}(z,\bar z)$ are the spectral flowed primaries in the $SL(2)/U(1)$ and $SU(2)/U(1)$ coset CFTs, respectively. $\eta, \bar \eta$ and $\eta',\bar \eta'$ are the spectral flow parameters in the $\mathcal{N}=2$ $SL(2)/U(1)$ and $SU(2)/U(1)$. The holomorphic weight of the $\mathcal{N}=2$ $SL(2)_k/U(1)$ coset primary $V^{sl,\eta}_{j,m}(z)$ is
\ie\label{slweight}
 {-j (j+1) +(m+\eta)^2 \over k }+{\eta^2\over2} ,
\fe
while the holomorphic weight of the $\mathcal{N}=2$ $SU(2)_k/U(1)$ coset primary $V^{su,\eta'}_{j',m'}(z)$ is
\ie\label{suweight}
 {j' (j'+1) -(m'+\eta')^2 \over k }+{\eta'^2\over2} ,
\fe
 We have the identification $V^-_{R,\ell} = V^+_{R,k-2-\ell}$. 

The correlator of interest is
\ie
& \left\langle V_{R,\ell}^+(z,\bar z) V_{R,\ell}^+(0) V_{R,\ell}^-(1) V_{R,\ell}^-(\infty) \right\rangle,
\fe
where the operators are arranged so that the $\widetilde{\mathbb{Z}_k}$ charge is conserved. 
The $SL(2)/U(1)$ part of the correlator was determined in \cite{Chang:2014jta}, using Ribault and Teschner's relation \cite{Ribault:2005wp} between the bosonic $SL(2)$ WZW and Liouville correlators. The result is of the form (see (3.37) and (3.39) of \cite{Chang:2014jta})\footnote{Note that the identity block does not show up in the cigar CFT four-point function because the identity operator is non-normalizable.  This can also be understood from the normalization when compared with the K3 CFT discussed in Section \ref{Sec:Cigar}.}
\ie\label{bpsr}
&{\cal N} |z|^{{(\ell+1)^2\over k} +{1\over2} } |1-z|^{\ell+{3\over2}-{(\ell+1)^2\over k}} 
\\
& \times \int_0^\infty {dP\over 2\pi} C(\A_1,\A_2, {Q\over 2}+iP) C(\A_3,\A_4, {Q\over 2}-iP) |F^{Vir}(h_1,h_2,h_3,h_4;h_P;z)|^2.
\fe
Here $F^{Vir}(h_1,\cdots;h_P;z)$ is the Virasoro conformal block with central charge $c=1+6Q^2$. ${\cal N}$ is a normalization constant. $Q$ is the background charge of a corresponding bosonic Liouville theory, and $\A_i$ are the exponents labeling Liouville primaries of weight $h_i = \alpha_i ( Q- \alpha_i)$. They are related to $k$ (labeling the $A_{k-1}$ cigar theory) and $\ell$ (labeling the BPS primaries) by
\ie\label{qkl}
& Q=b+{1\over b}, ~~~b^2 = {1\over k},
\\
& \A_1=\A_2 = {\ell+2\over 2} b,~~~ \A_3=\A_4 = {k-\ell\over 2}b,
\\
& h_1 = h_2 = {(\ell+2)(2k-\ell)\over 4k},
\\
& h_3 = h_4 = {(k+\ell+2)(k-\ell)\over 4k}.
\fe
Note that the Liouville background charge $Q$ is {\it not} the same as the background of the asymptotic linear dilaton in the original cigar CFT (which is $1/\sqrt{k}$). The weight of the  intermediate continuous state in the Liouville theory is
\ie\label{intermediate}
h_P = \alpha_P (Q- \alpha_P),~~~\alpha_P = {Q\over 2} + i P,~~~P\in \mathbb{R}.
\fe
$C(\A_1,\A_2,\A_3)$ is the structure constant of Liouville theory \cite{Zamolodchikov:1995aa,Teschner:1995yf},
\ie\label{liouville3}
C(\A_1,\A_2,\A_3) = \widetilde\mu^{Q-\sum \A_i\over b} {\Upsilon_0 \prod_{i=1}^3 \Upsilon(2\A_i)\over \Upsilon(\sum\A_i-Q) \Upsilon(\A_1+\A_2-\A_3) \Upsilon(\A_2+\A_3-\A_1) \Upsilon(\A_3+\A_1-\A_2)},
\fe
where $\tilde \m =\pi \m \C(b^2) b^{2-2b^2}$ is the dual cosmological constant to $\m$ with $\C(x)=\Gamma(x)/\Gamma(1-x)$, $\Upsilon_0\equiv \Upsilon'(0)$, and 
\ie
\Upsilon(\A) = {\Gamma_2(Q/2|b,b^{-1})^2 \over  \Gamma_2(\A |b,b^{-1}) \Gamma_2(Q-\A |b,b^{-1}) }.
\fe
Here $\Gamma_2(x|a_1,a_2)$ is the Barnes double Gamma function \cite{1901}. $\Upsilon(\A)$ has zeroes at 	$\A=-nb-m/b$ and $\A=(n+1)b+(m+1)/b$, for integer $n,m\geq 0$.

 The integration contour in \eqref{bpsr} is the standard one if $\A_i$ lie on the line ${Q\over 2}+i\mathbb{R}$. We need to analytically continue $\A_i$ to the real values given above. In doing so, the integral may pick up residues from poles in the Liouville structure constants. These residue contributions, if present, correspond to discrete intermediate state contributions \cite{Maldacena:2001km}. We will have more to say about these discrete intermediate state contributions to the four-point function \eqref{bpsr} in the $\mathcal{N}=4$ $A_{k-1}$ cigar CFT  in Section \ref{sec:mixed}.

\subsection{Four-Point Function of the $\mathcal{N}=4$ $A_1$ Cigar CFT}
\label{fpff}

Now we shall specialize to the $A_1$ theory (i.e. $k=2$). In this case, the asymptotic region of the cigar CFT is simply given by one bosonic linear dilaton $\mathbb{R}_\phi$, with background charge ${1\over \sqrt{2}}$, and 4 free fermions. Note that the non-BPS ${\cal N}=4$ character with $c=6$ (and necessarily with $SU(2)_R$ spin $j=0$) is identical to the oscillator partition function of one chiral boson and 4 free fermions. Thus, the non-BPS superconformal primaries of the ${\cal N}=4$ $A_1$ cigar CFT are in one-to-one correspondence with exponential operators in the bosonic part of the asymptotic linear dilaton CFT, of the form 
\ie
V_\A=e^{2\A\phi},~ \text{with}~ \A= {1\over 2\sqrt{2}}+iP, ~P\in\mathbb{R}.
\fe
 Importantly, these non-BPS primaries are labeled by the same quantum number, a real number $P$, as the intermediate Liouville primaries in  \eqref{bpsr}.

The result (\ref{bpsr}) that expresses the BPS four-point function in terms of Virasoro conformal blocks labeled by the Liouville primaries $V_\A$ then strongly suggests that in the $A_1$ theory, the ${\cal N}=4$ superconformal block decomposition is identical to the decomposition (\ref{bpsr}) in terms of Virasoro conformal blocks. Here, the Virasoro block is that of central charge 
\ie
c=1+6Q^2=28,
\fe
 with external weights $h_i=1$ ((\ref{qkl}) with $k=2$, $\ell=0$). Next, we want to relate the intermediate Liouville primary with weight $h_P$ to the corresponding $\mathcal{N}=4$ non-BPS primaries in the $A_1$ cigar CFT. The non-BPS ${\cal N}=4$ primary, in the $SL(2)/U(1)$ coset description, would be constructed from an $SL(2)$ primary of spin\footnote{The $\sqrt{2}$ is introduced to match with the convention in \eqref{bpsr}.}
 \ie\label{contcoset}
 j=-{1\over 2}-i\sqrt{2}P,~~  P\in\mathbb{R},
 \fe
  with conformal weight 
\ie\label{interh}
h = -{j(j+1)\over k} = {1\over 8} +P^2.
\fe
On the other hand, by the relation of Ribault and Teschner (see also (3.17) of \cite{Chang:2014jta}), the intermediate Liouville primary in (\ref{bpsr}) is labeled by the exponent $\A_P$ given by
\ie
& \A_P = -b j + {1\over 2b} = {Q\over 2}+iP.
\fe
Using \eqref{intermediate}, we obtain the weight of the intermediate Liouville primary in terms of $P$ labeling the $SL(2)_k/U(1)$ coset states in \eqref{contcoset},
\ie\label{interhp}
h_P={9\over 8}+P^2.
\fe
This leads us to identify the relation between the Virasoro primary weight $h_P$ and the weight of the non-BPS primary in the corresponding ${\cal N}=4$ superconformal block,
\ie
h_P=h+1.
\fe
Including the $z$-dependent prefactor in (\ref{bpsr}) in (the $k=2$, $\ell=0$ case), and matching the normalization in the $z\to 0$ limit, we then deduce the relation (\ref{idkey}).


\section{${\cal N}=2$ Superconformal Blocks}\label{sec:N=2block}

The $c=6$ ${\cal N}=4$ superconformal block with BPS external primaries is in fact identical to the chiral-anti-chiral channel superconformal block of the ${\cal N}=2$ subalgebra.\footnote{We thank Sarah Harrison for a discussion on this issue.} This follows from the fact that a non-BPS weight $h$ representation of the ${\cal N}=4$ SCA decomposes into an infinite series of ${\cal N}=2$ non-BPS representations of weight $h+{m^2\over 2}$ and $U(1)_R$ charge $m$ \cite{Eguchi:1988af}, with $m=0,1,\cdots$. By a similar contour argument as in Section~\ref{sec:review}, only the $U(1)_R$ neutral ${\cal N}=2$ primaries and their descendants can appear in the OPE of the external chiral operator $\phi^+$ and anti-chiral operator $\phi^-$,\footnote{One can in fact reach a more general statement based on $\cN=2$ SCA. The OPE of two (anti)chiral primaries with $U(1)_R$ charge $q_1$ and $q_2$ can only contain a primary (and descendants of) with $U(1)_R$ charge $q_3$ if $q_1\leq 0$ and $q_1+q_2-q_3 \leq 0$ or $q_1\geq 0$ and $q_1+q_2-q_3 \geq 0$. In particular when we consider the OPE of one chiral and one antichiral primaries with opposite $U(1)_R$ charges, only the $U(1)_R$ neutral primaries (and descendants) can appear.} hence the claim.

\begin{figure}
\centering
\includegraphics[scale=1]{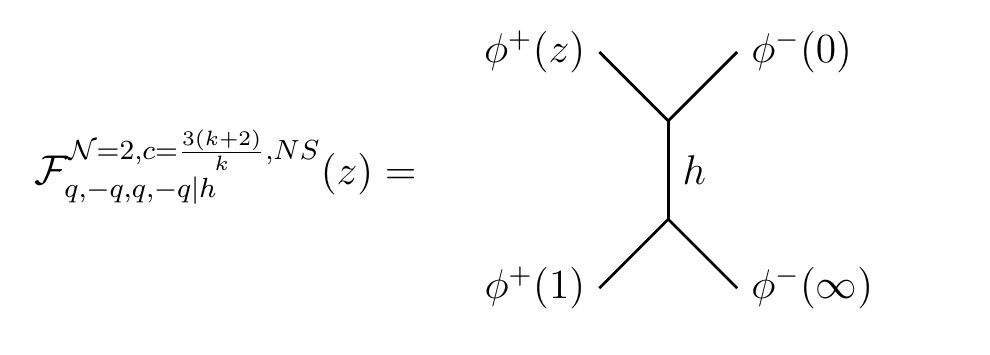}
\caption{The chiral-anti-chiral $c={3(k+2)\over k}$ NS $\mathcal{N}=2$ superconformal block with external chiral/anti-chiral primaries ${\phi}^\pm$ of weight ${|q|\over 2}$ and $U(1)_R$ charge $\pm q = \pm\left( {\ell+2\over k}\right)$, and intermediate $U(1)_R$ neutral  non-BPS primary of weight $h$.}
\end{figure}

More generally, one can extract the chiral-anti-chiral NS superconformal block of a general ${\cal N}=2$ SCA with central charge $c={3(k+2)\over k}$ from the ${\cal N}=2$ $SL(2)_k/U(1)$ cigar CFT. For instance, by a similar argument as in sections \ref{rite} and \ref{fpff}, one can show that the $c={3(k+2)\over k}$ ${\cal N}=2$ superconformal block with external chiral or anti-chiral operators of weight ${|q|\over 2}$ and $U(1)_R$ charge $q$, $-q$, $q$, $-q$,  with\footnote{Under spectral flow, the  NS sector chiral primaries are mapped to R sector ground states with R-charges
\ie
q={\ell+1\over k}-{1\over 2},\quad \ell=0,1,\cdots,k-2.
\fe } 
\ie
\label{qv}
q = {\ell+2\over k},~~~~\ell=0,1,\cdots,k-2,
\fe
and the internal $U(1)_R$ neutral non-BPS primary with weight $h$, 
is related to the bosonic Virasoro conformal block of central charge $c=13+6k+{6\over k}$ by 
\ie\label{nteq}
\boxed{
 {\cal F}^{{\cal N}=2, \,c={3(k+2)\over k}, \,NS}_{q,-q,q,-q|h}(z) = z^{(\ell+2) (k-\ell-2) \over 2k}(1-z)^{(\ell+2) (3k-2\ell-4) \over 2k} F^{Vir}_{c=13+6k+{6\over k}}(h_q, h_{-q}, h_{q}, h_{-q};h+{k+2\over 4};z),}
\fe
where
\ie
& h_q = {(\ell+2)(2k-\ell)\over 4k},~~~~ h_{-q} = {(k-\ell)(k+\ell+2)\over 4k} .
\fe
Note that in the special case when $k=2$ and $\ell=0$, the $\mathcal{N}=2$ block becomes identical  to the $\mathcal{N}=4$ block as argued above and \eqref{nteq} reduces to the claim \eqref{idkey}. 
The shift in the intermediate weight $h_P = h+ {k+2\over4}$ comes from the difference between $Q^2/4$ and $1/4k$, similar to \eqref{interh} and \eqref{interhp} in the $k=2$ case.  We have checked directly using  Mathematica that (\ref{nteq}) (and therefore \eqref{idkey} as a special case) holds up to level 4 superconformal descendants with various values of $q$ in \eqref{qv}.  We expect \eqref{nteq} to hold for (anti)chiral primaries with general $U(1)_R$ charges and central charge $c=3(k+2)/k$ by analytic continuation in $\ell$ and $k$. 

 The details of extracting the BPS ${\cal N}=2$ superconformal blocks of general central charge from the cigar CFT will be presented elsewhere.

\section{The Integrated Four-Point Functions}\label{sec:integrated}

In this section we discuss the integrated four-point function of ${1\over2}$-BPS operators, whose exact moduli dependence will be later incorporated into the bootstrap equations (see Section \ref{Sec:Gapa1111} and  Appendix \ref{app:Aijkl}).  The integrated sphere four-point functions $A_{ijkl}$ and $B_{ij,kl}$ are defined as \cite{Lin:2015dsa}\footnote{More precisely, this integral is defined by analytic continuation in $s,t$ from the region where it converges.}
\ie\label{4ptint}
& \int d^2z |z|^{-s-1}|1-z|^{-t-1}  \left\langle \phi^{RR}_i(z,\bar z) \phi^{RR}_j(0) \phi^{RR}_k(1) \phi^{RR}_\ell(\infty) \right\rangle 
\\
&= 2\pi \left( {\delta_{ij}\delta_{k\ell}\over s}+{\delta_{ik}\delta_{j\ell}\over t}  + {\delta_{i\ell}\delta_{jk}\over u}\right) + A_{ijk\ell} + B_{ij,k\ell} s + B_{ik,j\ell} t + B_{i\ell,jk} u
 + {\cal O}(s^2,t^2,u^2),
\fe
where $\phi_i^{RR}$ are the RR sector ${1\over2}$-BPS primaries of weight $({1\over 4},{1\over 4})$ that are related to NS-NS ${1\over2}$-BPS primaries ${\cal O}_i^{\pm\pm}$ by spectral flow, and the variables $s,t,u$ are subject to the constraint $s+t+u=0$. $A_{ijkl}$ by definition is symmetric in $(ijkl)$. $B_{ij,kl}$ is symmetric in $(ij)$, $(kl)$, and under the exchange $(ij)\leftrightarrow (kl)$. Furthermore, $B_{ij,kl}$ is subject to the constraint $B_{ij,kl} + B_{ik,lj} + B_{il,jk}=0$. $A_{ijk\ell}$ is also known as the tree-level $\mathcal{N}=4$ topological string amplitude \cite{Berkovits:1994vy,Antoniadis:2006mr}.

The first term in \eqref{4ptint} is related to the tree-level amplitude of tensor multiplets in type IIB string theory compactified by K3 at two-derivative order. In particular, it captures the Riemannian curvature of the Zamolodchikov metric on the K3 CFT moduli space. Moreover $A_{ijkl}$ and $B_{ij,kl}$ can be identified as the tree level amplitudes of tensor multiplets in the 6d $(2,0)$ supergravity at 4- and 6-derivative orders respectively. They can be obtained from the weak coupling limit of the exact results for the 4- and 6-derivative order tensor effective couplings determined in \cite{Kiritsis:2000zi,Lin:2015dsa}. In this paper, we will make use of
\ie\label{aexp}
A_{ijk\ell} = {1\over 16\pi^2} \left.{\partial^4\over \partial y^i \partial y^j\partial y^k \partial y^\ell}\right|_{y=0} \int_{\cal F} d^2\tau { \Theta_\Lambda(y|\tau,\bar\tau)\over \eta(\tau)^{24}},
\fe
where ${\cal F}$ is the fundamental domain of $PSL(2,\mathbb{Z})$ acting on the upper half plane, $\Lambda$ is the even unimodular lattice $\Gamma_{20,4}$ embedded in $\mathbb{R}^{20,4}$, which parameterizes the moduli of the K3 CFT, and the theta function $\Theta_\Lambda$ is defined to be
\ie
\Theta_\Lambda(y|\tau,\bar\tau) = e^{{\pi\over 2\tau_2} y^2} \sum_{\ell\in\Lambda} e^{\pi i \tau \ell_L^2 - \pi i \bar\tau \ell_R^2+2\pi i \ell_L\cdot y}.
\fe
Here $\ell_L$ and $\ell_R$ are the projection of the lattice vector $\ell$ onto the positive subspace $\mathbb{R}^{20}$ and negative subspace $\mathbb{R}^4$ respectively. The lattice inner product is defined as $\ell\circ\ell=\ell_L^2-\ell_R^2$. $y$ is an auxiliary vector in the $\mathbb{R}^{20}$, whose components are in correspondence with the 20 BPS multiplets of the K3 CFT. Note that in (\ref{aexp}), the integral is modular invariant only after taking the $y$-derivatives and restricting to $y=0$.

The expression (\ref{aexp}) is obtained from the weak coupling limit of (1.3) in \cite{Lin:2015dsa} (by decomposing $\Gamma_{21,5}=\Gamma_{20,4}\oplus\Gamma_{1,1}$, and taking a limit on the $\Gamma_{1,1}$). The normalization can be fixed by comparison with an explicit computation of twist field correlators in the $T^4/\mathbb{Z}_2$ free orbifold CFT, as shown in Appendix~\ref{app:T4}. There is an analogous formula for $B_{ij,kl}$ as an integral of ratios of modular forms over the moduli space of a genus two Riemann surface.

If we assume that all non-BPS primaries have scaling dimension above a gap $\Delta$,\footnote{Note that the assumption of a nonzero gap holds in the singular CFT limits where the K3 develops ADE type singularities, but obviously fails in the large volume limit.} one can derive an inequality between the integrated four-point function $A_{1111}$ of a single ${1\over 2}$-BPS primary $\phi_1$, and the four-point function $f(z,\bar z)$ itself evaluated at a given cross ratio, say $z={1\over 2}$, of the form (see Appendix~\ref{Sec:Aijkl1})
\ie
\label{bounda1111}
A_{1111} \leq 3A_0 + M(\Delta) \left[ f(1/2)  -f_0 \right].
\fe
Here $A_0$ and  $f_0$ are constants, and $M(\Delta)$ is a function of $\Delta$ that goes like $1/\Delta$ in the $\Delta\to 0$ limit. Since $A_{1111}$ is known as an exact function of the moduli, this inequality will provide a lower bound on $f({1\over 2})$ over the moduli space. In particular, it can be used to show that $f({1\over 2})$ diverges in the singular CFT limits.

\section{Special Loci on the K3 CFT Moduli Space}

Some loci on the moduli space of the K3 CFT are more familiar to us, such as near the free orbifold points\footnote{Free $T^4$ orbifold points on the moduli space of the K3 CFT fall in the following classes: $T^4/\bZ_2$, $T^4/\bZ_3$, $T^4/\bZ_4$ and $T^4/\bZ_6$. They share similar qualitative features and we will only discuss the $T^4/\bZ_2$ case in detail here.} and  where ADE singularities develop.  This section reviews certain properties of the K3 CFT at these special points, that will allow us to check the consistency of our bootstrap results in Section~\ref{Sec:Bootstrap}.  In fact, some of the examples we discuss here will saturate the bounds from bootstrap analysis.

\subsection{$T^4/\bZ_2$ Free Orbifold}
\label{Sec:FreeOrbifold}

\begin{figure}[t]
\centering
\subfigure{
\includegraphics[width=.4\textwidth]{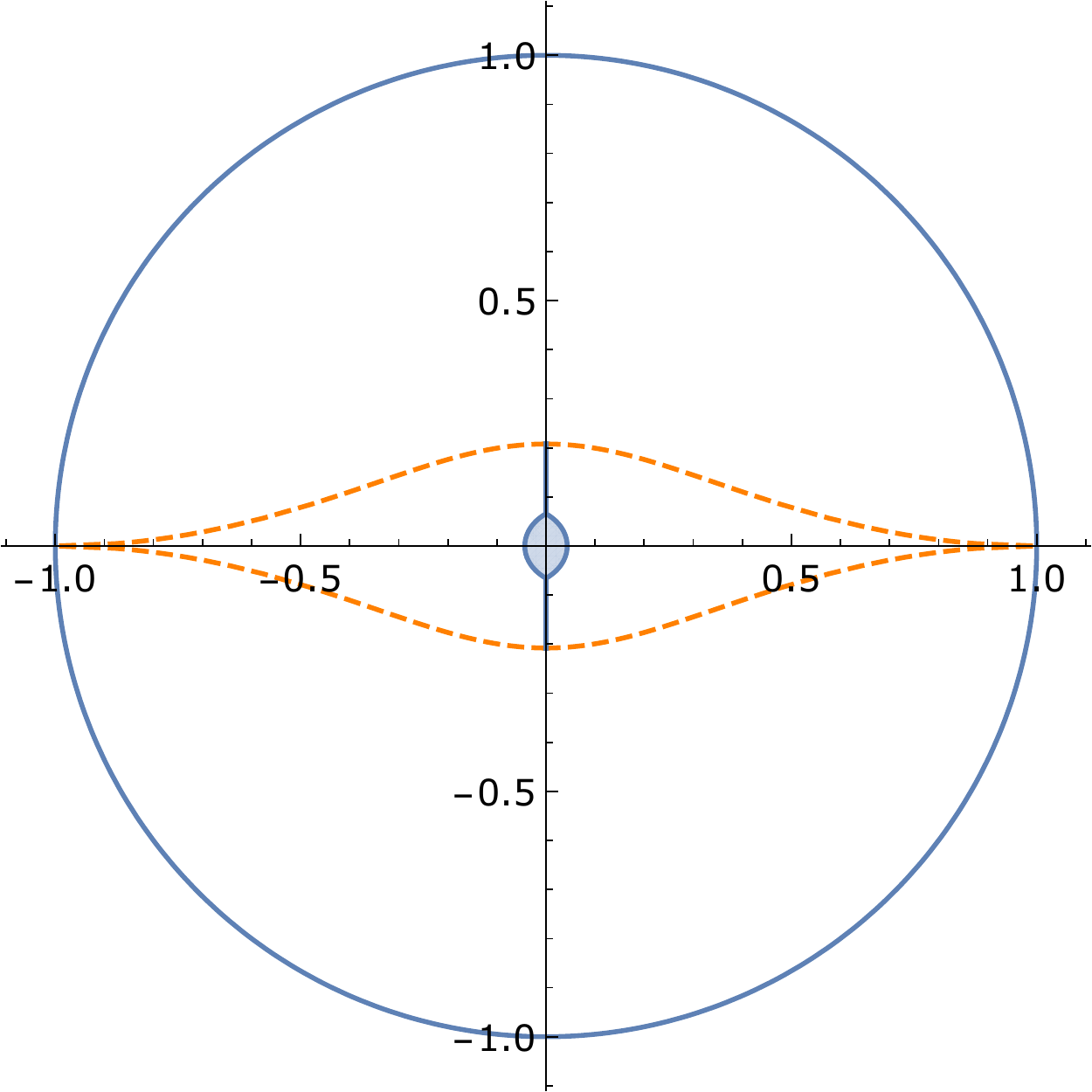}
}
~~
\subfigure{
\includegraphics[width=.4\textwidth]{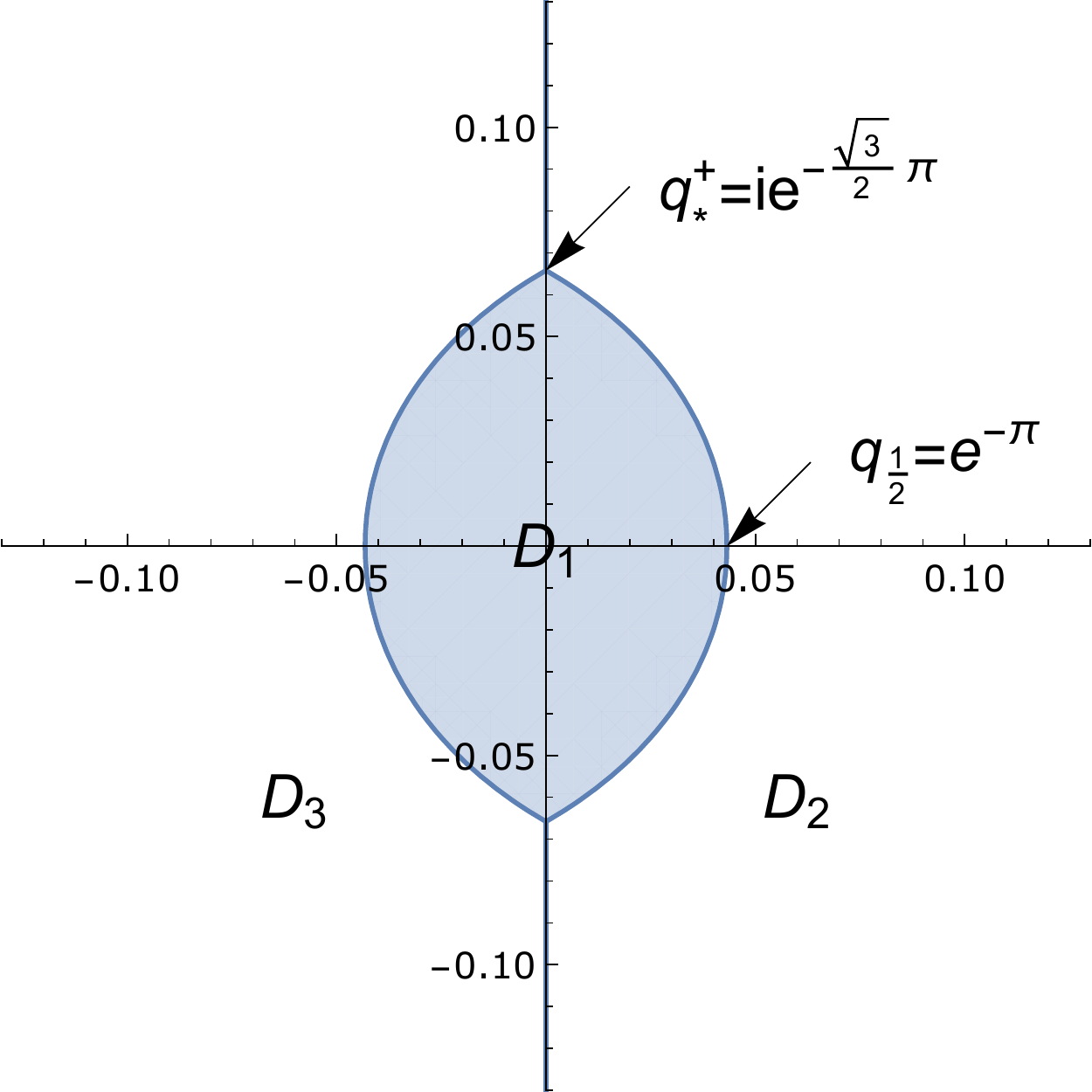}
}
\caption{The eye-shaped region bounded by the dashed line is the range of $q(z)$ under one branch of the $q$-map \eqref{qmap}.  The regions $D_1$, $D_2$ and $D_3$ each contains two fundamental domains of the $S_3$ crossing symmetry group.  See Appendix~\ref{Sec:qmap}.}
\label{fig:qmap}
\end{figure}

There is a locus on the K3 CFT moduli space that corresponds to the $\bZ_2$ free orbifold of a rectangular $T^4$ of radii $(R_1, R_2, R_3, R_4)$.  Let us first consider the twisted sector ground state in the RR sector $\sigma(z, \bar z)$, associated with one of the $\bZ_2$ fixed points.  Its OPE with itself will receive contributions from all states in the untwisted sector with even winding number\cite{Dijkgraaf:1987vp}, which has a gap of size $1/\max(R_i)^2$ (here we adopt the convention $\A'=2$).  The four-point function of $\sigma(z, \bar z)$ is \cite{Dixon:1986qv, Gluck:2005wr}
\ie\label{T4twist4pt}
f(z, \bar z) = {|z(1-z)|^{-1} \over |F(z)|^4} \sum_{(p_L, p_R) \in \Lambda} q(z)^{p_L^2 \over 2} \bar q(\bar z)^{p_R^2 \over 2}, 
\fe
where\footnote{Our convention for $\theta_3(q)$ is $
\theta_3(q) =  \sum_{n\in\mathbb{Z}} q^{ n^2}$, 
with $q=e^{i\pi\tau}$.
} $q(z) = \exp(i \pi \tau(z))$, $\tau(z) = i {F(1-z) / F(z)}$, $F(z) = {}_2F_1({1\over2},{1\over2},1|z) = [\theta_3(q(z))]^2$, and the lattice $\Lambda=\{(p^i_L,p^i_R)=\{({n^i\over R_i}+ {m^i R_i\over 2},{n^i\over R_i}- {m^iR_i\over 2})|n^i\in {\mathbb Z},m^i\in {2\mathbb Z}\}$ which is $\sqrt 2$ times the $(4,4)$ Narain lattice for a rectangular $T^4$ with different radii $R'_i=\sqrt{2}R_i$. Note again that the untwisted sector operators with odd winding numbers are absent in \eqref{T4twist4pt} due to the selection rule in the orbifold theory \cite{Dijkgraaf:1987vp}. The map $z \to q(z)$ is due to Zamolodchikov \cite{Zamolodchikov:1985ie,Zamolodchikov:1995aa} and is explained further in Appendix~\ref{Sec:qmap}.  The range of this $q$-map is shown in Figure~\ref{fig:qmap}.

The four-point function evaluated at $z = 1/2$ has a particularly simple expression
\ie
f(1/2, 1/2) = {4 \over |F(1/2)|^4} \prod_{i = 1}^4 | \theta_3(e^{- \pi {R }_i^2}) \theta_3(e^{- \pi / { R }_i^2}) | \geq 4.
\fe
The minimal value is achieved by a square $T^4$ at radius $R_i=1$ (or ${R'_i} = \sqrt{2}$). Note that $R_i=1$ is not the self-dual point for the $T^4$ since we set $\alpha'=2$.  Later in Section \ref{Sec:Gap} and Section \ref{Sec:Gapa1111}, we will compare the twisted sector four-point function with our bootstrap bounds on the gap in the spectrum.

Next let us consider the four-point function of untwisted sector operators. The NS sector ${1\over2}$-BPS operators in the untwisted sector can be built from the free fermions $\psi^{\A A}(z)$, which satisfy the OPE
\ie
\psi^{\A A} (z) \psi ^{\B B} (0 ) \sim { \epsilon^{\A\B} \epsilon^{AB} \over z}.
\fe
From the bilinears of $\psi^{\A A}$ we have either the $SU(2)_R$ current $\psi^{\A A}\psi^{\B B}\epsilon_{AB}$ which is an $\cN=4$ descendant of identity or the current $\psi^{\A A}\psi^{\B B}\epsilon_{\A\B}$ which is a weight $(1,0)$ non-BPS superconformal primary.

Consider a single ${1\over 2}$-BPS operator in the untwisted sector of the free orbifold theory  ${\cal O}^{\pm\pm}={1\over 2}\psi^{\pm A}\widetilde{\psi}^{\pm B}\epsilon_{AB}$. Its four-point function is,
\ie\label{ff4pt}
&f(z,\bar z)=\la{\cal O}^{++}(z,\bar z){\cal O}^{--}(0){\cal O}^{++}(1){\cal O'}^{--}(\infty)\ra
={1\over z\bar z}+1-{1\over 2 z}-{1\over 2\bar z}.
\fe
In the OPE between $\mathcal{O}^{++}$ and $\mathcal{O}^{--}$, the lowest non-identity primary is $\epsilon_{AB}\epsilon_{CD} : \psi^{+ A}\widetilde{\psi}^{+ B}\psi^{- C}\widetilde{\psi}^{-D}:$ of weight (1,1). This will show up as a special example in Section \ref{Sec:Gap} and Section \ref{Sec:Gapa1111} when we study the bootstrap constraint on the gap in the spectrum. Note that the integrated four-point function $A_{1111}$ at the free orbifold point $T^4/\mathbb{Z}_2$ is zero as can be checked explicitly from \eqref{4ptint} and \eqref{explicitAijkl}.

More generally, we can consider two ${1\over 2}$-BPS operators $\phi_1^{\pm\pm}$ and $\phi_2^{\pm\pm}$ in the untwisted sector,
\ie
& \phi_1^{\pm\pm} \equiv \psi^{\pm A} \widetilde\psi^{\pm B} M_{AB}, \quad \ \phi_2^{\pm\pm} \equiv \psi^{\pm A} \widetilde\psi^{\pm B} \overline M_{AB},
\fe
where $M_{AB}$ and $\overline M_{AB}$ are some independent general $2 \times 2$ complex matrices. 
Below we will show that if the identity block is absent in the OPE of a ${1\over 2}$-BPS primary $\phi_i^{\pm\pm}$ in the untwisted sector with itself, the $(1,0)$ non-BPS primary must appear in the OPE of $\phi_i^{\pm\pm}$ with  any other ${1\over 2}$-BPS primary $\phi_j^{\pm\pm}$ in the untwisted sector if the identity block  appears there. The OPE coefficient of the identity block in the $\phi_1\phi_1$ OPE is proportional to $\det(M)$, whereas that in the $\phi_1\phi_2$ OPE is proportional to $\epsilon^{AB}\epsilon^{CD}M_{AC}\overline M_{BD}$. Therefore, we require $\det(M)=0$ but $\epsilon^{AB}\epsilon^{CD}M_{AC}\bar M_{BD}\neq 0$ to exclude the identity in the $\phi_1\phi_1$ OPE but not in the $\phi_1\phi_2$ OPE. If the $(1,0)$ primary is absent in the $\phi_1\phi_2$ channel, we require $\epsilon^{CD}M_{AC}\bar M_{BD}\propto  \epsilon_{AB}$  with a nonzero proportionality constant. This is in contradiction with $\det(M)=0$.

In this case, the lowest primary in the $\phi_i\phi_i$ OPE would be a $(1,1)$ non-BPS primary which combines the holomorphic $(1,0)$ primary with its antiholomorphic counterpart.
In other words, if the $\phi_i\phi_i$ channel does not contain identity whereas the $\phi_i\phi_j$ channel contains identity, $\Delta_{gap}=1$ in the $\phi_i \phi_j$ channel and $\Delta_{gap}=2$ in the $\phi_i\phi_i$ channel. As we will see in subsection~\ref{sec:mixed}, if we take $\phi_j$ to be the complex conjugate of $\phi_i$, this corresponds to a special kink on the boundary of the numerical bound for the $\la \phi\phi\bar\phi\bar\phi \ra$ correlator.

\subsection{$\mathcal{N}=4$ $A_{k-1}$ Cigar CFT}
\label{Sec:Cigar}

We have already introduced the $\mathcal{N}=4$ $A_{k-1}$ cigar CFT in Section \ref{sec:block}. Here we will focus on its continuous spectrum and divergent OPE coefficients.

We will consider the RR sector ${1\over2}$-BPS primaries $V^+_{R,\ell}$ and $V^-_{R,\ell}$ (\eqref{VR+} and \eqref{VR-}) \cite{Aharony:2003vk, Aharony:2004xn, Lin:2015dsa} with $\widetilde{\mathbb{Z}_k}$ charge $(\ell+1)$. 
Here $\ell$ ranges from 0 to $\lfloor{k-2\over2} \rfloor$. For $\ell$ between $\lfloor{k-2\over2} \rfloor+1$ and $k-2$ we  use the identification $V^-_{R,\ell}= V^+_{R,k-2-\ell}$.

\subsubsection*{Continuum in the Cigar CFT}

As already mentioned in \eqref{Ddivcigar}, in the OPE between $ V^+_{R,\ell}$ and $ V^-_{R,\ell}$, there is a continuum of delta function normalizable non-BPS primaries above  
\ie\label{phibphi}
\Delta^{\phi\bar\phi}_{cont} = {1\over2k}.
\fe
Here we have adopted the notation that will be used in subsection~\ref{sec:mixed} where we denote $V^+_{R,\ell}$ by $\phi$ and $V^-_{R,\ell}$ by $\bar\phi$.

Let us move on to  the lowest weight operator that lies at the bottom of the continuum in the OPE between $ V^+_{R,\ell}$ and $ V^+_{R,\ell}$. This operator can be factorized into the $SL(2)_k/U(1)$ and $SU(2)_k/U(1)$ parts. Let us denote the lowest holomorphic weights of the operators in the two parts by $h^{sl}$ and $h^{su}$, respectively. 

$h^{sl}$ can be determined by studying the four-point function \eqref{bpsr} together with the fusion rule in the $\mathcal{N}=2$ $SU(2)_k/U(1)$ coset.
The leading $z$ power in \eqref{bpsr} is
\ie\label{zpower}
\left[ {(\ell+1)^2\over 2k}+{1\over 4} \right]+ h_P- h_1-h_2,~~~\text{with}~~
h_P={Q^2\over 4},~h_1=h_2 = {(\ell+2)(2k-\ell)\over 4k},
\fe
where we have used \eqref{qkl} and $h_P= \alpha_P (Q-\alpha_P)$ with $\alpha_P=Q/2$ for the lowest dimension state in the continuum. Recall that $Q=\sqrt{k } +{1\over\sqrt{k}}$ is the background charge of the corresponding bosonic Liouville theory in the Ribault-Teschner relation.  Writing the four-point function \eqref{bpsr} in the conformal block expansion, \eqref{zpower} is the power of $z$ in the $\mathcal{N}=4$ superconformal block with intermediate state being the bottom state in the continuum and external states being $V^{sl,(-{1\over2} , -{1\over2} )} _{{\ell\over 2} , {\ell+2\over 2}, {\ell+2\over 2}}$ (the $SL(2)/U(1)$ part of $V_{R,\ell}^+$). The holomorphic weight of the latter is given by \eqref{slweight} to be ${1\over 4k } +{1\over8}$. Hence,
\ie
h^{sl} = \left[ {(\ell+1)^2\over 2k}+{1\over 4} \right]+ {(k+1)^2\over 4k} - {(\ell+2)(2k-\ell)\over 2k} + 2\left( {1\over 4k } +{1\over8}\right).
\fe

As for the $SU(2)/U(1)$ part, the lowest dimension intermediate operator in the OPE between two $V^{su,({1\over2}, {1\over2})} _{ {\ell\over 2} ,{\ell\over 2} ,{\ell\over2}}$ is $V^{su ,(1,1)} _{ \ell,\ell ,\ell}$, whose holomorphic weight is given by \eqref{suweight},
\ie
h^{su}  =  - {\ell+1\over k } +{1\over 2}.
\fe
Adding $h^{sl}$ and $h^{su}$ together, we obtain the lowest scaling dimension $\Delta_{cont}^{\phi\phi}$ in the continuum of the OPE channel between $V^+_{R,\ell}$ and $V^+_{R,\ell}$,
\ie\label{ppcont}
\Delta_{cont}^{\phi\phi} =  2 (h^{sl}+h^{su} )  = {(k-2\ell -1)^2\over 2k}.
\fe
As we will show below, in addition to the continuum, there are generally discrete states contributing to the four-point function \eqref{bpsr} of the cigar CFT with divergent structure constant when normalized properly.

\subsubsection*{Discrete Non-BPS Primaries}
As mentioned in Section \ref{sec:block}, the discrete state contributions come from the poles in the Liouville structure constants $C(\A_1,\A_2,\A_P)$ when we analytically continue the external states, labeled by their exponents $\alpha_i$, from ${Q\over 2} + i \mathbb{R}$ to their actual values on the real line given in \eqref{qkl} \cite{Maldacena:2001km}.  The relevant factor in the Liouville structure constant is $\Upsilon(\alpha_1 + \A_2 -\A_P)$ in the denominator of \eqref{liouville3},\footnote{The factor $\Upsilon(\alpha_1 + \A_2 +\A_P -Q)$ in \eqref{liouville3} will give other discrete states with the same weights. The structure constant $C(\A_3 ,\A_4,{Q\over 2}- iP)$ yields an identical analysis with $\ell$ replaced by $k-2-\ell$, and hence gives the same set of poles. } where $\Upsilon(x)$ has zeroes at
\ie\label{pole}
x= - {n\over \sqrt{k}} - m \sqrt{k} ,~~~\text{and}~~~x= {n+1\over \sqrt{k} } + (m+1) \sqrt{k},~~~n,m\in \mathbb{Z}_{\ge0}.
\fe 
The argument of $\Upsilon(\A_1+\A_2-\A_P)$ is deformed from $Q/2+i\mathbb{R}$ to  ${\ell+2\over \sqrt{k}} -{Q\over 2} +i\mathbb{R}$. By noting that $Q=\sqrt{k}+{1\over \sqrt{k}}$, the question of identifying the poles is equivalent to asking whether the interval  
\ie
\, \left( {1\over \sqrt{k} } \left( \ell+ {3\over 2} - {k\over 2}\right) \,,\, {1\over \sqrt{k}} \left( {k\over 2} +{1\over2}\right) \right)
\fe
contains any of the poles in \eqref{pole}. It is not hard to see that the only possible poles in \eqref{pole} that lie in the above interval are 
\ie
x=-{n\over\sqrt{k}},~~~~n = 0,1,\cdots, \Big\lfloor \,{k\over2}-\ell -2 \,\Big\rfloor.
\fe
Note that $k\ge 4$ for these poles to contribute.\footnote{For $k=3$ and $\ell=0$, the pole lies precisely at the new contour but the contribution to the four-point function is cancelled by poles from other factors in the Liouville structure constant. In any case, the potential discrete state lies at the bottom of the continuum and therefore does not affect the distinction between $\Delta_{discrete}$ with $\Delta_{cont}$.}  
These poles occur at
\ie
{1\over \sqrt{k} } \left( \ell+ {3\over 2} - {k\over 2}\right) + i P = -{ n\over\sqrt{k}},
\fe
or, in other words,
\ie
P  = i {1\over \sqrt{k}} \left( \ell+{3-k \over2} +n  \right).
\fe
The imaginary shift of the momentum shifts the scaling dimension of the discrete non-BPS primary of question from the continuum gap by the amount of $2P^2$, to
\ie
{ (k-2\ell -1)^2\over 2k } +2P^2 = 2(n+1) - {2(n+1) (2+2\ell +n) \over k}.
\fe
The lowest scaling dimension $\Delta^{\phi\phi}_{discrete}$ of such a discrete state (with divergent structure constant) is given by choosing $n=0$,
\ie
\Delta^{\phi\phi}_{discrete} = 2 - {4 (1+\ell)\over k},~~~\text{for}~~k\ge4.
\fe

\subsubsection*{The Normalization of Structure Constants}

We now argue these discrete non-BPS operators, when viewed as a limit of those in the K3 CFT (that is described by the cigar CFT near a singularity), have divergent structure constants with the external ${1\over 2}$-BPS primaries.

Let us first clarify the normalization  of operators in the cigar CFT versus in the K3 CFT.  In comparing the cigar CFT correlators to the K3 CFT correlators, there is a divergent normalization factor involving the length $L$ of the cigar. That is, let $V$ be some operator in the cigar CFT, then an $n$-point function $\la VV\cdots V\ra$ in the cigar CFT of order 1 really scales like $1/L$ when viewed as part of the K3 CFT in the singular limit. In particular, the two-point function $\la VV\ra$  goes like $1/L$, thus the normalized operator in the K3 CFT is $\phi \sim \sqrt{L} V$, so that $\la\phi \phi \phi \phi \ra$ goes like $L$, which diverges in the infinite $L$ limit, for generic cross ratio.

The discrete non-BPS states discussed above contribute to the four-point function \eqref{bpsr} by an amount that is a finite fraction of the continuum contribution, and both diverge in the singular cigar CFT limit. Consequently, these discrete states in the OPE of two ${1\over2}$-BPS operators $\phi^{RR}$ have divergent structure coefficients in this limit.

\section{Bootstrap Constraints on the K3 CFT Spectrum: Gap}
\label{Sec:Bootstrap}

\subsection{Crossing Equation for the BPS Four-Point Function}

Let us consider the four-point function $f(z, \bar z) \equiv \langle \phi^{RR}(z, \bar z) \phi^{RR}(0) \phi^{RR}(1) \phi^{RR}(\infty) \rangle$ of identical R sector ground states (the four-point function in the NS sector is related by spectral flow).  Decomposed into $c = 6$ ${\cal N} = 4$ R sector superconformal blocks ${\cal F}^R_h(z)$ (in the $z\to 0$ channel),
\ie
f(z,\bar z) = \sum_{h_L,h_R} C_{h_L,h_R}^2 {\cal F}^R_{h_L}(z)\overline{{\cal F}^R_{h_R}(z)},
\fe
where
\ie
{\cal F}^R_h(z) = z^{1\over 2}(1-z)^{1\over 2} { F}^{Vir}_{c=28}(1,1,1,1;h+1;z),
\fe
and ${F}^{Vir}_c(h_1,h_2,h_3,h_4;h;z)$ is the sphere four-point conformal block of the Virasoro algebra of central charge $c$.  Crossing symmetry relates the decomposition in the $z \to 0$ channel to that in the $z \to 1$ channel
\ie
0 = \sum_{h_L,h_R} C_{h_L,h_R}^2 \left[ {\cal F}^R_{h_L}(z)\overline{{\cal F}^R_{h_R}(z)} - {\cal F}^R_{h_L}(1-z)\overline{{\cal F}^R_{h_R}(1-z)} \right].
\fe
This is equivalent to the statement that
\ie
0 = \sum_{\Delta, s} C_{h_L,h_R}^2 \A[ {\cal F}^R_{h_L}(z)\overline{{\cal F}^R_{h_R}(z)} - {\cal F}^R_{h_L}(1-z)\overline{{\cal F}^R_{h_R}(1-z)} ]
\fe
for all possible linear functionals $\A$ \cite{Rattazzi:2008pe}.  In particular, we can pick our basis of linear functionals to consist of derivatives evaluated at the crossing symmetric point
\ie 
\A_{m, n} = 
\partial_z^m \bar\partial_z^n \big|_{z = 1/2}.
\fe
Since $\A_{m, n} [{\cal H}_{\Delta, s}(z, \bar z)]$ trivially vanishes for $m + n$ even, we want to consider functionals that are linear combinations of $\A_{m, n}$ for $m + n$ odd.  Restricting to this subset of functionals, the crossing equation becomes
\ie\label{crossing}
0 = \sum_{\Delta, s} C_{h_L,h_R}^2 \A[{\cal H}_{\Delta, s}(z, \bar z)],
\fe
where for convenience we define 
\ie
{\cal H}_{\Delta, s}(z, \bar z) \equiv {\cal F}^R_{h_L}(z)\overline{{\cal F}^R_{h_R}(z)}.  
\fe

Using the crossing equation, we will constrain the spectrum of intermediate primaries appearing in the $\phi^{RR} \phi^{RR}$ OPE, by finding functionals that have certain positivity properties. In particular, we will be interested in bounding the gap in the non-BPS spectrum, as well as the lowest scaling dimension in the continuum of the spectrum in the singular K3 limits.

\subsection{The Gap in the Non-BPS Spectrum as a Function of $f(1/2)$}
\label{Sec:Gap}

We first bound the gap in the non-BPS spectrum in the OPE of identical BPS operators.  Fix a $\widehat \Delta_{gap}$, and search for a nonzero functional $\A$ satisfying\footnote{Here and henceforth, the unitarity bound $\Delta \geq s$ is implicit.  That is, positivity is enforced for $\Delta > \max(s, \widehat\Delta_{gap})$.
}
\ie
\A[{\cal H}_{\Delta, s}(z, \bar z)] > 0 \quad \text{for} \quad \Delta = s = 0 ~ \text{and} ~ \Delta > \widehat\Delta_{gap}, \quad s \in 2\bZ,
\fe
If such a functional exists, then there must be a contribution to the four-point function from a primary with scaling dimension below $\widehat\Delta_{gap}$ that is not the identity.  In other words, we obtain an upper bound on the gap in the spectrum,
\ie
\widehat\Delta_{gap} \ge \Delta_{gap}.
\fe
The search of positive functionals can be effectively implemented using semidefinite programming \cite{Poland:2011ey,Kos:2013tga,Kos:2014ab,Simmons-Duffin:2015qma}, and the optimal bound is obtained by minimizing $\widehat\Delta_{gap}$.

Over certain singular loci on the moduli space of the K3 CFT, for example, the $\mathcal{N}=4$ cigar CFT points, 
 the four-point function at generic cross ratios diverge (away from the singular loci, the primary operators are always taken to be normalized by the two-point function).  Since the four-point function is unbounded above on the moduli space of the K3 CFT, this motivates us to look for a more refined $\widehat\Delta_{gap}$ that depends on the four-point function.  Let us first discuss how to improve $\widehat\Delta_{gap}$ using the four-point function evaluated at the crossing symmetric point $f(1/2)$.
 In the next section, we will explore an alternative, which is to bound $\Delta_{gap}$ conditioned on the integrated four-point function $A_{1111}$, whose dependence on the K3 CFT moduli is explicitly known (see Section~\ref{sec:integrated}).  In Appendix~\ref{sec:inequalityforAandfhalf}, we bound $f(1/2)$ below by $A_{1111}$.

The information of $f(1/2)$ can be easily incorporated into semidefinite programming.  Define ${\cal H}'_{\Delta, s}(z, \bar z) \equiv {\cal F}^R_{h_L}(z) \overline{{\cal F}^R_{h_R}(z)} - f(1/2) \delta_{\Delta, 0}$, so that $\A_{m, n} [{\cal H}'_{\Delta, s}(z, \bar z)] = \A_{m, n} [{\cal H}_{\Delta, s}(z, \bar z)]$ for $m + n$ odd as before, and
\ie
0 = \sum_{\Delta, s} \A_{0, 0} [{\cal H}'_{\Delta, s}(z, \bar z)]
\fe
is equivalent to the conformal block decomposition of $f(1/2)$.  An optimal $\widehat\Delta_{gap}$ can be obtained by scanning over functionals acting on ${\cal H}'_{\Delta, s}(z, \bar z)$, except that now the functionals are linear combinations of $\A_{m, n}$ with $m + n$ odd as well as $m = n = 0$.

\begin{table}[t]
\begin{center}
\begin{tabular}{| c | c | c |}
\hline
Derivative order $d$ & $\widehat\Delta_{gap}$ & ${f(1/2)_{min}}$
\\\hline\hline
8 & 2.04892 & 2.97672
\\
10 & 2.03414 & 2.98401
\\
12 & 2.01089 & 2.99507
\\
14 & 2.01080 & 2.99513
\\
16 & 2.00449 & 2.99806
\\
18 & 2.00408 & 2.99823
\\
20 & 2.00179 & 2.99923
\\
22 & 2.00134 &
\\
24 & 2.00063 &
\\
26 & 2.00056 &
\\
28 & 2.00030 &
\\
30 & 2.00024 &
\\\hline\hline
$T^4/\bZ_2$ free orbifold: untwisted sector & $\Delta_{gap} = 2$ & $f(1/2) = 3$
\\\hline
\end{tabular}
\end{center}
\caption{The bound on the gap in the identical primary OPE, and the minimal value of the four-point function evaluated at the crossing symmetric point, as the derivative order of the basis of functionals is increased.  Also shown are the values of the untwisted sector correlator at the $T^4/\bZ_2$ free orbifold point computed in Section~\ref{Sec:FreeOrbifold}, which within numerical error saturate the bounds.}
\label{Table:gap}
\end{table}

\paragraph{A word on numerics.}  The results of semidefinite programing depend on a set of parameters.  The conformal block is evaluated to $q^N$ order using Zamolodchikov's recurrence relations (see Appendix~\ref{Sec:qmap}) \cite{Zamolodchikov:1985ie,Zamolodchikov:1995aa}, and we scan over functionals that are linear combinations of derivatives evaluated at the crossing symmetric point, up to $d$ derivative orders, namely, $\A_{m, n}$ for $m + n \leq d$.  Moreover, the positivity condition is in practice only imposed for spins lying in a finite range $s \leq s_{max}$ (but for {\it all} scaling dimensions $\Delta \geq \widehat\Delta_{gap}$).  The truncation on spin is justified by the unitarity bound $\Delta \geq s$ and the convergence rate of the sum over intermediate states in the four-point function \cite{Pappadopulo:2012jk}.  There are subtle interplays between these parameters.  For example, if we go up to $d$ derivative order, then we need $N$ to be larger than $d$; empirically we find that $N = d + 10$ gives a good approximation that is stable as $N$ is further increased.  Also, as $d$ is increased, $s_{max}$ should also be increased, otherwise the bound may violate physical examples \cite{Caracciolo:2014aa}.  The default setting in this paper is $N = 30$, $s_{max} = 40$, and up to $d = 20$, unless noted otherwise.

\begin{figure}[t]
\centering
\includegraphics[width=\textwidth]{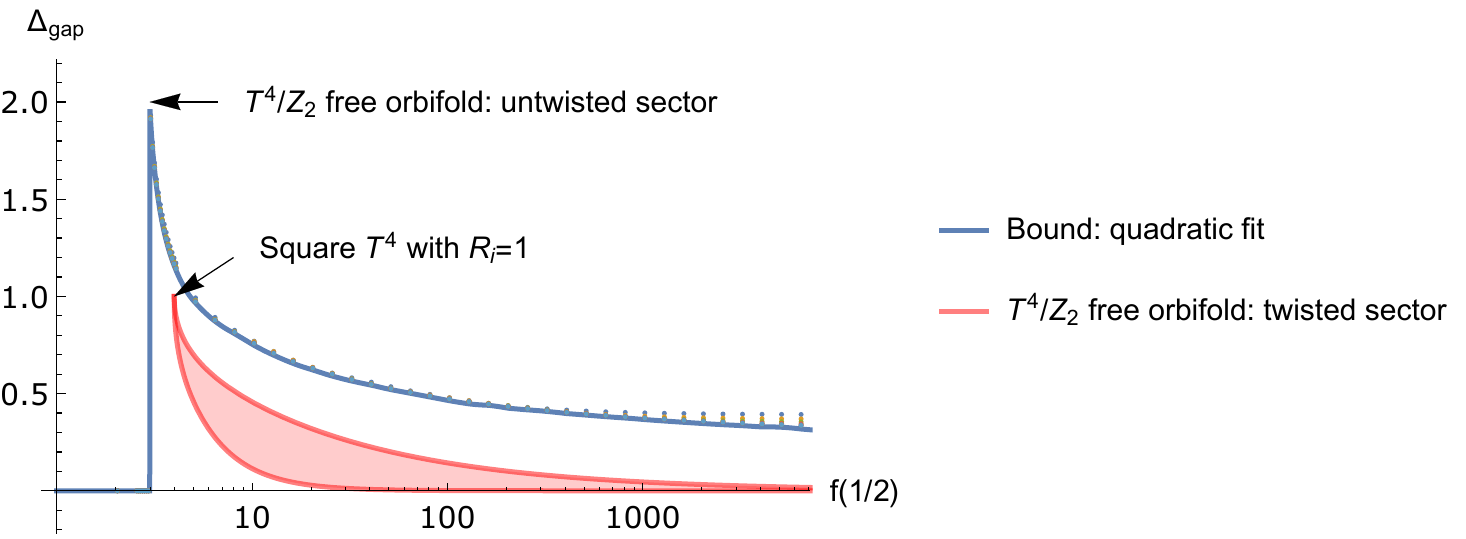}
\caption{The dots indicate the upper bound $\widehat\Delta_{gap}$ on the gap versus $f(1/2)$, the four-point function evaluated at the crossing symmetric point, at derivative orders ranging from 8 to 20.  The solid line plots the extrapolation to infinite order using a quadratic fit.  The minimal $f(1/2)$ and maximal gap are simultaneously saturated by an untwisted sector correlator at the free orbifold point. The shaded region represents the gap in the OPE of twist fields at a fixed point of $T^4/\bZ_2$ with a rectangular $T^4$, where the minimal $f(1/2)$ and maximal gap are achieved by a square $T^4$ at radii $R_i=1$ ($1/\sqrt{2}$ times the self-dual radius). }
\label{Fig:Fhalf}
\end{figure}

\paragraph{Numerical results.}
The first two columns of Table~\ref{Table:gap} show the numerical results for the optimal $\widehat\Delta_{gap}$ without the information of $f(1/2)$, for up to $d = 30$ derivative orders.  The conformal block is evaluated to $q^{40}$ order to accommodate the high derivative orders.  Within numerical error, $\widehat\Delta_{gap}$ approaches $2$ as we increase the derivative order.  This bound is saturated by a free fermion correlator at the free orbifold point, as was explained in Section~\ref{Sec:FreeOrbifold}.

After incorporating the information of $f(1/2)$ (reverting to the default setting of parameters),  we find that $f(1/2)$ less than a certain threshold $f(1/2)_{min}$ is completely ruled out ($\widehat\Delta_{gap} = 0$).  Above this threshold, $\widehat\Delta_{gap}$ starts from $\widehat\Delta_{gap} \approx 2$ at $f(1/2)=f(1/2)_{min}$ and then monotonically decreases.  Table~\ref{Table:gap} shows the values of $f(1/2)_{min}$, which seem to asymptote to $f(1/2)_{min} \approx 3$ at infinite derivative order.  Figure~\ref{Fig:Fhalf} plots the dependence of $\widehat\Delta_{gap}$ on $f(1/2)$.  It is observed that the limiting value $\widehat\Delta_{gap}$ as $f(1/2) \to \infty$ is approximately equal to another quantity $\widehat\Delta_{crt} \approx 1/4$ that we will introduce in the next section. 
 Note that for smaller values of $f(1/2)$, the numerical bound  $\widehat\Delta_{gap}$ appears to converge exponentially with the derivative order $d$, while for larger values of $f(1/2)$ the convergence is much slower and we  extrapolate the bound to infinite $d$ using a quadratic fit. There seems to be a crossover between the exponential convergence and power law convergence as $f(1/2)$ increases.  Since $\widehat\Delta_{gap}$ approaches $\widehat\Delta_{crt}$ in the large $f(1/2)$ limit,  a quadratic fit (rather than, for example, a linear fit) is justified in  this limit  as it works well for the latter (see Table \ref{Tab:DeltaCritical}).

The value $f(1/2)_{min} \approx 3$ with ${\Delta}_{gap} \approx 2$ agrees with the four point function \eqref{ff4pt} of untwisted sector BPS primaries at the $T^4/{\mathbb Z}_2$ orbifold point where the numerical bound on the gap is saturated.
Furthermore, it appears that the gap in the OPE of the twisted field $\sigma(z,\bar z)$ at the orbifold point lies close to, but does not quite saturate the numerical bound.  It remains to be understood whether our numerical bound can be further improved or there exist other operators in the OPE of BPS primaries at other points on the moduli space that saturate the bound.  
 

%

\subsection{The Gap in the Non-BPS Spectrum as a Function of $A_{1111}$}
\label{Sec:Gapa1111}

A more desirable constraint to impose is the integrated four-point function $A_{1111}$, since its dependence on the K3 CFT moduli is explicitly known (see Section~\ref{sec:integrated}).  Using crossing symmetry, $A_{1111}$ can be decomposed into a sum of conformal blocks integrated over the cross ratio in some finite domain.  We then incorporate the equation
\ie
0 &= (3 A_0 - A_{1111}) + 3 \sum_{{\rm non-BPS}~{\cal O}} C^2_{11{\cal O}} A(\Delta, s)
\fe
into bootstrap, where the integrated blocks are
\ie
A(\Delta, s) &=  \int_{\cal D} {d^2z\over |z(1-z)|}{\cal F}^R_{\Delta + s \over 2}(z) \overline{{\cal F}^R_{\Delta - s \over 2}(z)},
\quad A_0 = \lim_{\Delta \to 0} \left[ A(\Delta, 0) - {2\pi \over \Delta} \right].
\fe
Using semidefinite programming, if we can find a set of coefficients $a > 0$ and $a_{m, n}$ such that ($m+n$ odd)
\ie
\label{A1111BS}
& a (3 A_0 - A_{1111}) + \sum_{m, n} a_{m, n} \A_{m, n} [{\cal H}_{0}(z, \bar z)] > 0,
\\
& 3 a A(\Delta, s) + \sum_{m, n} a_{m, n} \A_{m, n} [{\cal H}_{\Delta, s}(z, \bar z)] > 0 \quad \text{for} \quad \Delta > \widehat\Delta_{gap}, \quad s \in 2\bZ
\fe
are satisfied, then the gap in the non-BPS spectrum $\Delta_{gap}$ must be bounded above by $\widehat\Delta_{gap}$.

\begin{figure}[t]
\centering
\subfigure{
\includegraphics[width=.4\textwidth]{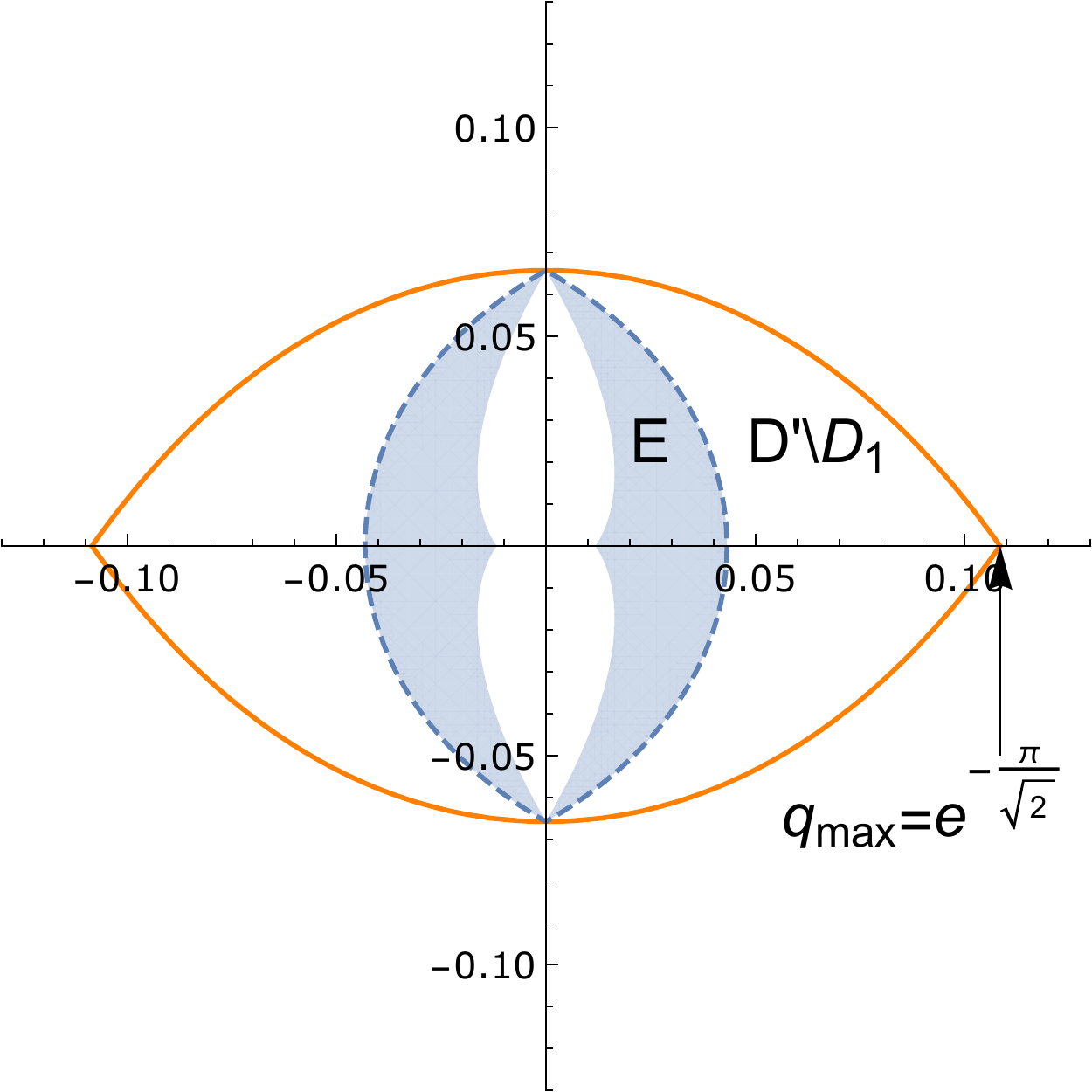}
}
~~
\subfigure{
\includegraphics[width=.4\textwidth]{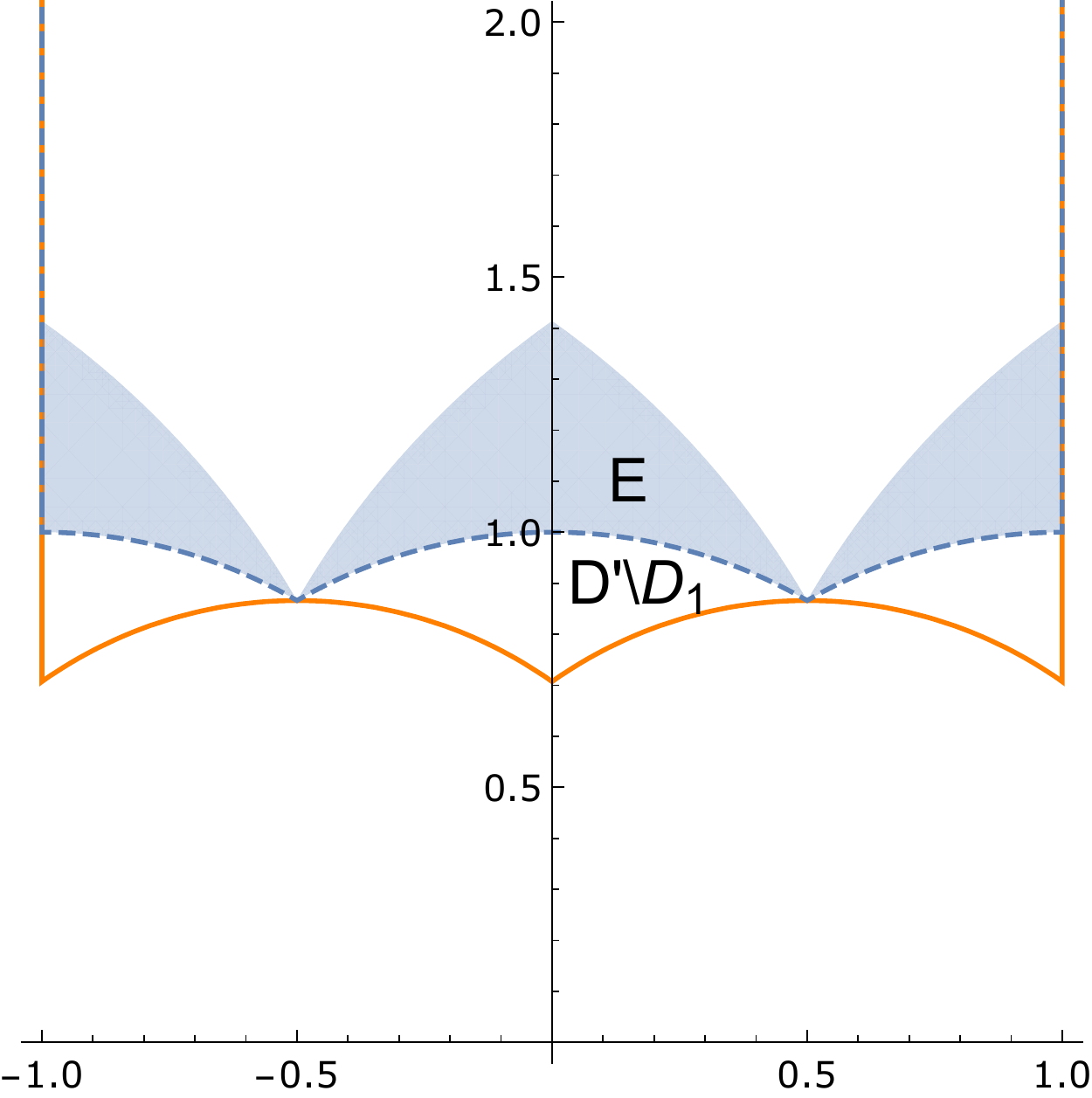}
}
\caption{The integration region ${\cal D} = D' \setminus E$.  The left is in the $q$-plane, and the right in the $\tau$-plane.  The entire region enclosed by the solid line is $D'$.  The region between the solid and dashed lines is $D' \setminus D_1$, and the shaded region is its image $E$ under $z \to 1-z$ for the right half and $z \to 1/z$ for the left half.  The entire unshaded region inside solid line is the integration region $\cal D$.  See Appendix~\ref{Sec:Aijkl1}.}
\label{fig:ellipse}
\end{figure}

However, the region of integration $\cal D$ has to be carefully chosen so that the integrated blocks obey certain positivity properties at large weights, otherwise the bound cannot be improved below $\widehat\Delta_{gap} \approx 2$.  More specifically, $\cal D$ should contain two fundamental domains of the $S_3$ crossing symmetry group, and have a maximal $|q(z)|$ value on the real axis.  See Appendix~\ref{Sec:Aijkl1} for a detailed discussion and a specific choice of ${\cal D}$, and Figure~\ref{fig:ellipse} for an illustration.


\begin{figure}[t]
\centering
\includegraphics[width=\textwidth]{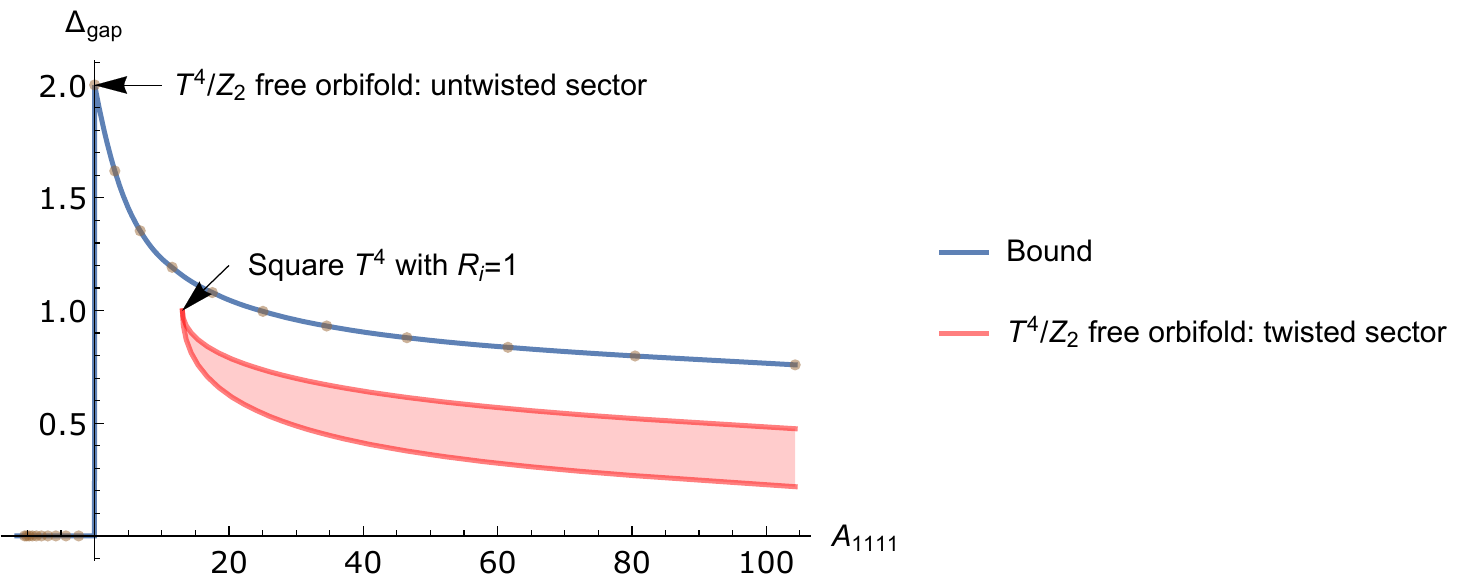}
\caption{ The solid line shows the upper bound $\widehat\Delta_{gap}$ on the gap versus the integrated four-point function $A_{1111}$, at 20 derivative order, which we observe to already stabilize with increment of the derivative order in the range of $A_{1111}$ shown here; the dots are the actual data points.  The minimal $A_{1111}$ and maximal gap are simultaneously saturated by an untwisted sector correlator at the free orbifold point.  The shaded region represents the gap in the OPE of twist fields at a fixed point of $T^4/\bZ_2$ with a rectangular $T^4$, where the minimal $A_{1111}$ and maximal gap are achieved by a square $T^4$ at radii $R_i=1$ ($1/\sqrt{2}$ times the self-dual radius). }
\label{Fig:A1111}
\end{figure}

Figure~\ref{Fig:A1111} shows the dependence of the numerical bound $\widehat\Delta_{gap}$ on $A_{1111}$; the data points are bounds obtained at 20 derivative order, which we observe to already stabilize with incrementing the derivative order.  We verified by testing that the bounds are not sensitive to the choice of $\cal D$.\footnote{We found that for a given ``good'' choice of ${\cal D}$ (see Appendix~\ref{Sec:Aijkl1} for restrictions on $\cal D$), there is a minimum derivative order $d_*$ below which the bound is the same as that without the input of $A_{1111}$, namely $\widehat\Delta_{gap} \approx 2$.  Above $d_*$, the bound suddenly exhibits the nontrivial dependence on $A_{1111}$ that is shown in Figure~\ref{Fig:A1111}.  The choice of $\cal D$ given in Appendix~\ref{Sec:Aijkl1} is made for simplicity, and has $d_* = 16$; other choices may give smaller $d_*$.  However, the bound is not sensitive to the choice of $\cal D$, as long as we look at derivative orders larger than the respective $d_*$.
}
The results indicate that $A_{1111}$ must be non-negative.  Above $A_{1111} \approx 0$, $\widehat\Delta_{gap}$ starts from $\approx 2$ and monotonically decreases with $A_{1111}$.
The point $A_{1111} \approx 0$ and $\widehat{\Delta}_{gap} \approx 2$ is saturated by the integrated four-point function \eqref{ff4pt} of untwisted sector BPS primaries at the $T^4/{\mathbb Z}_2$ free orbifold point.  In the limit $A_{1111} \to \infty$, $\widehat\Delta_{gap}$ approaches $\widehat\Delta_{crt} \approx 1/4$, a quantity we define in the next section.\footnote{Since $A_{1111}$ is bounded above by $f(1/2)$ assuming a finite gap \eqref{bounda1111}, and we already observed that $\widehat\Delta_{gap} \to \widehat\Delta_{crt} \approx 1/4$ in the large $f(1/2)$ limit, it follows that $\widehat\Delta_{gap} \to \widehat\Delta_{crt} \approx 1/4$ in the large $A_{1111}$ limit as well.
}
 Note that  $A_{1111}$ is related to the tree-level $H^4$ coefficient in the 6d (2,0) supergravity effective action of IIB string theory compactified on K3.  The consistency of string theory requires that this coefficient be non-negative, because otherwise it leads to superluminal propagation \cite{Adams:2006aa}.  Amusingly, here this non-negativity follows from unitarity constraints on the CFT correlator.  Again, the gap in the OPE of the twisted field $\sigma(z,\bar z)$ at the orbifold point lies close to, but does not quite saturate the numerical bound.

\subsection{Constraints on the OPE of Two Different ${1\over 2}$-BPS Operators}\label{sec:mixed}

\begin{figure}[t]
\centering
\includegraphics[width=.75\textwidth]{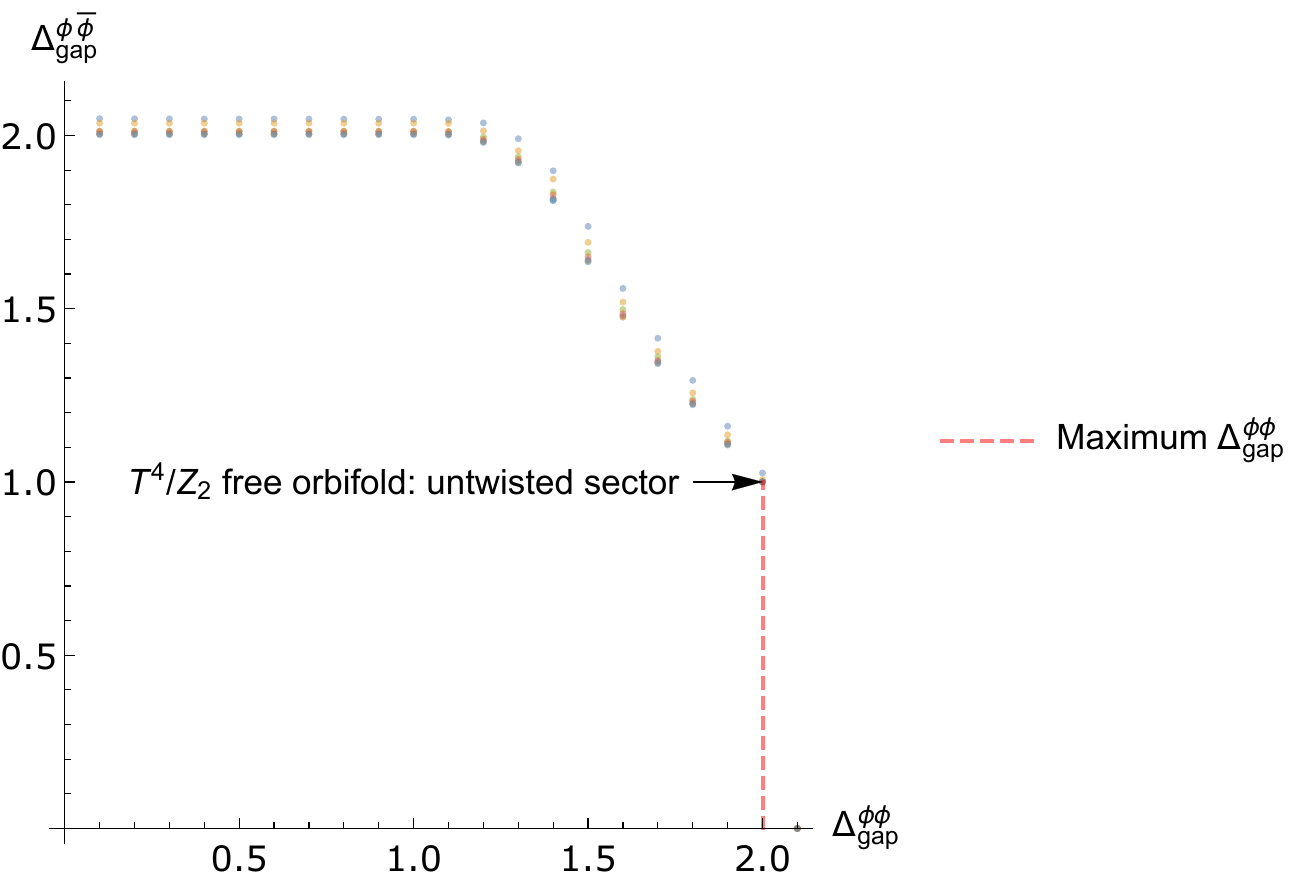}
\caption{ The dots indicate upper bounds $( \widehat\Delta_{gap}^{\phi\phi} ,\widehat\Delta_{gap}^{\phi\bar\phi})$ on the gap in the respective OPEs, at derivative orders ranging from 8 to 20.  We find that $\Delta^{\phi\phi}_{gap}$ is bounded above by 2, beyond which $\Delta^{\phi\bar\phi}_{gap} = 0$.  The point $(2, 1)$ is realized by an untwisted sector correlator at the $T^4/\bZ_2$ free orbifold point. }
\label{Fig:MixedGap}
\end{figure}

By considering the four-point function $\la \phi^{RR}\phi^{RR}\bar\phi^{RR}\bar\phi^{RR} \ra$ of two different RR sector ${1\over 2}$-BPS primaries $\phi^{RR}$ and $\bar\phi^{RR}$, we will be able to detect the gap $\Delta_{gap}$ and $\Delta_{crt}$
in two different OPEs.  The two RR primaries are chosen so that the identity block only appears in the $\phi^{RR}\times\bar\phi^{RR}$ OPE but not in $\phi^{RR}\times \phi^{RR}$ or $\bar\phi^{RR} \times \bar\phi^{RR}$.  Taking $\phi^{RR}$ and $\bar\phi^{RR}$ to be complex conjugates of each other, the two crossing equations are\footnote{This is not what is usually meant by ``mixed correlator bootstrap'', where the crossing equation for $\la \phi\phi\bar\phi\bar\phi \ra$, $\la \phi\phi\phi\phi \ra$, $\la \bar\phi\bar\phi\bar\phi\bar\phi \ra$ are all considered at the same time as in \cite{Kos:2014ab}.
}
\ie
0 &= \sum_{{\cal O} \in \phi \times \bar\phi} |C_{\phi\bar\phi{\cal O}}|^2 \left[ {\cal F}^R_{h_L}(z)\overline{{\cal F}^R_{h_R}(z)} - {\cal F}^R_{h_L}(1-z)\overline{{\cal F}^R_{h_R}(1-z)} \right],
\\
0 &= \sum_{{\cal O} \in \phi \times \bar\phi}  (-1)^s |C_{\phi\bar\phi{\cal O}}|^2 {\cal F}^R_{h_L}(z)\overline{{\cal F}^R_{h_R}(z)} 
- \sum_{{\cal O} \in \phi \times \phi} |C_{\phi\phi{\cal O}}|^2 {\cal F}^R_{h_L}(1-z)\overline{{\cal F}^R_{h_R}(1-z)}.
\fe
By defining ${\cal G}_{\Delta, s}^\pm(z, \bar z) = {\cal F}^R_{h_L}(z)\overline{{\cal F}^R_{h_R}(z)} \pm {\cal F}^R_{h_L}(1-z)\overline{{\cal F}^R_{h_R}(1-z)}$, and the vectors
\ie
\vec V_{\Delta, s}^{\phi\bar\phi}(z, \bar z) = \begin{pmatrix}
{\cal G}^-_{\Delta, s} (z, \bar z) \\ (-1)^s {\cal G}^-_{\Delta, s} (z, \bar z) \\ (-1)^s {\cal G}^+_{\Delta, s} (z, \bar z)
\end{pmatrix}, \quad
\vec V_{\Delta, s}^{\phi\phi} = \begin{pmatrix}
0 \\ {\cal G}^-_{\Delta, s} (z, \bar z) \\ - {\cal G}^+_{\Delta, s} (z, \bar z)
\end{pmatrix},
\fe
we can write the crossing equations compactly as
\ie
\label{VecCrossing}
\vec 0 &= \sum_{{\cal O} \in \phi \times \bar\phi} |C_{\phi\bar\phi{\cal O}}|^2 \vec V_{\Delta, s}^{\phi\bar\phi}(z, \bar z) +  \sum_{{\cal O} \in \phi \times \phi} |C_{\phi\phi{\cal O}}|^2 \vec V_{\Delta, s}^{\phi\phi}(z, \bar z).
\fe
By symmetry, only odd derivative order functionals act nontrivially on ${\cal G}^-_{\Delta, s}(z, \bar z)$, and only even derivative order ones act nontrivially on ${\cal G}^+_{\Delta, s}(z, \bar z)$.

To bound the gap in the two channels, we seek linear functionals $\vec\A$ such that
\ie
& \vec\A \cdot \vec V_{\Delta, s}^{\phi\bar\phi} > 0 \quad \text{for} \quad \Delta = s = 0 ~\text{and}~ \Delta > \widehat\Delta_{gap}^{\phi\bar\phi}, \quad s \in \bZ,
\\
& \vec\A \cdot \vec V_{\Delta, s}^{\phi\phi} > 0 \quad \text{for} \quad \Delta = s = 0 ~\text{and}~ \Delta > \widehat\Delta_{gap}^{\phi\phi}, \quad s \in 2 \bZ,
\fe
for some $(\widehat\Delta_{gap}^{\phi\bar\phi}, \widehat\Delta_{gap}^{\phi\phi})$.  Note that only even integer spin primaries appear in the $\phi^{RR} \times \phi^{RR}$ OPE.  The crossing equation \eqref{VecCrossing} implies that
\ie
\text{\it either} \quad \widehat\Delta_{gap}^{\phi\bar\phi} \ge \Delta^{\phi\bar\phi}_{gap} \quad \text{\it or} \quad \widehat\Delta_{gap}^{\phi\phi} \ge \Delta^{\phi\phi}_{gap}.
\fe

Figure~\ref{Fig:MixedGap} shows the numerical results for the allowed region of $(\Delta_{gap}^{\phi\bar\phi}, \Delta_{gap}^{\phi\phi})$.  We find that both $\Delta^{\phi\bar\phi}_{gap}$ and $\Delta^{\phi\phi}_{gap}$ are bounded above by $\approx 2$,  
and the point with $(\Delta^{\phi\phi}_{gap}, \Delta^{\phi\bar\phi}_{gap}) \approx (2, 1)$ is realized by the OPE of untwisted sector primaries at the $T^4 / \bZ_2$ free orbifold point.

\section{Bootstrap Constraints on the Critical Dimension $\widehat \Delta_{crt}$}
\label{sec:critical}


Over certain singular loci on the moduli space of the K3 CFT, the following two phenomena can occur:
\begin{itemize}
\item The density of states diverges, leading to a continuum in the spectrum. 
\item The structure constants of some discrete states diverge.
\end{itemize}

At the singular loci, some components of the integrated four-point function $A_{ijk\ell}$ diverge.  The latter may occur in two different ways: (1) The four-point function remains finite at generic cross ratio $z$, with divergent contribution to $A_{ijk\ell}$ localized at $z = 0, 1, \infty$ due to a vanishing gap in the spectrum.  This occurs in the large volume limit.  (2)  The gap in the spectrum remains finite (i.e., away from the large volume limit), but the whole four-point function diverges at generic $z$.  This is demonstrated in Appendix~\ref{sec:inequalityforAandfhalf}. 


In higher dimensions, there exist absolute upper bounds on OPE coefficients coming from crossing symmetry and unitarity \cite{Caracciolo:2009bx}. In the following subsections, we take a moment to study these bounds.
Our discussion will motivate us to introduce a critical dimension $\widehat\Delta_{crt}$, which is roughly the dimension above which OPE bounds exist.\footnote{These are relative bounds, namely, the OPE coefficients above $\widehat\Delta_{crt}$ are bounded by the OPE coefficients below $\widehat\Delta_{crt}$, in contrast to the absolute bounds in \cite{Caracciolo:2009bx}.  We define $\widehat\Delta_{crt}$ more rigorously in~(\ref{eq:rigorousdefinitionofdeltacritical}) below.}  

Let $\Delta_{crt}$ be the lowest scaling dimension at which either a continuum develops or an OPE coefficient diverges.  For example, at the $\cN=4$ $A_1$ cigar CFT point, there is a continuum of states starting from $\Delta_{crt}=1/4$. We show in Appendix~\ref{sec:deltacriticalanddivergence} that  
\be
\Delta_{crt} \equiv \text{min} (\Delta_{cont}, \Delta_{discrete}) \leq \widehat\Delta_{crt}
\ee
in the notations of Section \ref{Sec:Cigar}.  In the following, we describe how to use crossing symmetry to derive a numerical upper bound on $\widehat\Delta_{crt}$ that is universal across the moduli space.  We will see that $\widehat\Delta_{crt}>0$, so that it is possible to have unbounded contributions to the conformal block expansion from operators below $\widehat\Delta_{crt}$.

\subsection{A Simple Analytic Bound on OPE Coefficients and $\widehat \Delta_{crt}$}

We begin with a simple analytic bound on OPE coefficients. Consider a four-point function of scalars $\phi$ with dimension $\Delta_\phi$, in any number of spacetime dimensions $d$.  For the moment, we set $z=\bar z=x$.  The four-point function can be written as a positive linear combination of ``scaling blocks" $x^{\Delta-2\Delta_\phi}$,\footnote{Here we adopt the convention, common in 2$d$, where $(z\bar z)^{-2\Delta_\phi}$ is included in the conformal blocks.}
\be
f(z=x,\bar z = x) &= \sum_\Delta p_{\Delta} x^{\Delta-2\Delta_\phi},\qquad p_{\Delta} \geq 0.
\ee
Positivity of $p_{\Delta}$ is a consequence of unitarity.
The expansion in scaling blocks ignores relations between primaries and descendants due to conformal symmetry.

Crossing symmetry implies
\be
f(x) &= f(1-x)\nn\\
-(x^{-2\Delta_\phi}-(1-x)^{-2\Delta_\phi}) &= \sum_{\Delta> 0} p_{\Delta} \left(x^{\Delta-2\Delta_\phi}-(1-x)^{\Delta-2\Delta_\phi}\right)\nn\\
1 &= \sum_{\Delta>0} p_{\Delta} \left(\frac{x^{\Delta-2\Delta_\phi}-(1-x)^{\Delta-2\Delta_\phi}}{-x^{-2\Delta_\phi}+(1-x)^{-2\Delta_\phi}}\right),
\label{eq:crossingforscalingblocks}
\ee
where in the second line we separated out the contribution of the unit operator on the left hand side, and on the last line we divided by it.  Evaluating (\ref{eq:crossingforscalingblocks}) at $x=\frac 1 2$, we obtain
\be
1 &= \sum_{\Delta>0} p_{\Delta} \left(\frac 1 2\right)^{\Delta} \frac{\Delta-2\Delta_\phi}{2\Delta_\phi}.
\ee
In particular, suppose all operators have dimension $\Delta\geq 2\Delta_\phi$. (This happens, for example, in the 2$d$ and 3$d$ Ising models).  Then we obtain an upper bound on the contribution of any individual scaling block
\be
\label{eq:firstboundonOPE}
p_{\Delta}\left(\frac 1 2\right)^{\Delta-2\Delta_\phi} &\leq \frac{2^{1+2\Delta_\phi}\Delta_\phi}{\Delta-2\Delta_\phi}.
\ee
When all $\Delta$ are bounded away from $2\Delta_\phi$, there is also an upper bound on the contribution of multiple blocks, and also on the value of the four-point function itself at $x=\frac 1 2$,
\be
f\left(\frac 1 2\right) &\leq \frac{2^{1+2\Delta_\phi}\Delta_\phi}{\Delta_{min}-2\Delta_\phi},
\ee
where $\Delta_{min}$ is the lowest dimension appearing in the conformal block expansion.
As we show in section~\ref{sec:inequalityforAandfhalf}, if the four-point function is bounded at $x=\frac 1 2$, it is bounded everywhere by a known function of $z$.

To obtain (\ref{eq:firstboundonOPE}), we had to assume that only operators with dimension $\Delta\geq 2\Delta_\phi$ appear in the four-point function.  
When operators lie below $2\Delta_\phi$, it may be possible to have unbounded contributions to the conformal block expansion.\footnote{A simple toy example using scaling blocks is
\be
\label{eq:toylargeexample}
\frac{1}{|z|^{2\Delta_\phi}} + \frac{1}{|1-z|^{2\Delta_\phi}} + P
\ee
where $P$ can be arbitrarily large.  This expression is crossing-symmetric and has a positive expansion in scaling blocks.  Because there exists a scaling block with $\Delta=2\Delta_\phi$, namely the constant $P$, the four-point function can be arbitrarily large. (However, this example does not have a positive expansion in conformal blocks.) We thank Petr Kravchuk for this example.} Let $\widehat\Delta_{crt}$ be the dimension above which general bounds on OPE coefficients exist. We have shown $\widehat\Delta_{crt}\leq 2\Delta_\phi$.

\subsection{Improved Analytic Bounds on $\widehat\Delta_{crt}$}

There are two ways to obtain stronger bound on OPE coefficients and $\widehat\Delta_{crt}$.  Firstly, we can include more information about conformal symmetry by writing the four-point function as a positive sum over more sophisticated blocks.
For example, in any spacetime dimension, we have
\be
f(x) &= x^{-2\Delta_\phi}\sum p'_\Delta \rho(x)^\Delta,\qquad p'_\Delta\geq 0,
\ee
where
\be
\rho(x) &\equiv \frac{x}{(1+\sqrt{1-x})^2}
\ee
is the radial coordinate of \cite{Pappadopulo:2012jk,Hogervorst:2013sma}.  Evaluating the crossing equation at $x=\frac 1 2$ then gives
\be
\label{analyticbound}
p'_\Delta \rho\left(\frac 1 2\right)^\Delta &\leq \frac{\Delta-\sqrt 2 \Delta_\phi}{\sqrt 2 \Delta_\phi}.
\ee
This implies that OPE bounds exist whenever $\Delta\geq \sqrt 2 \Delta_\phi$.  In other words,\footnote{The estimates $2\Delta_\phi$ (coming from $x$ blocks) and $\sqrt 2 \Delta_\phi$ (coming from $\rho$ blocks) are the same as the reflection-symmetric points in the discussion of \cite{Kim:2015oca}.}
\be
\label{eq:rhoboundondeltacritical}
\widehat\Delta_{crt}\leq \sqrt 2 \Delta_\phi, \qquad(d\geq 2).
\ee

In two-dimensional theories, we can write the four-point function in terms of a positive expansion in $q^\Delta$, where $q$ is the elliptic nome \cite{Zamolodchikov:1985ie,Zamolodchikov:1995aa,Maldacena:2015iua}.  This leads to stronger bounds on OPE coefficients and the result
\be
\widehat\Delta_{crt} &\leq \frac{\pi-3}{12\pi}c + \frac{4\Delta_\phi}{\pi},\qquad(d=2),
\ee
where $c$ is the central charge. (This bound is worse than (\ref{eq:rhoboundondeltacritical}) when $\Delta_\phi$ is small compared to $c$.)

The best possible OPE bound comes from using the full conformal block expansion --- either global blocks in $d>2$ or the appropriate Virasoro blocks in 2$d$.

\subsection{Numerical Bounds on $\widehat\Delta_{crt}$}

The second way to improve these bounds is to consider more general linear functionals, other than simply evaluating the crossing equation at $x=\frac 1 2$.  Consider the conformal block expansion
\be
f(z,\bar z) &= \sum_{\Delta,s} p_{\Delta,s} \mathcal{F}_{\Delta,s}(z,\bar z).
\ee
Fix a dimension $\widehat \Delta$ and search for a nonzero functional $\alpha$ with the property
\be
\label{eq:positivityfordeltacrt}
\alpha[\mathcal{F}_{\Delta,s}(z,\bar z)-\mathcal{F}_{\Delta,s}(1-z,1-\bar z)] > 0 \quad \textrm{for}\quad
\Delta \geq \max(\textrm{unitarity bound},\ \widehat \Delta),\quad s \in 2\mathbb{Z}.
\ee
This is the same procedure as placing upper bounds on $\Delta_{gap}$, with exception that we {\it do not impose positivity for $\alpha$ acting on the unit operator $\mathcal{F}_{0,0}$}.  In fact, it is sometimes helpful to use the normalization condition
\be
\label{eq:normalizationconditionfordeltacrt}
\alpha[\mathcal{F}_{0,0}(z,\bar z)-\mathcal{F}_{0,0}(1-z,1-\bar z)] = -1.
\ee
Now suppose there exists $\alpha$ exists satisfying (\ref{eq:positivityfordeltacrt}), (\ref{eq:normalizationconditionfordeltacrt}), and suppose further also that only operators with dimension $\Delta\geq \widehat \Delta$ appear in the conformal block expansion.  Then we find a general OPE bound
\be
p_{\Delta,s} \leq \alpha[\mathcal{F}_{\Delta,s}(z,\bar z)-\mathcal{F}_{\Delta,s}(1-z,1-\bar z)]^{-1}.
\ee

We can now give a more rigorous definition of $\widehat\Delta_{crt}$:
\be
\label{eq:rigorousdefinitionofdeltacritical}
\widehat\Delta_{crt} &\equiv \textrm{the smallest $\widehat \Delta$ such that there exists nonzero $\alpha$ satisfying (\ref{eq:positivityfordeltacrt})}.
\ee
If all operators in the conformal block expansion have dimension above $\widehat\Delta_{crt}$, then their OPE coefficients obey universal bounds.  By contrast, if some operators are above and some operators are below, then the contributions above are bounded in terms of the contributions below.  See Appendix~\ref{sec:deltacriticalanddivergence} for a more detailed discussion.

\subsection{$\widehat\Delta_{crt}$ in 2, 3, and 4 Spacetime Dimensions}

In higher dimensional theories, we will use a slightly modified definition of $\widehat \Delta_{crt}$.  The reason is that the stress-tensor always appears in the conformal block expansion, so it is nonsensical to impose that spin-2 operators must have dimension greater than $d$.  The same is true in 2d theories when using global $SL(2,\mathbb{R})\times SL(2,\mathbb{R})$ conformal blocks.  By contrast, Virasoro blocks include the contribution of the stress tensor, so the constraint (\ref{eq:positivityfordeltacrt}) makes sense in that case.

In higher dimensions (and for global blocks in 2d), we instead define
\be
&\widehat\Delta_{crt}^{scalar} \equiv \textrm{the smallest $\widehat \Delta$ such that there exists nonzero $\alpha$ satisfying}\nonumber\\
&\alpha[\mathcal{F}_{\Delta,s}(z,\bar z)-\mathcal{F}_{\Delta,s}(1-z,1-\bar z)] > 0 \quad \textrm{for}\quad
\Delta \geq \begin{cases}
 \widehat \Delta & s=0\\
 \textrm{unitarity bound} & s \geq 0.
 \end{cases}
\ee
The quantity $\widehat\Delta_{crt}^{scalar}$ agrees with $\widehat\Delta_{crt}$ when $\widehat\Delta_{crt} \leq d$, and may differ when $\widehat\Delta_{crt} > d$.

\begin{figure}[!ht]
\begin{center}
\subfigure{
\includegraphics[width=0.45\textwidth]{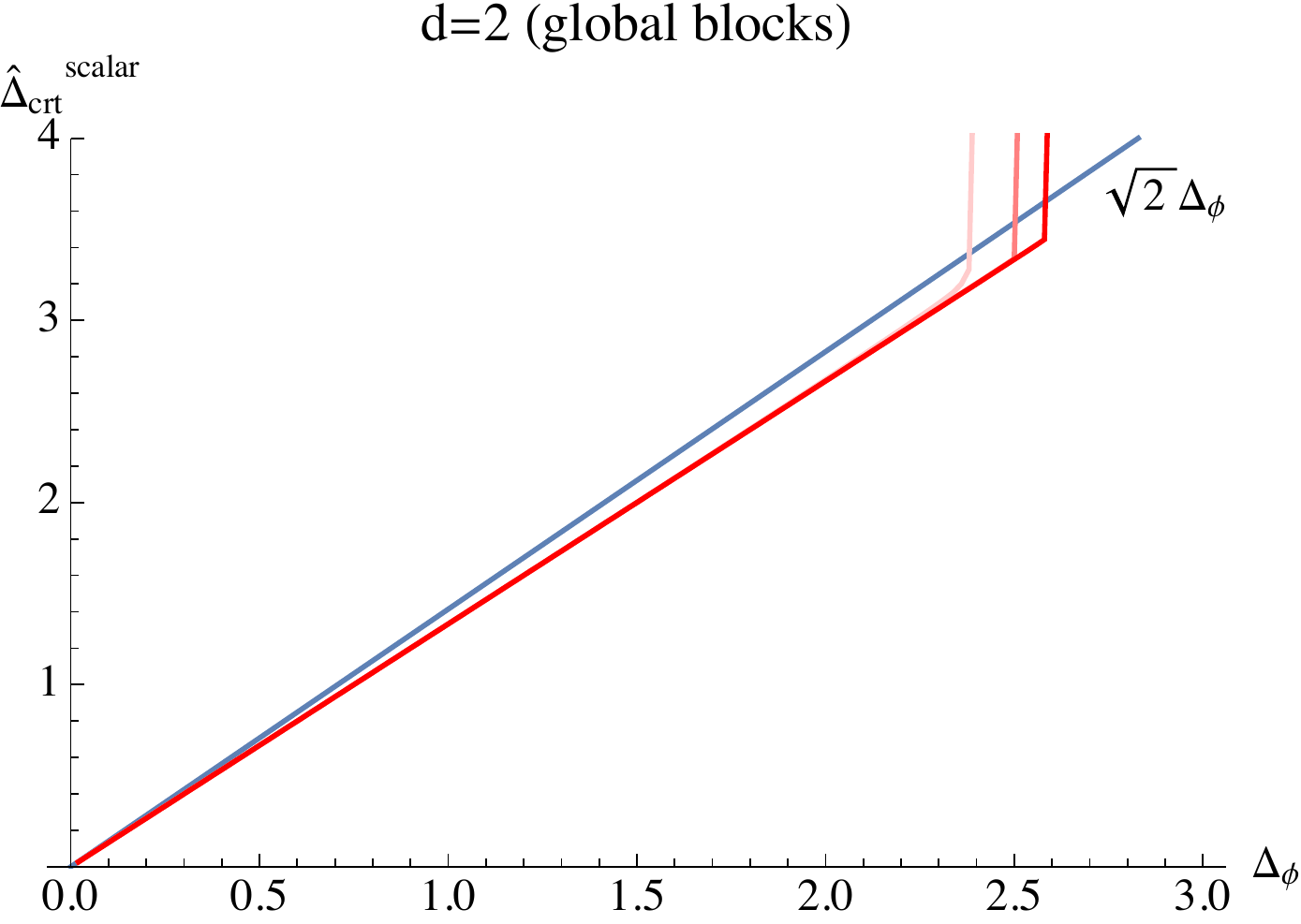}
}
\subfigure{
\includegraphics[width=0.45\textwidth]{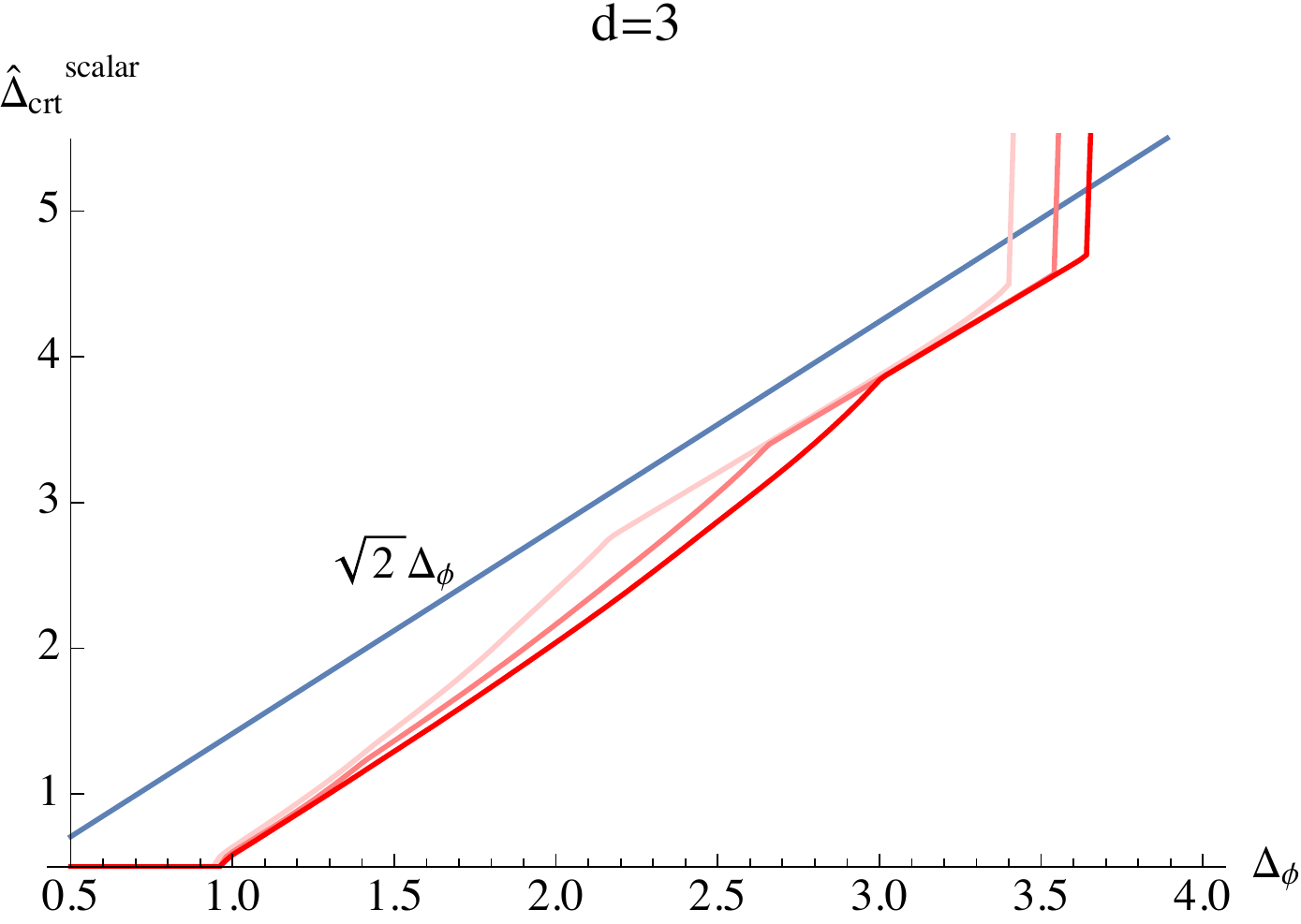}
}
\subfigure{
\includegraphics[width=0.45\textwidth]{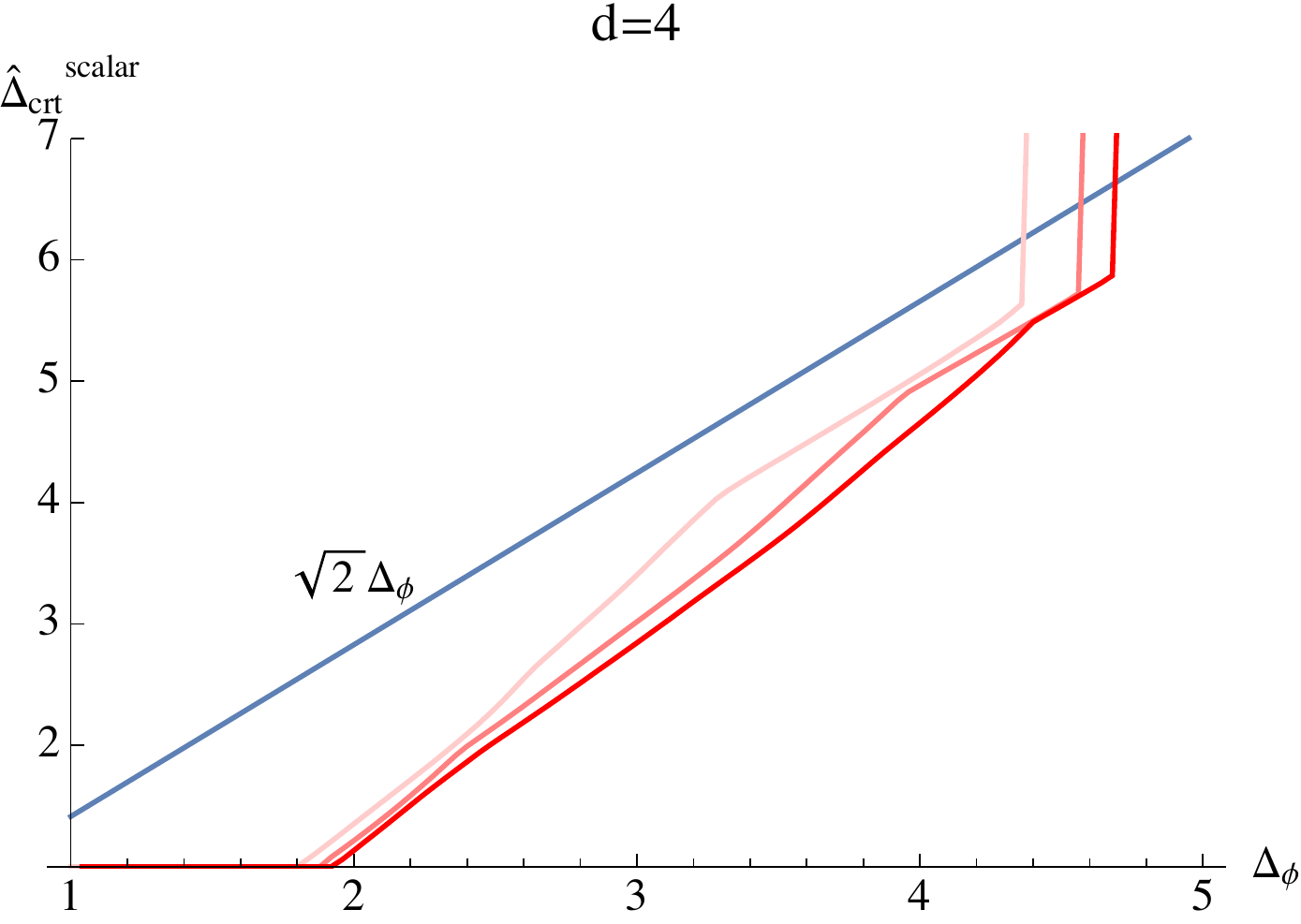}
}
\end{center}
\caption{Upper bounds on $\widehat\Delta_{crt}^{scalar}$ as a function of $\Delta_\phi$ in 2 dimensions (using global conformal blocks), 3 dimensions, and 4 dimensions.  The blue line shows the analytic bound $\sqrt 2 \Delta_\phi$ on $\widehat\Delta_{crt}$.  The red bounds are computed numerically with derivative order $12,20,28$, with the darkest line and strongest bound corresponding to derivative order 28.  
For $\Delta_\phi\lesssim 1$ in 3d and $\Delta_\phi\lesssim 2$ in 4d, the red bounds meet at the unitary bounds, thus giving universal OPE bounds in this range of $\Delta_\phi$.}
\label{fig:deltacrthigherd}
\end{figure}

We plot $\widehat\Delta_{crt}^{scalar}$ in 2 dimensions (using global blocks), 3 dimensions, and 4 dimensions in Figure~\ref{fig:deltacrthigherd}.  In all cases, the bounds are consistent with the analytic estimate $\widehat \Delta_{crt} \leq \sqrt 2 \Delta_\phi$ in the regime $\widehat\Delta_{crt} < {d}$, where $\widehat \Delta_{crt}$ and $\widehat \Delta_{crt}^{scalar}$ agree.  Beyond this regime, $\widehat \Delta_{crt}^{scalar}$ eventually jumps to a large value, and we have not explored its behavior.

Interestingly, in 3d and 4d, there are ranges of $\Delta_\phi$ where $\widehat\Delta_{crt}$ coincides with the unitarity bound: roughly $\Delta_\phi \lesssim 1$ in 3d and $\Delta_\phi \lesssim 2$ in 4d.  For $\Delta_\phi$ in this range, there always exist universal bounds on OPE coefficients and the size of the four-point function, independent of any assumptions about which operators appear in the four-point function.  Outside of these special cases, $\widehat\Delta_{crt}$ is nontrivial.\footnote{The fact that there are universal OPE bounds  when $\Delta_\phi\lesssim 1.7$ in 4d was mentioned in \cite{Caracciolo:2009bx}.  We thank Petr Kravchuk for pointing this out.} 

\subsection{$\widehat\Delta_{crt}$ for the K3 CFT}

Now, let us finally return to the K3 CFT.  Table~\ref{Tab:DeltaCritical} shows the numerical results for $\widehat\Delta_{crt}$ for several derivative orders, where we use the $\mathcal{N}=4$ conformal blocks appropriate to the K3 CFT.  Our results show rigorously that $\Delta_{crt}$ in the K3 CFT must lie below $0.29321$, at every point on the moduli space.  By extrapolating to infinite order, we find that $\widehat\Delta_{crt}$ 
is saturated, within numerical error, by the $A_1$ cigar whose continuum lies above $\Delta_{crt} = 1/4$.

As in Section~\ref{sec:mixed}, we can consider a correlator $\la \phi^{RR}\phi^{RR}\bar\phi^{RR}\bar\phi^{RR} \ra$ for two different RR-sector ${1 \over 2}$-BPS operators that are complex conjugate of each other, and bound the divergent operator of the lowest scaling dimension in the $\phi^{RR} \times \bar\phi^{RR}$ and $\phi^{RR} \times \phi^{RR}$ channels.  We fix $(\widehat\Delta_{crt}^{\phi\bar\phi}, \widehat\Delta_{crt}^{\phi\phi})$, and search for nonzero functionals $\vec\A$ that satisfy
\ie
& \vec\A \cdot \vec V_{\Delta, s}^{\phi\bar\phi} > 0 \quad \text{for} \quad \Delta > \widehat\Delta_{crt}^{\phi\bar\phi}, \quad s \in \bZ,
\\
& \vec\A \cdot \vec V_{\Delta, s}^{\phi\phi} > 0 \quad \text{for} \quad \Delta > \widehat\Delta_{crt}^{\phi\phi}, \quad s \in 2 \bZ.
\fe
If such a functional exists, then
\ie
\text{\it either} \quad \widehat\Delta_{crt}^{\phi\bar\phi} \ge \Delta^{\phi\bar\phi}_{div} \quad \text{\it or} \quad \widehat\Delta_{crt}^{\phi\phi} \ge \Delta^{\phi\phi}_{div}.
\fe

\begin{figure}[t]
\centering
\includegraphics[width=.9\textwidth]{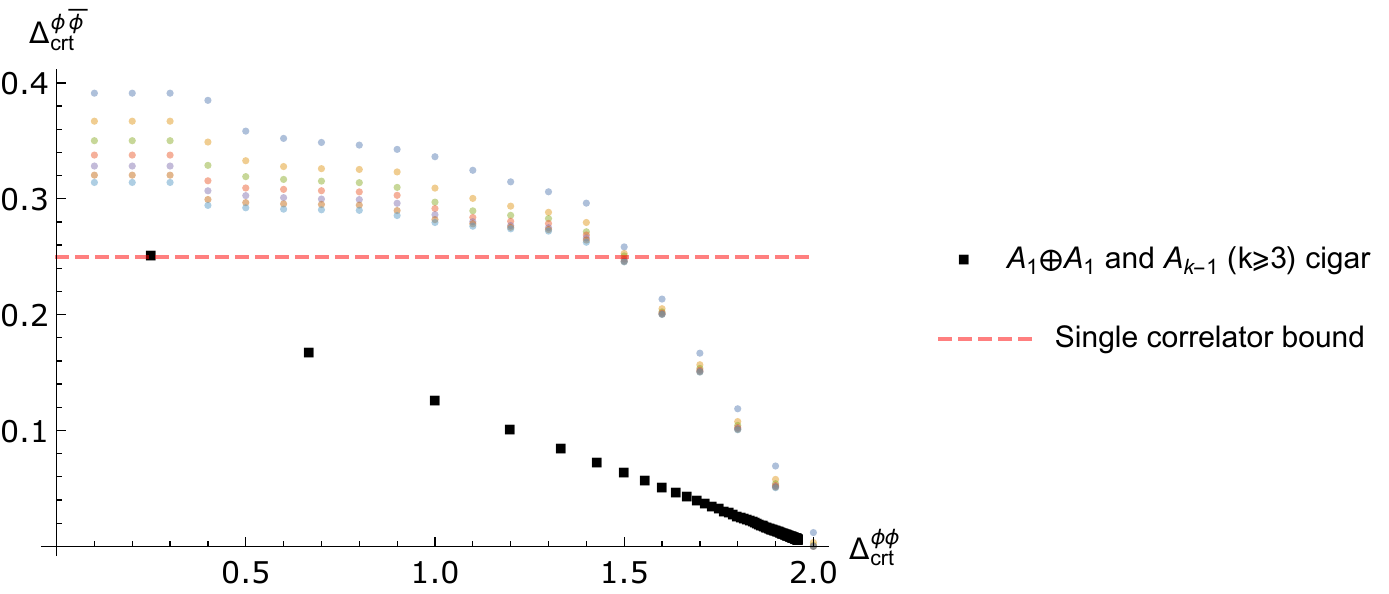}
\caption{The circle dots indicate upper bounds $( \widehat\Delta_{crt}^{\phi\phi}, \widehat\Delta_{crt}^{\phi\bar\phi})$ on the divergent operator of the lowest scaling dimension in the respective OPEs, at derivative orders ranging from 8 to 20.  
At infinite order, the bound cannot be worse than the single correlator bound 0.25 indicated by the dashed line.  We also find that $\Delta^{\phi\phi}_{crt}$ is bounded above by 2, beyond which $\Delta^{\phi\bar\phi}_{crt} = 0$.  The square dots indicate the values for the $A_1 \oplus A_1$ (at $(1/4, 1/4)$) and $A_{k-1}$ ($k \geq 3$) cigar theories. }
\label{Fig:MixedCrit}
\end{figure}

Figure~\ref{Fig:MixedCrit} shows the allowed region of $(\Delta_{crt}^{\phi\bar\phi}, \Delta_{crt}^{\phi\phi})$ obtained at various derivative orders.  For any fixed $\Delta^{\phi\phi}_{crt}$, the bound on $\Delta^{\phi\bar\phi}_{crt}$ cannot be worse than the single correlator bound $\Delta^{\phi\bar\phi}_{crt} \lesssim 0.25$.  For $\Delta^{\phi\phi}_{crt} \lesssim 1.5$, extrapolating to infinite order gives bounds on $\Delta^{\phi\bar\phi}_{crt}$ that lie close to the single correlator bound.  For $\Delta^{\phi\phi}_{crt} \gtrsim 1.5$, the bound on $\Delta^{\phi\bar\phi}_{crt}$ decreases until it reaches 0 at $\Delta^{\phi\phi}_{crt} \approx 2$.

\begin{table}[h]
\begin{center}
\vspace{.3in}
\begin{minipage}[c]{0.4\textwidth}
\begin{tabular}[c]{|c | c |}
\hline
Derivative order $d$ & $\widehat\Delta_{crt}$
\\\hline\hline
8 & 0.39111
\\
10 & 0.36693
\\
12 & 0.35011
\\
14 & 0.33768
\\
16 & 0.32822
\\
18 & 0.32037
\\
20 & 0.31407
\\
22 & 0.30886
\\
24 & 0.30447
\\
26 & 0.30075
\\
28 & 0.29742
\\
30 & 0.29321
\\\hline\hline
quadratic fit & 0.252
\\\hline\hline
$A_1$ cigar & $0.25$
\\\hline
\end{tabular}
\end{minipage}
\begin{minipage}[c]{0.5\textwidth}
\includegraphics[width=\textwidth]{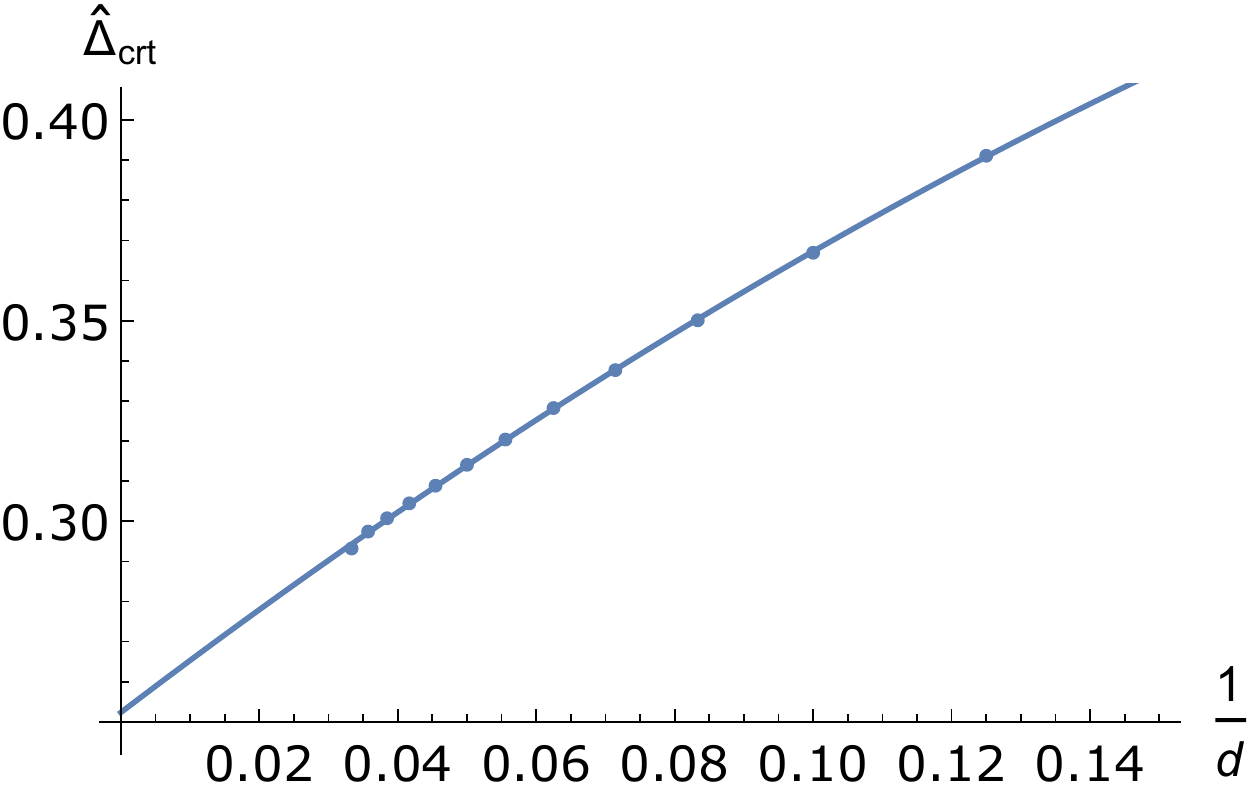}
\end{minipage}
\end{center}
\caption{Upper bound $\widehat\Delta_{crt}$ on the divergent operator of the lowest scaling dimension, as the derivative order is increased, as well as the extrapolation to infinite order using a quadratic fit.  Also shown is the value of $\Delta_{crt}$ for the $A_1$ cigar. }
\label{Tab:DeltaCritical}
\end{table}

\paragraph{$A_{k-1}$ Cigar CFT}

Let us comment on where the $A_{k-1}$ cigar CFTs analyzed in Section \ref{Sec:Cigar} sit in Figure \ref{Fig:MixedCrit}. For the cigar CFT, we take $\phi^{RR}$ and $\bar\phi^{RR}$ to be RR sector ${1\over2}$-BPS primaries $V^+_{R,\ell}$ and $V^-_{R,\ell}$ (\eqref{VR+} and \eqref{VR-}). The continua of the $A_{k-1}$ cigar CFT in $\phi^{RR} \times \phi^{RR}$ and $\phi^{RR}\times\bar\phi^{RR}$ start at $\Delta^{\phi\phi}_{cont} = (k-2\ell-1)^2/ 2k$ and $\Delta^{\phi\bar\phi}_{cont} =1/2k$, respectively (see \eqref{ppcont} and \eqref{phibphi}). For $k\ge 4$, there are discrete state contributions to the four-point function in the channel $\phi^{RR}\times \phi^{RR}$ starting at $\Delta_{discrete} = 2 - 4(1+\ell)/k$. As argued in Section \ref{Sec:Cigar}, their OPE coefficients are divergent when compared with a generic K3 CFT.  Since $\Delta_{crt}$ is defined as the lowest scaling dimension such that \textit{either} a continuous spectrum appears or the structure constants of some states in the discrete spectrum diverge, we have
\ie
\Delta_{crt}^{\phi\phi} = \text{min}\left( \Delta_{cont}^{\phi\phi} , \Delta_{discrete}^{\phi\phi} \right)
= 
\begin{cases}
{(k-2\ell +1)^2\over 2k},~~~\quad~~~\text{if}~~k=2,3,\\
2-  {4 (1+\ell)\over k},~~~~~\,~~\text{if}~~k\ge 4,
\end{cases}
\fe
in the OPE channel between $V_{R,\ell}^+$ and $V_{R,\ell}^+$ in the $A_{k-1}$ cigar CFT. On the other hand, in the OPE channel between $V_{R,\ell}^+$ and $V_{R,\ell}^-$,  $\Delta_{crt}^{\phi\bar\phi} = \Delta_{cont}^{\phi\bar\phi}=1/2k$ as in \eqref{phibphi}.  We would like to emphasize that the presence of these R-charge non-singlet discrete states below the continuum is crucial for the consistency with the bootstrap bound derived from the crossing equations.

In Figure \ref{Fig:MixedCrit}, the point  $(1/4, 1/4)$ in the OPE of $\phi\phi$ and $\phi\bar\phi$ can be realized at an $A_1\oplus A_1$ point on the moduli space, and the other black dots at $A_{k-1}$ points with $k\geq 3$ which asymptote to $(2, 0)$ at large $k$.\footnote{The minimal resolution of an ADE singularity of rank $\m$ gives $\m$ exceptional divisors which are dual to self-dual elements of $H^{1,1}(K3)$, thus $\m \leq 19$.  In particular, the K3 surface can develop an $A_{k}$ singularity only for $k\leq 19$.  However our bound on $\widehat\Delta_{crt}$ is insensitive to the identity superconformal block contribution, and applies to noncompact theories as well, such as nonlinear sigma model on ALE spaces \cite{Anselmi:1993sm} and the ${\cal N}=4$ cigar CFTs. } 

\section{The Large Volume Limit}

In this section we consider the gap in the OPE of ${1\over 2}$-BPS operators in the large volume regime of the K3 CFT. Based on unitarity constraints on the superconformal block decomposition of the BPS 4-point function (but without making direct use of the crossing equation), we will derive an upper bound on the gap, which remains nontrivial in the large volume regime, and leads to an interesting inequality that relates the first nonzero eigenvalue of the scalar Laplacian on the K3 to an integral constructed from a harmonic 2-form, and data of the lattice $\Gamma_{19,3}$ that parameterize the K3 moduli.  The eigenvalues of the Laplacian on K3 can be studied using the explicit numerical metric in \cite{Douglas:aa,Headrick:2009jz}.

\subsection{Parameterization of the K3 Moduli}

The quantum moduli space of the K3 CFT can be parameterized by the embedding of the lattice $\Gamma_{20,4}$ into $\mathbb{R}^{20,4}$, or equivalently, the choice of a positive 4-dimensional hyperplane in the span of $\Gamma_{20,4}$. Let us write $\Gamma_{20,4}$ as $\Gamma_{1,1}\oplus \Gamma_{19,3}$, with the $\Gamma_{19,3}$ identified with the cohomology lattice $H^2(K3,\mathbb{Z})$ \cite{Aspinwall:1996mn}. Let $u,v$ be a pair of null basis vectors of the $\Gamma_{1,1}$, with $u^2=v^2=0$, $u\cdot v=1$. Let $\Omega_i$ ($i=1,2,3$) be a triplet of $H^2(K3,\mathbb{R})$ classes associated with the hyperk\"ahler structure of the K3 surface, normalized so that $\Omega_i\cdot\Omega_j=\delta_{ij}$. We will denote by $B$ the cohomology class of a flat $B$-field, and by $V$ the volume of the K3 surface (more precisely it is $(2\pi)^4$ times the volume in units of $\A'^2$). An orthonormal basis of the 4-dimensional positive hyperplane is \cite{Aspinwall:1996mn}
\ie
& E_0 = {(V-{B^2\over 2}) u + v + B\over \sqrt{2V}},
\\
& E_i = - B\cdot\Omega_i u + \Omega_i,~~~i=1,2,3.
\fe
Now an orthonormal basis of the 20-dimensional negative subspace can be constructed as
\ie
& e_0 = {-(V+{B^2\over 2}) u + v + B\over \sqrt{2V}},
\\
& e_\A = B\cdot W_\A u + W_\A, ~~~\A=1,\cdots,19,
\fe
where $W_\A\in {\rm span}(\Gamma_{19,3})$ are a set of orthonormal vectors that are orthogonal to $\Omega_i$, and correspond to a basis of anti-self-dual harmonic 2-forms on the K3 surface.

A general lattice vector of $\Gamma_{20,4}$ can be written as
\ie
\ell = n u + m v + \A,
\fe
where $\A\in H^2(K3,\mathbb{Z})\simeq \Gamma_{19,3}$. Let $\A_+$ be the self-dual projection of $\A$, or equivalently, $\A_+ = \sum_{i=1}^3(\A\cdot\Omega_i ) \Omega_i$. We have
\ie
& \ell\circ\ell = -\ell_L^2 + \ell_R^2 = \A^2 + 2nm,
\fe
and
\ie
\ell_R^2 &= (\ell\cdot E_0)^2 + \sum_{i=1}^3 (\ell\cdot E_i)^2
\\
&= (\A - m B)_+^2 + {\left[ \A\cdot B + n + m (V-{B^2\over 2}) \right]^2 \over 2V}.
\fe
We can now write the theta function 
\ie
\Theta_{\Gamma_{20,4}}(\tau,\bar\tau|y) = e^{{\pi\over 2\tau_2}y^2} \sum_{n,m\in\mathbb{Z},~\A\in\Gamma_{19,3}} q^{\ell_L^2\over 2} \bar q^{\ell_R^2\over 2} e^{2\pi i \ell_L\cdot y},
\fe
where $y\in \mathbb{R}^{20}$, and $\ell_L\cdot y\equiv \sum_{a=0}^{19} (-\ell\cdot e_a) y_a$. 
In the large volume $V$ limit, we can restrict to the sum to $m=0$ term, and replace the summation over $n$ by an integral. The integrated 4-point function of BPS operators \eqref{aexp} associated with deformations of $\Gamma_{19,3}$ (as opposed to the overall volume modulus, parameterizing the embedding of $\Gamma_{1,1}$) becomes
\ie
A_{\A\B\C\D} \to {\sqrt{V} \over 16\pi^2} \int_{\cal F} {d^2\tau \over \tau_2^{1\over 2} \eta(\tau)^{24}} \left.{\partial^4\over \partial y^\A\partial y^\B\partial y^\C\partial y^\D}\right|_{y=0}\Theta_{19,3}(\tau,\bar\tau| y).
\fe
Note that this result does not apply to the integrated 4-point function of the BPS operator associated to the volume modulus, which in fact vanishes in the large volume limit.

\subsection{Bounding the First Nonzero Eigenvalue of the Scalar Laplacian on K3}

Let us write the four-point function of a given ${1\over 2}$-BPS, weight $({1\over 4},{1\over 4})$ operator in the RR sector $\phi^{RR}$, which is related to a weight $({1\over 2},{1\over 2})$ NS-NS primary by spectral flow, as
\ie
\left\langle \phi^{RR}(z,\bar z) \phi^{RR}(0) \phi^{RR}(1) \phi^{RR}(\infty) \right\rangle = f(z,\bar z).
\fe
We have
\ie
A &\equiv \lim_{\epsilon\to 0} \int_{|z|,|1-z|,|z|^{-1}>\epsilon} {d^2z\over |z(1-z)|} f(z,\bar z)  + 6 \pi \ln\epsilon
\\
&= {1\over 16\pi^2} \left.{\partial^4\over\partial y^4}\right|_{y=0} \int_{\cal F} d^2\tau {\Theta_{\Gamma_{20,4}}(\tau,\bar\tau|y)\over \eta(\tau)^{24}}.
\fe
$f(z,\bar z)$ admits a conformal block decomposition (in the $z\to 0$ channel) of the form
\ie
f(z,\bar z) = |{\cal F}^R_0(z)|^2 + \sum_{h_L,h_R} C_{h_L,h_R}^2 {\cal F}^R_{h_L}(z)\overline{{\cal F}^R_{h_R}(z)},
\fe
where according to our claim \eqref{idkey} 
\ie
{\cal F}^R_h(z) = z^{1\over 2}(1-z)^{1\over 2} { F}^{Vir}_{c=28}(1,1,1,1;h+1;z),
\fe
and ${\cal F}^{Vir}_c(h_1,h_2,h_3,h_4;h;z)$ is the sphere four-point conformal block of the Virasoro algebra of central charge $c$. We can write
\ie
{\cal F}^R_h(z) = (z(1-z))^{-{1\over 3}} \theta_3(q)^{-2} g_h(q),
\fe
where the function $g_h(q)$ takes the form
\ie
g_h(q) = q^{h-{1\over 6}} \sum_{n=0}^\infty a_n q^n,~~~~ a_n\geq 0.
\fe
Positivity of the $a_n$ follows from reflection positivity of the theory on the pillowcase \cite{Maldacena:2015iua}.
In particular, we learn that ${\cal F}^R_h(z)$ obeys the inequality
\ie
\left| {{\cal F}^R_h(z)\over (z(1-z))^{-{1\over 3}} \theta_3(q)^{-2} q^{h-{1\over 6}}} \right| \leq  {{\cal F}^R_h(z_*)\over (z_*(1-z_*))^{-{1\over 3}} \theta_3(q_*)^{-2} q_*^{h-{1\over 6}}} ,
\fe
for $|q(z)|\leq q(z_*)\equiv q_*$, $0<z_*<1$ and $0<q_*<1$.

In the large volume limit, $A$ is dominated by the contribution from light non-BPS operators in the OPE, integrated near $z=0$, 1 or $\infty$. Let us assume that there is a gap $\Delta_0$ in the spectrum of non-BPS (scalar) primaries. We can write in this limit
\ie
A &\approx 3 \sum_{\Delta_0\leq \Delta \leq \Lambda} C_\Delta^2 \int_{|z|<\delta} {d^2z\over |z(1-z)|} \left| {\cal F}^R_{\Delta\over 2}(z) \right|^2 \leq {6\pi\over\Delta_0} 2^{4\over 3} \sum_{\Delta_0\leq \Delta\leq \Lambda} C_\Delta^2 \left[ {{\cal F}^R_{\Delta\over 2}(z_*)\over (z_*(1-z_*))^{-{1\over 3}} \theta_3(q_*)^{-2} q_*^{{\Lambda\over 2}-{1\over 6}}} \right]^2
\\
& \leq {6\pi\over \Delta_0} 2^{4\over 3}  (z_*(1-z_*))^{2\over 3} \theta_3(q_*)^{4} q_*^{-{\Lambda}+{1\over 3}} \left[ f(z_*) - |{\cal F}^R_0(z_*)|^2\right].
\fe
In the first approximation, we have dropped finite contributions that are unimportant in the large volume limit, where $A$ diverges like $V^{1\over 2}$, while $\Delta_0$ goes to zero like $V^{-{1\over 2}}$. Here $\Lambda$ is a cutoff on the operator dimension that can be made small but finite, and $\delta$($\leq z_*$) is a small positive number. Taking $\Lambda$ to zero {\it after} taking the large volume limit, we derive the bound (which holds only in the large volume limit)
\ie
\Delta_0 A \leq 6\pi (z_*(1-z_*))^{2\over 3} \theta_3(q_*)^{4} (16q_*)^{1\over 3} \left[ f(z_*) - |{\cal F}^R_0(z_*)|^2\right].
\fe
One might be attempted to take $z_*$ to be small, but $f(z_*)$ diverges in the small $z_*$ limit. In practice, we can simply choose $z_*={1\over 2}$, and arrive at the large volume bound
\ie\label{lvb}
\Delta_0 A \leq 6\pi \theta_3(q_{1\over 2})^{4} (q_{1\over 2})^{1\over 3} \left[ f(1/2) - |{\cal F}^R_0(1/2)|^2\right],
\fe
where $q_{1\over 2} \equiv q(z={1\over 2})=e^{-\pi}$. Note that for generic Einstein metric on the K3, the four-point function $f({1\over 2})$ remains finite in the infinite volume limit. In this limit, we can identify $\Delta_0 = \lambda_1/2$, where $\lambda_1$ is the first nonzero eigenvalue of the scalar Laplacian on the K3 surface, in units of $\A'$.\footnote{It is known \cite{Cheng1975, Li1980} that
\ie
{\pi^2\over 4d^2}\leq \lambda_1 \leq {4\pi^2\over d^2},
\fe
where $d$ is the diameter of the K3. The compatibility with our large volume bound then demands an inequality relating the diameter of the K3 to $f(1/2)$ and $A$.
}

Let $\omega=\omega_{i\bar j} dz^i dz^{\bar j}$ be a harmonic $(1,1)$-form that is orthogonal to the K\"ahler form, normalized such that $V^{-1}\int_{K3} \sqrt{g} \omega_{i\bar j} \omega^{i\bar j}=1$. Let ${\cal O}_\omega^{\pm\pm}$ be the BPS primary associated with the corresponding moduli deformation. We have for instance ${\cal O}^{++}_\omega\approx \omega_{i\bar j} \psi^i \widetilde\psi^{\bar j}$, ${\cal O}^{--}_\omega\approx \omega_{i\bar j} \psi^{\bar j} \widetilde\psi^i$ in the large volume limit.  The 4-point function of the corresponding $\phi_\omega^{RR}$ evaluated at $z={1\over 2}$ is
\ie
f_\omega(1/2) \approx {1\over V} \int_{K3}\sqrt{g} \left[ 5(\omega^2)^2 - 4 \omega^4 \right],
\fe
where $\omega^2 \equiv \omega_{i\bar j}\omega^{i\bar j}$, $\omega^4 = \omega_{i\bar j}\omega^{k\bar j}\omega_{k\bar \ell}\omega_{i\bar \ell}$. Thus, we derive the following upper bound on $\lambda_1$,
\ie
\lambda_1 \leq {192\pi^3 \theta_3(q_{1\over 2})^{4} (q_{1\over 2})^{1\over 3} \left[ f_\omega({1\over 2}) - |{\cal F}^R_0({1\over 2})|^2\right]\over \sqrt{V} \int_{\cal F} d^2\tau \,\tau_2^{-{1\over 2}}\eta(\tau)^{-24} \Theta_{19,3}^{\omega}(\tau,\bar\tau) },
\fe
with
\ie
\Theta_{19,3}^{\omega}(\tau,\bar\tau) \equiv \left. {\partial^4\over \partial y_\omega^4}\right|_{y_\omega=0} \Theta_{19,3}(\tau,\bar\tau|y_\omega e_\omega),
\fe
where $e_\omega$ is the unit vector in $\mathbb{R}^{20}$ associated with the deformation ${\cal O}_\omega$.

The upper bound (\ref{lvb}) was derived by consideration of the 4-point function of a single ${1\over 2}$-BPS primary ${\cal O}_\omega$, and applies to the gap in the OPE of ${\cal O}_\omega$ with itself. We see that in the large volume limit, a light scalar non-BPS operator must appear in such an OPE, provided that $\omega$ is not proportional to the K\"ahler form, so that $A$ scales like $\sqrt{V}$. As noted earlier, if we take $\omega$ to be the K\"ahler form $J$ itself, the corresponding BPS operator ${\cal O}_J$ would have an integrated 4-point function $A$ that vanishes in the large volume limit instead, and we cannot deduce the existence of a light operator in the OPE of ${\cal O}_J$ with itself.

\section{Summary and Discussion}

Let us summarize the main results of this paper.

\begin{enumerate}
\item By analyzing the ${\cal N}=4$ $A_1$ cigar CFT, we found an exact relation between the BPS four-point $c=6$ ${\cal N}=4$ superconformal block and the bosonic Virasoro conformal block of central charge $c=28$. Further, a class of BPS ${\cal N}=2$ superconformal blocks with central charge $c={3(k+2)\over k}$ are identified, up to a simple known factor, with Virasoro blocks of central charge $c=13+6k+{6\over k}$ and shifted weights.

\item We derived a lower bound on the four-point function of a ${1\over 2}$-BPS primary by the integrated four-point function $A_{1111}$, assuming the existence of a gap in the spectrum. We also determined $A_{ijkl}$ as an exact function of the K3 CFT moduli (parameterized by the embedding of the lattice $\Gamma_{20,4}$).

\item We found an upper bound on the lowest dimension non-BPS primary appearing in the OPE of two identical ${1\over 2}$-BPS primaries, as a function of the BPS four-point function evaluated at the cross ratio $z={1\over 2}$, and as a function of $A_{1111}$ (thus  a known function on the moduli space of the K3 CFT).  Both vary monotonously from 2 to ${1\over 4}$, and interpolate between the the untwisted sector of the free orbifold CFT and the $A_1$ cigar CFT. It is also observed that $A_{1111}$ must be non-negative from the bootstrap constraints (see Figure \ref{Fig:A1111}), which is consistent with the superluminal bound on the  $H^4$ coefficient in the 6d (2,0) supergravity coming from IIB string theory compactified on K3. 

\item Bounding the contribution to the BPS four-point function by contributions from non-BPS primaries of scaling dimension below $\widehat\Delta_{crt}$, and assuming the boundedness of the OPE coefficients, we deduce that a continuum in the spectrum develops near the ADE singular points on the K3 CFT moduli space, and find numerically that $\widehat\Delta_{crt}$ agrees with the gap below the continuum in the $A_1$ cigar CFT, namely ${1\over 4}$.

\item We explored the possibility of the appearance of either a continuum or divergent contribution from discrete non-BPS operators in the OPE of two distinct ${1\over 2}$-BPS operators, near a singular point of the moduli space where the BPS four-point function diverges (beyond the $A_1$ case). The bootstrap bounds we found are consistent with the spectrum and OPE of the $\mathcal{N}=4$ $A_{k-1}$ cigar theory, and know about the appearance of discrete non-BPS primaries in the OPE below the continuum gap.

\item For general CFTs in 2,3,4 spacetime dimensions, we derived a crude analytic bound $\widehat \Delta_{crt}\le \sqrt{2} \Delta_\phi$,  where $\Delta_\phi$ is  the scaling dimension of the external scalar operator. It was observed (see Figure \ref{fig:deltacrthigherd}) from the stronger numerical bounds on $\widehat\Delta_{crt}$ that they  meet at the unitarity bounds for  $\Delta_\phi\lesssim 1$ in 3 spacetime dimensions and $\Delta_\phi \lesssim 2$ in 4 spacetime dimensions, thus providing  universal upper bounds on the four-point functions for this range of external operator dimension.

\item Independently of the crossing equation, but using nonetheless unitarity and exact results of the integrated BPS four-point function, we derived in the large volume regime a bound that is meaningful in classical geometry, namely an upper bound on the first nonzero eigenvalue of the scalar Laplacian on K3 surface, that depends on the moduli of Einstein metrics on K3 (parameterized by the embedding of the lattice $\Gamma_{19,3}$) and an integral constructed out of a harmonic 2-form on the K3. 
\end{enumerate}
\bigskip

While we have exhibited some of the powers of the crossing equation based on the full ${\cal N}=4$ superconformal algebra, clearly much more can be said regarding the non-BPS spectrum and OPEs in the K3 CFT over the entire moduli space. We would like to understand to what extent our bootstrap bounds can be saturated, away from free orbifold and cigar points in the moduli space. In particular, it would be interesting to compare with results from conformal perturbation theory.

Apart from a few basic vanishing results, the OPEs of the ${1\over 4}$-BPS primaries remain largely unexplored. Neither have we investigated the torus correlation functions, which should provide further constraints on the non-BPS spectrum. Note that there are certain integrated torus four-point functions, analogous to $A_{ijkl}$ and $B_{ij,kl}$, that can be determined as exact functions of the moduli, by expanding the result of \cite{Lin:2015dsa} perturbatively in the type IIB string coupling.

There are a number of important generalizations of our bootstrap analysis that will be left to future work. One of them is to derive bootstrap bounds on the non-BPS spectrum of $(2,2)$ superconformal theories, with input from the known chiral ring relations. To do so, we will need to extend the results of section \ref{sec:N=2block} to ones that express a more general set of BPS ${\cal N}=2$ superconformal blocks in terms of Virasoro conformal blocks (of a different central charge and shifted weights). These relations can be extracted from BPS correlators of the ${\cal N}=2$ $SL(2)_k/U(1)$ cigar CFT (or the T-dual ${\cal N}=2$ Liouville theory \cite{Hori:2001ax}), and will be presented in detail elsewhere.  

Another generalization would be to extend our analysis to $(4,4)$ superconformal theories of higher central charge, namely $c=6k'$ for $k'\geq 2$, and use it to understand the appearance of a continuous spectrum in the D1-D5 CFT at various singular points on its moduli space. There is conceivably a generalization of our relation between the $c=6$ ${\cal N}=4$ block and bosonic Virasoro blocks, to the $k'\geq 2$ case. This is currently under investigation.

Finally, our numerical bounds on $\widehat\Delta_{crt}$ seem to allow for the possibility of having an arbitrarily large four-point function  when $\Delta_\phi\gtrsim 1$ in 3 spacetime dimensions and $\Delta_\phi\gtrsim 2$ in 4 spacetime dimensions. We are not aware of  an example of such a CFT.  It is conceivable that such a CFT will be ruled out by unitarity constraints from other correlation functions, but this remains to be seen.

\section*{Acknowledgments}

We would like to thank Chris Beem, Clay C\'ordova, Thomas Dumitrescu, Matthew Headrick, Christoph Keller, Petr Kravchuk, Sarah Harrison, Juan Maldacena, Hirosi Ooguri, Nati Seiberg, Steve Shenker, Cumrun Vafa, Shing-Tung Yau, and Alexander Zhiboedov for discussions.  We would like to thank the workshop ``From Scattering Amplitudes to the Conformal Bootstrap" at Aspen Center for Physics, the Simons Summer Workshop in Mathematics and Physics 2015, and the workshop Amplitudes in Asia 2015, for hospitality during the course of this work.  DSD is supported by DOE grant DE-SC0009988 and a William D. Loughlin Membership at the Institute for Advanced Study. YW is supported in part by the U.S. Department of Energy under grant Contract Number DE-SC00012567. XY is supported by a Simons Investigator Award from the Simons Foundation, and in part by DOE grant
DE-FG02-91ER40654.

\appendix

\section{The Integrated Four-Point Function $A_{ijkl}$ at the $T^4/\mathbb{Z}_2$ CFT Orbifold Point}\label{app:T4}

In this Appendix we compare the proposed exact formula for the integrated four-point function $A_{ijkl}$ to explicit computation of the four-point function of twist fields in the $T^4/\mathbb{Z}_2$ free orbifold CFT. The twist fields of the latter are associated with the 16 $\mathbb{Z}_2$ fixed points on the $T^4$. We will focus on the case where $i,j,k,l$ label the same $\mathbb{Z}_2$ fixed point (denote by $i=j=k=l=1$). The result as given in \cite{Gluck:2005wr} is
\ie\label{aone}
A_{1111} = 6\pi^2 \int_{\cal F} d^2\tau \sum_{\ell\in \widetilde\Gamma_{4,4}} \exp\left( i\pi\tau \ell_L^2 - i\pi \bar\tau \ell_R^2 \right).
\fe
Here $\tau$ is related to the cross ratio $z$ by the mapping $\tau = i F(1-z)/F(z)$, $F={}_2F_1({1\over 2},{1\over 2};1;z)$. $\widetilde\Gamma_{4,4}$ is the Narain lattice associated with the $T^4$ with all radii {\it rescaled by} $\sqrt{2}$. The factor 6 comes from the integration over the fundamental domain of $\Gamma(2)$, which consists of 6 copies of the $PSL(2,\mathbb{Z})$ fundamental domain ${\cal F}$. Note that the $\ell=0$ term in the lattice sum leads to a divergent integral, which is regularized by analytic continuation in the Mandelstam variables $s,t,u$ and then dropping the polar terms in the $s,t,u\to 0$ limit as before.


We will take the original $T^4$ (before orbifolding) to be a rectangular torus with radii $R_i$, $i=1,\cdots,4$. To compare (\ref{aone}) with our exact formula for $A_{ijkl}$ as a function of the K3 moduli, we need to construct the lattice embedding $\Gamma_{20,4}\subset \mathbb{R}^{20,4}$ that corresponds to the $T^4/\mathbb{Z}_2$ CFT orbifold, as follows.
We will write $\mathbb{R}^{20,4}=(\mathbb{R}^{1,1})^{\oplus 4}\oplus \mathbb{R}^{16}$. Let $(u_i,v_i)$ be pairs of null vectors in the four $\mathbb{R}^{1,1}$ factors, such that $u_i\cdot v_i=1$. Denote by $w^L$ and $w^R$ the projection of a vector $w\in \mathbb{R}^{20,4}$ in the positive and negative subspaces, $\mathbb{R}^{20}$ and $\mathbb{R}^4$ respectively. We can write $|u^L_i|=|u^R_i|=\sqrt{\A'_h\over 2}{1\over  R_i^h}$, $|v^L_i|=|v^R_i|= \sqrt{1\over 2 \A'_h}{ R_i^h}$. 
Note that, importantly, $R_i^h$ are {\it not} to be identified with $R_i$. Rather, they are related by (see (2.5), (2.6) and footnote 2 of \cite{Bergman:1999kq})
\ie
{R_i\over \sqrt{\A'}} = {\sqrt{2R_1^h R_2^h R_3^h R_4^h} \over \sqrt{\A'_h} R_i^h}.
\fe
Let $A_i$ be the following vectors in the $\mathbb{R}^{16}$,
\ie
& A_1 = {1\over 2} (1, 1, 1, 1, 1, 1, 1, 1, 0, 0, 0, 0, 0, 0, 0, 0),
\\
&A_2 = {1\over 2} (1, 1, 1, 1, 0, 0, 0, 0, 1, 1, 1, 1, 0, 0, 0, 0),
\\
& A_3 = {1\over 2} (1, 1, 0, 0, 1, 1, 0, 0, 1, 1, 0, 0, 1, 1, 0, 0),
\\
& A_4 = {1\over 2} (1, 0, 1, 0, 1, 0, 1, 0, 1, 0, 1, 0, 1, 0, 1, 0).
\fe
Note that $A_i\cdot A_j = 1 + \delta_{ij}$. Let $\Gamma_{16}$ be the root plus chiral spinor weight lattice of $SO(32)$ embedded in the $\mathbb{R}^{16}$, generated by the root vectors $(0,\cdots,0,1,-1,0,\cdots,0)$, and $(\pm {1\over 2},\cdots,\pm{1\over 2})$ with even number of minuses. Now $\Gamma_{20,4}$ can be constructed as the span of the following generators
\ie
u_i,~~~ A_i + v_i - \sum_{j=1}^4 {A_i\cdot A_j\over 2} u_j,~~~\vec \ell - \sum_{j=1}^{4} ( \ell\cdot A_j) u_j,~~\ell\in\Gamma_{16}.
\fe
One can verify that this lattice is indeed even and unimodular.\footnote{This lattice can also be used to describe the compactification of $SO(32)$ heterotic string on a rectangular $T^4$ with radii $R_i^h$ and Wilson line turned on.  This can be seen from the large $R_i^h$ limit, where $u_i$ and $v_i$ are approximations to primitive lattice vectors. Note that in the opposite limit, say small $R_1^h$, ${u_1\over 2}$ and $2v_1$ are approximations to primitive lattice vectors. This means that the T-dual $E_8\times E_8$ heterotic string lives on a circle of radius $\widetilde R_1^h = {\A'_h\over 2 R_1^h}$. Note that the T-duality on all four circles of the heterotic $T^4$, taking $R_i^h\to {\A'_h\over 2 R_i^h}$, is equivalent to sending $R_i \to {\A'\over R_i}$, namely T-dualizing all four directions of the $T^4/\mathbb{Z}_2$ orbifold, in the type IIA dual.}

In the large $R_i$ limit, we can approximate the theta function of $\Gamma_{20,4}$ as
\ie
\theta_\Lambda(y|\tau,\bar\tau) \approx {\prod_{i=1}^4 R^h_i \over {\A'_h}^2\tau_2^2} \theta_{\Gamma_{16}}(y|\tau,\bar\tau) = {R_1R_2R_3R_4\over 4 {\A'}^2 \tau_2^2}  \theta_{\Gamma_{16}}(y|\tau,\bar\tau).
\fe
Note that the $\bar\tau$-dependence of $\theta_{\Gamma_{16}}$ is entirely through the factor $e^{{\pi\over 2\tau_2} y^2}$. 
We can then evaluate the integral
\ie
\int_{\cal F}{d^2\tau\over \tau_2^2} \left. {\theta_{\Gamma_{16}}(y|\tau,\bar\tau)\over \eta(\tau)^{24}}\right|_{y^4}
& =\int_{\cal F} d^2\tau  \left.{4i\over \pi y^2}  {\partial\over\partial\overline\tau} {\theta_{\Gamma_{16}}(y|\tau,\bar\tau)\over \eta(\tau)^{24}} \right|_{y^4}
\\
&= \oint_{\partial {\cal F}} d\tau \left.{4\over \pi y^2} {\theta_{\Gamma_{16}}(y|\tau,\bar\tau)\over \eta(\tau)^{24}}\right|_{y^4}
\\
&= - {4\over \pi y^2} \left. {\widetilde\theta_{\Gamma_{16}}(y|\tau)\over \eta(\tau)^{24}} \right|_{y^4 q^0}.
\fe
Here $d^2\tau\equiv 2d\tau_1d\tau_2$.
In the last line, the holomorphic function $\widetilde\theta_{\Gamma_{16}}(y|\tau)$ is $\theta_{\Gamma_{16}}$ with the $e^{{\pi\over 2\tau_2} y^2}$ factor dropped, due to the $\tau_2\to\infty$ limit taken in going to the boundary of ${\cal F}$. Furthermore, only the $y^4$ term is kept in the Laurent expansion in $y$, and in particular the constant term 1 in the lattice sum in $\widetilde\theta_{\Gamma_{16}}$ does not contribute. The only contribution comes from the terms of order $q$ in $\widetilde\theta_{\Gamma_{16}}(y|\tau)$, 
giving
\ie
\left.\int_{\cal F}{d^2\tau\over \tau_2^2} {\theta_{\Gamma_{16}}(y|\tau,\bar\tau)\over \eta(\tau)^{24}}\right|_{y^4} = -{4\over \pi y^2}\left. \widetilde\theta_{\Gamma_{16}}(y|\tau)\right|_{q^1 y^4}.
\fe
In particular,
\ie\label{tta}
\left.{\partial^4\over \partial y_1^4}\right|_{y=0} \int_{\cal F}{d^2\tau\over \tau_2^2} {\theta_{\Gamma_{16}}(y|\tau,\bar\tau)\over \eta(\tau)^{24}} = {4\over \pi} (2\pi)^6 {4!\over 6!}\cdot 60 = 2^{10}\pi^5.
\fe
The factor 60 comes from the sum of $(E_a\cdot \hat e_1)^6$ for all root vectors $E_a$ of $so(32)$, with $\hat e_1=(1,0,\cdots,0)$.

Note that in the large radii limit, the four-point function of twist fields at a given cross ratio in the free orbifold CFT diverges like the volume, as is $A_{ijkl}$.\footnote{This is to be contrasted with the large volume limit of a smooth K3, where the four-point function of BPS operators remain finite at generic cross ratio, while $A_{ijkl}$ diverges like the square root of volume.} Comparison with (\ref{aone}) then fixes the overall normalization of $A_{ijk\ell}$ as a function of moduli to be that of (\ref{aexp}).

\section{Conformal Blocks under the $q$-Map}
\label{Sec:qmap}

The four-punctured sphere can be uniformized by a map $T^2 / \bZ_2 \to S^2$ \cite{Zamolodchikov:1985ie,Zamolodchikov:1995aa,Maldacena:2015iua}.  The complex moduli $\tau$ of the $T^2$ is related to the cross ratio $z$ of the four punctures by a map
\ie
z \to \tau(z) \equiv {i F(1-z) \over F(z)}, \quad F(z) = {}_2F_1({1/2}, {1/2}, 1 | z).
\fe
Because $\tau$ lies in the upper half plane, the ``nome'' defined as
\ie
\label{qmap}
q(z) \equiv \exp(i\pi\tau(z))
\fe
has the property that its value lies inside the unit disk.  We shall simply refer to this map $z \to q(z)$ as the $q$-map.  The $q$-map has a branch cut at $(1, \infty)$; the value of $q(z)$ covered by one branch is shown in Figure~\ref{qmap}, and crossing to other branches brings us outside this eye-shaped region.  Also shown are the regions $D_1$, $D_2$, $D_3$ defined by
\ie\label{regions}
& D_1: ~~ |z|<1, ~~{\rm Re}\, z<{1\over 2},
\\
& D_2: ~~ |z-1|<1, ~~{\rm Re}\, z>{1\over 2},
\\
& D_3: ~~ |z|>1, ~~|1-z|>1,
\fe
each of which contains two fundamental domains of the $S_3$ crossing symmetry group.

The holomorphic Virasoro block for a four-point function $\langle {\cal O}_1(z) {\cal O}_2(0) {\cal O}_3(1) {\cal O}_4(\infty) \rangle$ with central charge $c$, external weights $h_i$, and intermediate weight $h$ has the following representation
\ie
F_c^{Vir}(h_i, h | z) = {(16 q)^{h - {c-1 \over 24}} x^{{c-1 \over 24} - h_1 - h_2}} (1-x)^{{c-1 \over 24} - h_1 - h_3} [\theta_3(q)]^{{c-1 \over 8} - 4(h_1 + h_2 + h_3 + h_4)} H( \lambda_i^2, h | q).
\fe
If we define
\ie
c = 1 + 6Q^2, \quad Q = b + {1 \over b}, \quad h_{m,n} = {Q^2\over 4} - \lambda_{m,n}^2, \quad \lambda_{m,n} = {1\over 2} ({m\over b} + nb),
\fe
then $H( \lambda_i^2, h | q)$ satisfies Zamolodchikov's recurrence relation \cite{Zamolodchikov:1985ie,Zamolodchikov:1995aa} 
\ie\label{recH}
H(\lambda_i^2, h | q) = 1 + \sum_{m,n\geq 1} {q^{mn} R_{m,n}(\{\lambda_i\}) \over h - h_{m,n} }
H(\lambda_i^2, h_{m,n} + mn|q),
\fe
where $h_{m,n}$ are the conformal weights of degenerate representations of the Virasoro algebra, and $R_{m,n}(\{\lambda_i \})$ are given by
\ie
R_{m,n}(\{\lambda_i \}) = 2 {\prod_{r,s} (\lambda_1+\lambda_2 - \lambda_{r,s}) (\lambda_1-\lambda_2 - \lambda_{r,s}) (\lambda_3+\lambda_4 - \lambda_{r,s}) (\lambda_3-\lambda_4 - \lambda_{r,s}) \over \prod_{k,\ell}' \lambda_{k,\ell} }.
\fe
The product of $(r,s)$ is taken over
\ie\label{rsrange}
& r = -m+1, -m+3, \cdots, m-1,
\\
& s = -n+1, -n+3, \cdots, n-1,
\fe
and the product of $(k,\ell)$ is taken over
\ie
& k = -m+1, -m+2, \cdots, m,
\\
& \ell = -n+1, -n+2, \cdots, n,
\fe
{\it excluding} $(k,\ell)=(0,0)$ and $(k,\ell)=(m,n)$.
Since $H(\lambda_i^2, h | q) \to 1$ as the intermediate weight $h \to \infty$, the prefactor multiplying $H(\lambda_i^2, h | q)$ gives the large $h$ asymptotics of the conformal block.  The superconformal block ${\cal F}^R_h(z)$ which is related to the Virasoro conformal block via \eqref{idkey} also has the same large $h$ asymptotics.



\section{More on the Integrated Four-Point Function $A_{ijk\ell}$}\label{app:Aijkl}
\label{Sec:Aijkl}

The purpose of this appendix is the explain how knowing the value of the integrated four-point function $A_{ijk\ell}$ can improve the bootstrap bounds on the spectrum.  We first explain the problem with naively incorporating $A_{ijk\ell}$ into semidefinite programming, and then discuss two solutions.  The first way is to cleverly use crossing symmetry to choose an appropriate region over which to integrate the conformal blocks.  The second way is to use $A_{1111}$ indirectly by bounding it above by the four-point function evaluated at the crossing symmetric point, $f(1/2)$, and incorporate $f(1/2)$ into semidefinite programming instead.

\subsection{Conformal Block Expansion}
\label{Sec:Aijkl1}

We can write the integrated four-point function $A_{ijk\ell}$ as
\ie\label{explicitAijkl}
A_{ijk\ell} &=\lim_{\epsilon\to 0} \int_{|z|,|1-z|,|z|^{-1}>\epsilon} {d^2z\over |z(1-z)|}\left\langle \phi^{RR}_i(z,\bar z) \phi^{RR}_j(0) \phi^{RR}_k(1) \phi^{RR}_\ell(\infty) \right\rangle 
\\
& \hspace{3in} + 2\pi \ln\epsilon \left ({\delta_{ij}\delta_{k\ell}}+{\delta_{ik}\delta_{j\ell}}  + {\delta_{i\ell}\delta_{jk}}\right).
\fe
In expressing the four-point function of the ${1\over 2}$-BPS operators in terms of conformal blocks, we would like the divergence in the $z$-integral to appear in the identity conformal block alone, so that the regularization can be performed on the identity block contribution alone. This can be achieved by dividing the integral over the $z$-plane into the contributions from three regions $D_1$, $D_2$ and $D_3$ defined in \eqref{regions}.  Note that regions $D_2$ and $D_3$ can be mapped from $D_1$ by $z\mapsto 1/z$ and $z\mapsto 1/(1-z)$, respectively. We have
\ie
A_{ijk\ell} &= \int_{D_1} {d^2z\over |z(1-z)|} \Bigg\{ \delta_{ij}\delta_{k\ell} \left[ |{\cal F}^R_0(z)|^2 - {1\over |z|} \right] + \sum_{{\rm non-BPS}~{\cal O}} C_{ij{\cal O}} C_{k\ell{\cal O}} {\cal F}^R_{h_L}(z) \overline{{\cal F}^R_{h_R}(z)} \Bigg\} \, - \delta_{ij}\delta_{k\ell} C_0
\\
& \hspace{2in} + (j\leftrightarrow k) + (j\leftrightarrow \ell),
\fe
where the constant $C_0$ is given by
\ie
C_0 = \lim_{\epsilon\to 0} \int_{\epsilon<|z|<1,~{\rm Re} z<{1\over 2}} {d^2z\over |z|^2|1-z|} +2\pi\ln\epsilon \approx -1.43907.
\fe
Now the integral in the domain $D_1$ can be performed term by term in the summation over superconformal blocks.  Define the constant $A_0$ and the function $A(h_L,h_R)$ by
\ie
& A_0 = \int_{D_1} {d^2z\over |z(1-z)|}\left[ |{\cal F}^R_0(z)|^2 - {1\over |z|} \right] - C_0,
\\
& A(\Delta, s) =  \int_{D_1} {d^2z\over |z(1-z)|}{\cal F}^R_{\Delta + s \over 2}(z) \overline{{\cal F}^R_{\Delta - s \over 2}(z)}.
\fe
$A_0$ can also be obtained as a limit of $A(\Delta, 0)$ by
\ie
A_0 = \lim_{\Delta \to 0} \left[ A(\Delta, 0) - {2\pi \over \Delta} \right].
\fe
We can now write
\ie
A_{ijk\ell} &= \left[ \delta_{ij}\delta_{k\ell} A_0 + \sum_{{\rm non-BPS}~{\cal O}} C_{ij{\cal O}}C_{k\ell{\cal O}} A(\Delta, s) \right] + (j\leftrightarrow k) + (j\leftrightarrow \ell).
\fe

Let us examine this equation for identical external operators
\ie
0 &= (3 A_0 - A_{1111}) + 3 \sum_{{\rm non-BPS}~{\cal O}} C^2_{11{\cal O}} A(\Delta, s).
\fe
It takes the same form as the equations corresponding to acting linear functionals $\A_{m, n} = \partial^m \bar\partial^n |_{z = 1/2}$ on the crossing equation (see Section~\ref{Sec:Bootstrap})
\ie
0 &= \A_{m, n} ({\cal H}_{0}(z, \bar z)) +  \sum_{{\rm non-BPS}~{\cal O}} C_{11\mathcal{O}}^2\A_{m, n} [{\cal H}_{\Delta, s}(z, \bar z)].
\fe
Clearly, if we can find a set of coefficients $a$ and $a_{m, n}$ such that
\ie
\label{A1111BS}
& a (3 A_0 - A_{1111}) + \sum_{m, n} a_{m, n} \A_{m, n} [{\cal H}_{0}(z, \bar z)] > 0,
\\
& 3 a A(\Delta, s) + \sum_{m, n} a_{m, n} \A_{m, n} [{\cal H}_{\Delta, s}(z, \bar z)] > 0 \quad \text{for} \quad \Delta > \widehat\Delta_{gap}, \quad s \in 2\bZ
\fe
are satisfied, then the gap in the non-BPS spectrum $\Delta_{gap}$ must be bounded above by $\widehat\Delta_{gap}$, in order to be consistent with the positivity of $C_{11{\cal O}}^2$.  

Despite the additional freedom of $a$, this naive incorporation of $A_{1111}$ does not improve the bound, for the following reasons.  As explained at the end of Appendix~\ref{Sec:qmap}, the holomorphic superconformal block ${\cal F}^R_h(z)$ asymptotes to $(16 q(z))^h$ at large $h$. 
This means that for any spin $s$, the integrated block $A(\Delta, s)$ at large $\Delta$ is dominated by the integration near the maximal value of $|q(z)|$ in the domain $D_1$, which is at  (see Figure~\ref{fig:qmap}),
\ie
z_*^\pm = {1 \over 2} \pm {\sqrt3 \over 2} i, ~~~\text{or}~~~ q(z_*^\pm) = \pm i e^{-{\sqrt3 \over 2}\pi}
\fe
 and therefore has the asymptotic behavior
\ie
A(\Delta, s) \sim (-1)^{s/2} (16 e^{-{\sqrt3 \over 2}\pi} )^\Delta,  \quad \Delta \gg |s|.
\fe
In comparison, $\A_{m, n} ({\cal H}_{\Delta, s}(z, \bar z)) \sim (16 e^{-\pi})^\Delta$ is dominated by $A(\Delta, s) \sim (-1)^{s/2} (16 e^{-{\sqrt3 \over 2} \pi})^\Delta$ at large $\Delta$, whose sign oscillates with $s$.  Thus positivity at large $\Delta$ forces $a = 0$, and $\widehat\Delta_{gap}$ cannot be improved despite specifying $A_{1111}$.


One may wonder if we can choose a different region (that also consists of two fundamental domains of the $S_3$ crossing group) to integrate in, so that the leading large $\Delta$ behavior of the integrated block is $(16 e^{-\pi})^\Delta$, same as $\A_{m, n} ({\cal H}_{\Delta, s}(z, \bar z))$.  This is not possible, because $z_*^\pm = {1 \over 2} \pm {\sqrt3 \over 2} i$ at most exchange with each other under crossing.  However, we can integrate over a larger region $D'$ whose maximum $|q|$ value is on the real axis (to avoid the sign oscillation), and map the extra region $D' \setminus D_1$ that needs to be subtracted off via crossing to a region $E$ inside $D_1$.  We thus have an equation for $A_{1111}$ related to the naive conformal block expansion by the replacement of $D_1 \to  D' \setminus E$ as the integration region.

We are free to choose $D'$, but in the end the bootstrap bound should not be sensitive to the choice.  Let $D'$ be symmetric under $q \to -q$ and $q \to \bar q$, so that it suffices to specify $D'$ in the first quadrant in the $q$-plane, or equivalently within the strip $0 \leq {\rm Re}\,\tau \leq {1 \over 2}$ in the $\tau$-plane (recall $q(z) = e^{i\pi\tau(z)}$).  In this strip, the region $D_1$ is bounded below by $|\tau| = 1$.  A choice of $D'$ is the region bounded below by the lower arc of $|\tau - {1 \over 2}| = {\sqrt3 \over 2}$, with $q_{max} = e^{-{\pi \over \sqrt2}}$.  The corresponding region $E$ is then the part of $D_1$ that satisfies $|\tau + 1 | \leq \sqrt3$.  See Figure~\ref{fig:ellipse}.



To perform semidefinite programming efficiently, it is desirable to factor out certain positive factors, including the exponential dependence on $\Delta$, and just work with polynomials.  Our strategy is to factor out $(16 e^{-\pi})^\Delta$, and approximate $(16 e^{\pi})^\Delta A(\Delta, s)$ by a rational function in $\Delta$, that works well up to a value beyond which $A(\Delta, s)$ is completely dominated by the asymptotic $(16 q_{max})^\Delta$ factor. 
We further demand $a > 0$, and that the rational approximation be strictly bounded above by the actual value, so that the bound can only be stronger as we improve the rational approximation to work well in a larger range of $\Delta$.



\subsection{An Inequality Relating $A_{1111}$ to the Four-point Function at $z={1\over 2}$}
\label{sec:inequalityforAandfhalf}

An alternative is to use $A_{1111}$ indirectly by bounding $A_{1111}$ above by the four-point function evaluated at the crossing symmetric point $f(1/2)$.  The conformal block evaluated at $z = {1\over2}$ has the same large $\Delta$ asymptotics $(16 e^{-\pi})^\Delta$ as $\A_{m, n} ({\cal H}_{0}(z, \bar z))$, and the sign does not oscillate with $s$.
The incorporation of $f(1/2)$ into bootstrap and the results are discussed in detail in Section~\ref{Sec:Gap}.  This section is devoted to proving the inequality between $A_{1111}$ and $f(1/2)$.

We can write the ${\cal N}=4$ superconformal block decomposition of the BPS four-point function $f(z,\bar z)$ in the form (see (7.8) of \cite{Maldacena:2015iua})
\ie
f(z,\bar z) =  
\left| \Lambda(z) \right|^2 \sum_{h_L,h_R} g_{h_L}(q) \overline{g_{h_R}(q)},
\fe
with $\Lambda(z) \equiv (z(1-z))^{-{1\over 3}} \theta_3(q)^{-2}$. The functions $g_h(q)$ take the form
\ie
& g_h(q) = q^{h-{1\over 6}}\sum_{n\geq 0} a_n q^n,
\fe
where, importantly, the coefficients $a_n$ are {\it non-negative}.

For a general complex cross ratio $z$, let $x$ be the real value between 0 and 1 such that $q(x) = |q(z)|$. Define $r = {\rm min}\{ x, 1-x\}$, and $q_r = q(r)$. Note that by crossing relation, $f(x)=f(x_0)$. We can then bound the four-point function at a generic cross ratio by
\ie
f(z, \bar z) &\leq \left| { \Lambda(z)\over \Lambda(x)} \right|^2  f(x) = \left| { \Lambda(z)\over \Lambda(x)} \right|^2  f(x_0) \leq \left| { \Lambda(z) \Lambda(r) \over \Lambda(x)\Lambda({1\over 2}) } \right|^2 \left| {q_{r} \over q_{1\over 2}} \right|^{-{1\over 3}} f({1\over 2}) .
\fe
We now make the assumption that the non-BPS operators have scaling dimensions above a nonzero gap $\Delta$. As before, we can write the integrated four-point function $A_{1111}$ as 3 times the contribution from an integral over the domain $D_1=\left\{ z\in\mathbb{C}: |z|<1, {\rm Re}\, z <{1\over 2}\right\}$, while regularizing the integral of the identity block contribution, in the form 
\ie
A_{1111} &=3 A_0 + 3 \int_{D_1} {d^2z\over |z(1-z)|}\left| \Lambda(z) \right|^2 \sum_{(h_L,h_R)\not=(0,0)} g_{h_L}(q) \overline{g_{h_R}(q)}
\\
&\leq 3A_0 + 3 \int_{D_1} {d^2z\over |z(1-z)|^{5\over 3}}\left|  \theta_3(q)  \right|^{-4} \left|{q\over q_{1\over2}}\right|^{\Delta-{1\over 3}} \sum_{(h_L,h_R)\not=(0,0)} g_{h_L}(q_{1\over2}) \overline{g_{h_R}(q_{1\over2})}
\\
&= 3A_0 + 3 \int_{D_1} {d^2z\over |z(1-z)|^{5\over 3}}\left|  \theta_3(q)  \right|^{-4} \left|{q\over q_{1\over2}}\right|^{\Delta-{1\over 3}}  {f(z_*,\bar z_*) - |{\cal F}^R_0(z_*)|^2 \over \left| \Lambda(z_*) \right|^2}
\\
&\leq 3A_0 + 3 \int_{D_1} {d^2z\over |z(1-z)|^{5\over 3}}\left|  \theta_3(q)  \right|^{-4} \left|{q\over q_{1\over2}}\right|^{\Delta-{1\over 3}} \left[ \left| { \Lambda(r_*) \over \Lambda(x_*)\Lambda({1\over 2}) } \right|^2 \left| {q_r \over q_{1\over 2}} \right|^{-{1\over 3}}{f({1\over 2}) } - { |{\cal F}^R_0(z_*)|^2 \over \left| \Lambda(z_*) \right|^2} \right]
\\
& = 3A_0 + M(\Delta)\left[   f({1\over 2}) -f_0\right].
\label{eq:inequalityforA}
\fe
Here $z_*={1+\sqrt{3} i\over 2}$ is the value of the cross ratio $z$ over the domain $D$ that achieves the maximal value of $|q(z)|$, $x_*\approx 0.653326$ is such that $q(x_*)=|q(z_*)|$, $r_*=1-x_*$, and $q_r\approx 0.0265799$. On the RHS of the inequality, $f_0$ is a constant defined by
\ie
f_0=
  \left| {  \Lambda(x_*)\Lambda({1\over 2})\over\Lambda(r_*)  } \right|^2 \left| {q_{1\over 2} \over q_r} \right|^{-{1\over 3}}{ |{\cal F}^R_0(z_*)|^2 \over \left| \Lambda(z_*) \right|^2} ,
  \fe
and the function $M(\Delta)$ is given by
\ie
M(\Delta) =  3\left| { \Lambda(r_*) \over \Lambda(x_*)\Lambda({1\over 2}) } \right|^2 \left| {q_r \over q_{1\over 2}} \right|^{-{1\over 3}} \int_D {d^2z\over |z(1-z)|^{5\over 3}}\left|  \theta_3(q)  \right|^{-4} \left|{q\over q_{1\over2}}\right|^{\Delta-{1\over 3}}    .
\fe 
Note that $M(\Delta)$ goes like $\Delta^{-1}$ in the $\Delta\to 0$ limit, with $\lim_{\Delta\to 0} \Delta M(\Delta)\approx 2.27548$.

\section{$\Delta_{crt}$ and the Divergence of the Integrated Four-Point Function $A_{1111}$}\label{app:critical}
\label{sec:deltacriticalanddivergence}

Recall that $\Delta_{crt}$ defined in Section \ref{sec:critical} is the lowest scaling dimension at which either   a continuous spectrum develops or the structure constant diverges, as the CFT is deformed to a singular point in its moduli space. In this appendix we will describe how to use crossing symmetry to bootstrap an upper bound on $\Delta_{crt}$  that is universal across the moduli space. In particular we will show that if the integrated four-point function $A_{1111}$ diverges somewhere on the moduli space, then  $\widehat\Delta_{crt}\ge\Delta_{crt}$ with $\widehat\Delta_{crt}$ defined in Section~\ref{sec:critical}.  
 
Consider the four-point function of the RR sector ${1\over2}$-BPS primaries $\phi_i^{RR}$ of weight $({1\over 4},{1\over4})$ that are  R-symmetry singlets ($i=1,\cdots,20$). Let us consider in particular the four-point function of the same operator, say, $\phi_1^{RR}$,
\ie
f(z,\bar z)&  \equiv \la \phi^{RR}_1(z,\bar z) \phi^{RR}_1(0) \phi^{RR}_1(1) \phi^{RR}_1(\infty)\ra = \sum_{\Delta} C_\Delta^2 F_\Delta(z,\bar z)\,,
\fe 
where we did not write out the sum over the spin explicitly but it will not affect the argument significantly. The conformal block has the following asymptotic growth
\ie
|F_\Delta(z,\bar z)|  \sim |16 q(z)|^{\Delta}\,,
\fe
for large $\Delta$.

The crossing equation takes the following expression,
\ie
\sum_{\Delta} C_\Delta^2 G_\Delta(z,\bar z)\equiv\sum_{\Delta} C_\Delta^2 \left[  F_\Delta (z,\bar z) - F_\Delta (1-z, 1-\bar z)\right]=0.
\fe
We will consider functionals $\mathcal{L}$ acting $G_\Delta(z,\bar z)$  with the following properties,
\ie
&L(\Delta) \equiv \cL(G_\Delta(z,\bar z)) >0,~~~\text{if}~~\Delta>\widehat\Delta_{crt},
\fe
for some $\widehat\Delta_{crt}$. 
Note that $\widehat\Delta_{crt}$ depends on the choice of the functional $\cL$. The significance of $\widehat\Delta_{crt}$ is that it implies the structure constants above $\widehat\Delta_{crt}$ are bounded  by those below,
\ie\label{critical}
\sum_{\Delta > \widehat\Delta_{crt} } C_{\Delta}^2 L(\Delta) = -\sum_{\Delta< \widehat\Delta_{crt} } C_{\Delta}^2L(\Delta).
\fe

Assuming that the integrated four-point function $A_{1111}$ diverges at some points on the moduli space, we will show that for any choice of the functional $\cL$, we always have
\ie\label{goal}
\widehat\Delta_{crt}\ge \Delta_{crt}.
\fe
In this way we can bootstrap an upper bound on $\Delta_{crt}$ by scanning through a large class of functionals $\cL$.

To prove our goal \eqref{goal}, we  assume that there exists a functional $\cL$ such that the associated $\widehat\Delta_{crt}<\Delta_{crt}$, and show that it leads to contradiction. By assumption the density of the spectrum is bounded and the structure constants are finite for $\Delta<\widehat\Delta_{crt}(<\Delta_{crt})$, hence the RHS of \eqref{critical} is finite,
\ie\label{finite}
\sum_{\Delta > \widehat\Delta_{crt} } C_{\Delta}^2 L(\Delta) = -\sum_{\Delta< \widehat\Delta_{crt} } C_{\Delta}^2L(\Delta) <\infty.
\fe
In the following we will try to bound the integrated four-point function
\ie
A_{1111} &=\lim_{\epsilon\to 0} \int_{|z|,|1-z|,|z|^{-1}>\epsilon} {d^2z\over |z(1-z)|}f(z,\bar z)
+ 6\pi \ln\epsilon ,
\fe
roughly by $\sum_{\Delta>\widehat\Delta_{crt}} C_\Delta^2 L(\Delta)$, which is finite by assumption, plus some other finite contributions. On the other hand, we know $A_{1111}$ diverges, for example, at the cigar CFT points, and hence the contradiction.

Let us now fill in the details of the proof.  As discussed in Appendix \ref{app:Aijkl}, in the  expression for $A_{1111}$, we  break the integral on the $z$-plane into three different regions $D_1,D_2,D_3$ \eqref{regions} that are mapped to each other under $z\to 1-z$ and $z\to 1/z$.  Since the four-point function is crossing symmetric, we can  focus on region $D_1$ alone. This has the advantage that the divergence in the $z$-integral only shows up in the identity block. We will cut a small disk around $z=0$ with radius $\epsilon_0$ and regularize the the contribution from the identity block.

We start by noting a bound on the conformal blocks. The functionals $\cL$ we consider are linear combination of powers of $\partial_z,\partial_{\bar z}$ evaluated at $z=1/2$. Therefore the asymptotic behavior of $L(\Delta)$ is the same as that of the conformal block $F_\Delta(z,\bar z)$ evaluated at $z=1/2$,
\ie
L(\Delta) \sim |16q_{1\over2}|^{\Delta},
\fe
where $q_{1\over2}  \equiv q(z=1/2).$ 
This  implies that there exists a moduli-independent constant $c_0$ and $\delta_0$ such that
\ie\label{boundblock}
|F_{\Delta}(z,\bar z) | \le c_0 L(\Delta),~~~~\text{for}~~\Delta \ge \widehat\Delta_{crt} + \delta_0,
\fe
if  $|q(z)|<|q_{1\over2}|$. 
We can always tune $\delta_0$ to be arbitrarily small by taking $c_0$ to be large. Note however that strictly at $\Delta=\widehat\Delta_{crt}$, we have $L(\widehat\Delta_{crt})=0$.

For $z$ in region I and $|q(z)|<|q_{1\over2}|$, from \eqref{boundblock} we have
\ie\label{bound}
|f(z,\bar z)| \le c_0 \sum_{\Delta \ge \widehat\Delta_{crt}+\delta_0} C_\Delta^2 L(\Delta) + \sum_{\Delta<\widehat\Delta_{crt} +\delta_0} C_\Delta^2 \text{max}\{ F_\Delta({1\over2} ) , F_\Delta(\epsilon_0)\}.
\fe
In particular it is true for $\epsilon_0 <z<1/2$.

Next we want to argue that \eqref{bound} is true for $z$ in the whole region I. First we note that we can write the four-point function as an expansion in $z,\bar z$ with \textit{non-negative} coefficients \cite{Hartman:2015lfa, Maldacena:2015iua}. By the Cauchy-Schwarz inequality, we have
\ie
| f(z,\bar z) | \le f( |z| ,|\bar z| ) .
\fe
Note that $|z|\in [ {1\over 2} ,1]$ for $z$ in region I but $|q(z)|>|q_{1\over2}|$. Next, by crossing symmetry, we have $f(|z|,|\bar z| ) = f( 1-|z| ,1-|\bar z|)$. We therefore arrive at the following bound
\ie
| f(z, \bar z) | \le f( 1- |z| , 1-|\bar z| ) \le c_0 \sum_{\Delta \ge \widehat\Delta_{crt}+\delta_0} C_\Delta^2 L(\Delta) + \sum_{\Delta<\widehat\Delta_{crt} +\delta_0} C_\Delta^2 \text{max}\{ F_\Delta({1\over2} ) , F_\Delta(\epsilon_0)\}
\fe
where we have used the fact that $1-|z|\in [\epsilon_0 ,{1\over2}]$ if $z$ is in region I with $|q(z)|>|q_{1\over2}|$. Hence the bound \eqref{bound} is true for all $z$ in region I. We can therefore bound the integrated four-point function as
\ie\label{bound1}
\Big| \int_{\text{region I}-\{ |z| =\epsilon_0\}} d^2 z |z|^{-s-1} |1-z|^{-t-1} f(z,\bar z)\Big|
\le c_1c_0 \sum_{\Delta> \widehat\Delta_{crt}+\delta_0} C_\Delta^2 L(\Delta) +\sum_{\Delta< \widehat\Delta_{crt}+\delta_0} C_\Delta^2 \tilde C(\Delta,\epsilon_0)
\fe
For integration inside the disk, we have to regularize the contribution from the identity block,
\ie\label{bound2}
&\Big| \int_{\{ |z| =\epsilon_0\}} d^2 z |z|^{-s-1} |1-z|^{-t-1} f(z,\bar z) - \text{reg.}\Big|
=c_2 + c_3 \sum_{\Delta\ge \Delta_{gap} >0} C_\Delta^2 F_\Delta(\epsilon_0) \\
& \le c_2 + c_3 \sum_{\Delta_{gap} < \Delta < \widehat\Delta_{crt} +\delta_0} C_\Delta^2 F_\Delta(\epsilon_0) + \tilde c_3 \sum_{\Delta > \widehat\Delta_{crt} +\delta_0 } C_\Delta^2 L(\Delta),
\fe
where we have assumed there is a gap in the spectrum. 
 $c_2$ is a moduli-independent constant coming from the regularized identity block contribution. 

Let us inspect every term in \eqref{bound1} and \eqref{bound2}.  First we tune $\delta_0$ such that $\widehat\Delta_{crt}+\delta_0$ is below $\Delta_{crt}$, possibly at the price of having larger $c_0$. After doing so, terms involving sums over $\Delta$ below $\widehat\Delta_{crt}+\delta_0$ are finite by our assumption that the density of the spectrum is bounded and the structure constants are finite for this range of $\Delta$. On the other hand, for terms involving sum of $\Delta$ above $\widehat\Delta_{crt}+\delta_0$, they are both of the form 
\ie
\sum_{\Delta > \widehat\Delta_{crt} +\delta_0} C_\Delta^2L(\Delta),
\fe
 which is bounded from above by the LHS of \eqref{finite}. Hence the LHS of \eqref{bound1} and \eqref{bound2} are both bounded. It follows that $A_{1111}<\infty$  under the assumption that $\widehat\Delta_{crt}<\Delta_{crt}$, which is a contradiction, say, at the cigar point. Thus we have proved our goal \eqref{goal}.

\bibliographystyle{JHEP}
\bibliography{K3refs}

\providecommand{\href}[2]{#2}\begingroup\raggedright\begin{thebibliography}{10}

\bibitem{polyakov1974nonhamiltonian}
A.~Polyakov, {\it Nonhamiltonian approach to conformal quantum field theory},
  {\em Zh. Eksp. Teor. Fiz} {\bf 66} (1974), no.~1 23--42.

\bibitem{Ferrara:1973yt}
S.~Ferrara, A.~F. Grillo, and R.~Gatto, {\it {Tensor representations of
  conformal algebra and conformally covariant operator product expansion}},
  {\em Annals Phys.} {\bf 76} (1973) 161--188.

\bibitem{Mack:1975jr}
G.~Mack, {\it {Duality in quantum field theory}},  {\em Nucl. Phys.} {\bf B118}
  (1977) 445--457.

\bibitem{Belavin:1984vu}
A.~A. Belavin, A.~M. Polyakov, and A.~B. Zamolodchikov, {\it {Infinite
  Conformal Symmetry in Two-Dimensional Quantum Field Theory}},  {\em Nucl.
  Phys.} {\bf B241} (1984) 333--380.

\bibitem{Knizhnik:1984nr}
V.~G. Knizhnik and A.~B. Zamolodchikov, {\it {Current Algebra and Wess-Zumino
  Model in Two-Dimensions}},  {\em Nucl. Phys.} {\bf B247} (1984) 83--103.

\bibitem{Gepner:1986wi}
D.~Gepner and E.~Witten, {\it {String Theory on Group Manifolds}},  {\em Nucl.
  Phys.} {\bf B278} (1986) 493.

\bibitem{Bouwknegt:1992wg}
P.~Bouwknegt and K.~Schoutens, {\it {W symmetry in conformal field theory}},
  {\em Phys. Rept.} {\bf 223} (1993) 183--276,
  [\href{http://arxiv.org/abs/hep-th/9210010}{{\tt hep-th/9210010}}].

\bibitem{Zamolodchikov:1995aa}
A.~B. Zamolodchikov and A.~B. Zamolodchikov, {\it {Structure constants and
  conformal bootstrap in Liouville field theory}},  {\em Nucl. Phys.} {\bf
  B477} (1996) 577--605, [\href{http://arxiv.org/abs/hep-th/9506136}{{\tt
  hep-th/9506136}}].

\bibitem{Teschner:1997ft}
J.~Teschner, {\it {On structure constants and fusion rules in the SL(2,C) /
  SU(2) WZNW model}},  {\em Nucl. Phys.} {\bf B546} (1999) 390--422,
  [\href{http://arxiv.org/abs/hep-th/9712256}{{\tt hep-th/9712256}}].

\bibitem{Teschner:1999ug}
J.~Teschner, {\it {Operator product expansion and factorization in the H+(3)
  WZNW model}},  {\em Nucl. Phys.} {\bf B571} (2000) 555--582,
  [\href{http://arxiv.org/abs/hep-th/9906215}{{\tt hep-th/9906215}}].

\bibitem{Teschner:2001rv}
J.~Teschner, {\it {Liouville theory revisited}},  {\em Class. Quant. Grav.}
  {\bf 18} (2001) R153--R222, [\href{http://arxiv.org/abs/hep-th/0104158}{{\tt
  hep-th/0104158}}].

\bibitem{Rattazzi:2008pe}
R.~Rattazzi, V.~S. Rychkov, E.~Tonni, and A.~Vichi, {\it {Bounding scalar
  operator dimensions in 4D CFT}},  {\em JHEP} {\bf 12} (2008) 031,
  [\href{http://arxiv.org/abs/0807.0004}{{\tt arXiv:0807.0004}}].

\bibitem{Rychkov:2009ij}
V.~S. Rychkov and A.~Vichi, {\it {Universal Constraints on Conformal Operator
  Dimensions}},  {\em Phys. Rev.} {\bf D80} (2009) 045006,
  [\href{http://arxiv.org/abs/0905.2211}{{\tt arXiv:0905.2211}}].

\bibitem{Poland:2010wg}
D.~Poland and D.~Simmons-Duffin, {\it {Bounds on 4D Conformal and
  Superconformal Field Theories}},  {\em JHEP} {\bf 05} (2011) 017,
  [\href{http://arxiv.org/abs/1009.2087}{{\tt arXiv:1009.2087}}].

\bibitem{Poland:2011ey}
D.~Poland, D.~Simmons-Duffin, and A.~Vichi, {\it {Carving Out the Space of 4D
  CFTs}},  {\em JHEP} {\bf 05} (2012) 110,
  [\href{http://arxiv.org/abs/1109.5176}{{\tt arXiv:1109.5176}}].

\bibitem{ElShowk:2012ht}
S.~El-Showk, M.~F. Paulos, D.~Poland, S.~Rychkov, D.~Simmons-Duffin, and
  A.~Vichi, {\it {Solving the 3D Ising Model with the Conformal Bootstrap}},
  {\em Phys. Rev.} {\bf D86} (2012) 025022,
  [\href{http://arxiv.org/abs/1203.6064}{{\tt arXiv:1203.6064}}].

\bibitem{Kos:2013tga}
F.~Kos, D.~Poland, and D.~Simmons-Duffin, {\it {Bootstrapping the $O(N)$ vector
  models}},  {\em JHEP} {\bf 06} (2014) 091,
  [\href{http://arxiv.org/abs/1307.6856}{{\tt arXiv:1307.6856}}].

\bibitem{Beem:2013qxa}
C.~Beem, L.~Rastelli, and B.~C. van Rees, {\it {The $\mathcal N=4$
  Superconformal Bootstrap}},  {\em Phys. Rev. Lett.} {\bf 111} (2013) 071601,
  [\href{http://arxiv.org/abs/1304.1803}{{\tt arXiv:1304.1803}}].

\bibitem{Beem:2014zpa}
C.~Beem, M.~Lemos, P.~Liendo, L.~Rastelli, and B.~C. van Rees, {\it {The
  ${\mathcal N}=2$ superconformal bootstrap}},
  \href{http://arxiv.org/abs/1412.7541}{{\tt arXiv:1412.7541}}.

\bibitem{El-Showk:2014dwa}
S.~El-Showk, M.~F. Paulos, D.~Poland, S.~Rychkov, D.~Simmons-Duffin, and
  A.~Vichi, {\it {Solving the 3d Ising Model with the Conformal Bootstrap II.
  c-Minimization and Precise Critical Exponents}},  {\em J. Stat. Phys.} {\bf
  157} (2014) 869, [\href{http://arxiv.org/abs/1403.4545}{{\tt
  arXiv:1403.4545}}].

\bibitem{Chester:2014fya}
S.~M. Chester, J.~Lee, S.~S. Pufu, and R.~Yacoby, {\it {The $ \mathcal{N}=8 $
  superconformal bootstrap in three dimensions}},  {\em JHEP} {\bf 09} (2014)
  143, [\href{http://arxiv.org/abs/1406.4814}{{\tt arXiv:1406.4814}}].

\bibitem{Chester:2014mea}
S.~M. Chester, J.~Lee, S.~S. Pufu, and R.~Yacoby, {\it {Exact Correlators of
  BPS Operators from the 3d Superconformal Bootstrap}},  {\em JHEP} {\bf 03}
  (2015) 130, [\href{http://arxiv.org/abs/1412.0334}{{\tt arXiv:1412.0334}}].

\bibitem{Chester:2014gqa}
S.~M. Chester, S.~S. Pufu, and R.~Yacoby, {\it {Bootstrapping $O(N)$ vector
  models in 4 $< d <$ 6}},  {\em Phys. Rev.} {\bf D91} (2015), no.~8 086014,
  [\href{http://arxiv.org/abs/1412.7746}{{\tt arXiv:1412.7746}}].

\bibitem{Bae:2014hia}
J.-B. Bae and S.-J. Rey, {\it {Conformal Bootstrap Approach to O(N) Fixed
  Points in Five Dimensions}},  \href{http://arxiv.org/abs/1412.6549}{{\tt
  arXiv:1412.6549}}.

\bibitem{Chester:2015qca}
S.~M. Chester, S.~Giombi, L.~V. Iliesiu, I.~R. Klebanov, S.~S. Pufu, and
  R.~Yacoby, {\it {Accidental Symmetries and the Conformal Bootstrap}},
  \href{http://arxiv.org/abs/1507.04424}{{\tt arXiv:1507.04424}}.

\bibitem{Iliesiu:2015qra}
L.~Iliesiu, F.~Kos, D.~Poland, S.~S. Pufu, D.~Simmons-Duffin, and R.~Yacoby,
  {\it {Bootstrapping 3D Fermions}},
  \href{http://arxiv.org/abs/1508.00012}{{\tt arXiv:1508.00012}}.

\bibitem{Kos:2015mba}
F.~Kos, D.~Poland, D.~Simmons-Duffin, and A.~Vichi, {\it {Bootstrapping the
  O(N) Archipelago}},  \href{http://arxiv.org/abs/1504.07997}{{\tt
  arXiv:1504.07997}}.

\bibitem{Beem:2015aoa}
C.~Beem, M.~Lemos, L.~Rastelli, and B.~C. van Rees, {\it {The $(2,0)$
  superconformal bootstrap}},  \href{http://arxiv.org/abs/1507.05637}{{\tt
  arXiv:1507.05637}}.

\bibitem{Lemos:2015awa}
M.~Lemos and P.~Liendo, {\it {Bootstrapping ${\mathcal N}=2$ chiral
  correlators}},  \href{http://arxiv.org/abs/1510.03866}{{\tt
  arXiv:1510.03866}}.

\bibitem{Hellerman:2009bu}
S.~Hellerman, {\it {A Universal Inequality for CFT and Quantum Gravity}},  {\em
  JHEP} {\bf 08} (2011) 130, [\href{http://arxiv.org/abs/0902.2790}{{\tt
  arXiv:0902.2790}}].

\bibitem{Keller:2012mr}
C.~A. Keller and H.~Ooguri, {\it {Modular Constraints on Calabi-Yau
  Compactifications}},  {\em Commun. Math. Phys.} {\bf 324} (2013) 107--127,
  [\href{http://arxiv.org/abs/1209.4649}{{\tt arXiv:1209.4649}}].

\bibitem{Fiset:2015pta}
M.-A. Fiset and J.~Walcher, {\it {Bounding the Heat Trace of a Calabi-Yau
  Manifold}},  {\em JHEP} {\bf 09} (2015) 124,
  [\href{http://arxiv.org/abs/1506.08407}{{\tt arXiv:1506.08407}}].

\bibitem{Harvey:2014nha}
J.~A. Harvey, S.~Lee, and S.~Murthy, {\it {Elliptic genera of ALE and ALF
  manifolds from gauged linear sigma models}},  {\em JHEP} {\bf 02} (2015) 110,
  [\href{http://arxiv.org/abs/1406.6342}{{\tt arXiv:1406.6342}}].

\bibitem{Seiberg:1988pf}
N.~Seiberg, {\it {Observations on the Moduli Space of Superconformal Field
  Theories}},  {\em Nucl. Phys.} {\bf B303} (1988) 286.

\bibitem{Eguchi:1988vra}
T.~Eguchi, H.~Ooguri, A.~Taormina, and S.-K. Yang, {\it {Superconformal
  Algebras and String Compactification on Manifolds with SU(N) Holonomy}},
  {\em Nucl. Phys.} {\bf B315} (1989) 193.

\bibitem{Cecotti:1990kz}
S.~Cecotti, {\it {N=2 Landau-Ginzburg versus Calabi-Yau sigma models:
  Nonperturbative aspects}},  {\em Int. J. Mod. Phys.} {\bf A6} (1991)
  1749--1814.

\bibitem{Cecotti:1991me}
S.~Cecotti and C.~Vafa, {\it {Topological antitopological fusion}},  {\em Nucl.
  Phys.} {\bf B367} (1991) 359--461.

\bibitem{Aspinwall:1994rg}
P.~S. Aspinwall and D.~R. Morrison, {\it {String theory on K3 surfaces}},
  \href{http://arxiv.org/abs/hep-th/9404151}{{\tt hep-th/9404151}}.

\bibitem{Nahm:1999ps}
W.~Nahm and K.~Wendland, {\it {A Hiker's guide to K3: Aspects of N=(4,4)
  superconformal field theory with central charge c = 6}},  {\em Commun. Math.
  Phys.} {\bf 216} (2001) 85--138,
  [\href{http://arxiv.org/abs/hep-th/9912067}{{\tt hep-th/9912067}}].

\bibitem{Gepner:1987qi}
D.~Gepner, {\it {Space-Time Supersymmetry in Compactified String Theory and
  Superconformal Models}},  {\em Nucl. Phys.} {\bf B296} (1988) 757.

\bibitem{Gaberdiel:2011fg}
M.~R. Gaberdiel, S.~Hohenegger, and R.~Volpato, {\it {Symmetries of K3 sigma
  models}},  {\em Commun. Num. Theor. Phys.} {\bf 6} (2012) 1--50,
  [\href{http://arxiv.org/abs/1106.4315}{{\tt arXiv:1106.4315}}].

\bibitem{Gaberdiel:2013psa}
M.~R. Gaberdiel, A.~Taormina, R.~Volpato, and K.~Wendland, {\it {A K3 sigma
  model with $\mathbb{Z}^8_2$ : $\mathbb{M}_{20}$ symmetry}},  {\em JHEP} {\bf
  02} (2014) 022, [\href{http://arxiv.org/abs/1309.4127}{{\tt
  arXiv:1309.4127}}].

\bibitem{Ooguri:1995wj}
H.~Ooguri and C.~Vafa, {\it {Two-dimensional black hole and singularities of CY
  manifolds}},  {\em Nucl. Phys.} {\bf B463} (1996) 55--72,
  [\href{http://arxiv.org/abs/hep-th/9511164}{{\tt hep-th/9511164}}].

\bibitem{Eguchi:2004ik}
T.~Eguchi and Y.~Sugawara, {\it {Conifold type singularities, N=2 Liouville and
  SL(2:R)/U(1) theories}},  {\em JHEP} {\bf 01} (2005) 027,
  [\href{http://arxiv.org/abs/hep-th/0411041}{{\tt hep-th/0411041}}].

\bibitem{Eguchi:2008ct}
T.~Eguchi, Y.~Sugawara, and A.~Taormina, {\it {Modular Forms and Elliptic
  Genera for ALE Spaces}},  in {\em {Workshop on Exploration of New Structures
  and Natural Constructions in Mathematical Physics: On the Occasion of
  Professor Akhiro Tsuchiya's Retirement Nagoya, Japan, March 5-8, 2007}},
  2008.
\newblock \href{http://arxiv.org/abs/0803.0377}{{\tt arXiv:0803.0377}}.

\bibitem{Ribault:2005wp}
S.~Ribault and J.~Teschner, {\it {H+(3)-WZNW correlators from Liouville
  theory}},  {\em JHEP} {\bf 06} (2005) 014,
  [\href{http://arxiv.org/abs/hep-th/0502048}{{\tt hep-th/0502048}}].

\bibitem{Chang:2014jta}
C.-M. Chang, Y.-H. Lin, S.-H. Shao, Y.~Wang, and X.~Yin, {\it {Little String
  Amplitudes (and the Unreasonable Effectiveness of 6D SYM)}},  {\em JHEP} {\bf
  12} (2014) 176, [\href{http://arxiv.org/abs/1407.7511}{{\tt
  arXiv:1407.7511}}].

\bibitem{Kiritsis:2000zi}
E.~Kiritsis, N.~A. Obers, and B.~Pioline, {\it {Heterotic / type II triality
  and instantons on K(3)}},  {\em JHEP} {\bf 01} (2000) 029,
  [\href{http://arxiv.org/abs/hep-th/0001083}{{\tt hep-th/0001083}}].

\bibitem{Lin:2015dsa}
Y.-H. Lin, S.-H. Shao, Y.~Wang, and X.~Yin, {\it {Supersymmetry Constraints and
  String Theory on K3}},  \href{http://arxiv.org/abs/1508.07305}{{\tt
  arXiv:1508.07305}}.

\bibitem{Zamolodchikov:1985ie}
A.~B. Zamolodchikov, {\it {CONFORMAL SYMMETRY IN TWO-DIMENSIONS: AN EXPLICIT
  RECURRENCE FORMULA FOR THE CONFORMAL PARTIAL WAVE AMPLITUDE}},  {\em Commun.
  Math. Phys.} {\bf 96} (1984) 419--422.

\bibitem{Aspinwall:1995zi}
P.~S. Aspinwall, {\it {Enhanced gauge symmetries and K3 surfaces}},  {\em Phys.
  Lett.} {\bf B357} (1995) 329--334,
  [\href{http://arxiv.org/abs/hep-th/9507012}{{\tt hep-th/9507012}}].

\bibitem{Hartman:2015lfa}
T.~Hartman, S.~Jain, and S.~Kundu, {\it {Causality Constraints in Conformal
  Field Theory}},  \href{http://arxiv.org/abs/1509.00014}{{\tt
  arXiv:1509.00014}}.

\bibitem{Maldacena:2015iua}
J.~Maldacena, D.~Simmons-Duffin, and A.~Zhiboedov, {\it {Looking for a bulk
  point}},  \href{http://arxiv.org/abs/1509.03612}{{\tt arXiv:1509.03612}}.

\bibitem{Schwimmer:1986mf}
A.~Schwimmer and N.~Seiberg, {\it {Comments on the N=2, N=3, N=4 Superconformal
  Algebras in Two-Dimensions}},  {\em Phys. Lett.} {\bf B184} (1987) 191.

\bibitem{Eguchi:1987sm}
T.~Eguchi and A.~Taormina, {\it {Unitary Representations of $N=4$
  Superconformal Algebra}},  {\em Phys. Lett.} {\bf B196} (1987) 75.

\bibitem{Eguchi:1987wf}
T.~Eguchi and A.~Taormina, {\it {Character Formulas for the $N=4$
  Superconformal Algebra}},  {\em Phys. Lett.} {\bf B200} (1988) 315.

\bibitem{deBoer:2008ss}
J.~de~Boer, J.~Manschot, K.~Papadodimas, and E.~Verlinde, {\it {The Chiral ring
  of AdS(3)/CFT(2) and the attractor mechanism}},  {\em JHEP} {\bf 03} (2009)
  030, [\href{http://arxiv.org/abs/0809.0507}{{\tt arXiv:0809.0507}}].

\bibitem{Baggio:2012rr}
M.~Baggio, J.~de~Boer, and K.~Papadodimas, {\it {A non-renormalization theorem
  for chiral primary 3-point functions}},  {\em JHEP} {\bf 07} (2012) 137,
  [\href{http://arxiv.org/abs/1203.1036}{{\tt arXiv:1203.1036}}].

\bibitem{Fitzpatrick:2014oza}
A.~L. Fitzpatrick, J.~Kaplan, Z.~U. Khandker, D.~Li, D.~Poland, and
  D.~Simmons-Duffin, {\it {Covariant Approaches to Superconformal Blocks}},
  {\em JHEP} {\bf 08} (2014) 129, [\href{http://arxiv.org/abs/1402.1167}{{\tt
  arXiv:1402.1167}}].

\bibitem{Khandker:2014mpa}
Z.~U. Khandker, D.~Li, D.~Poland, and D.~Simmons-Duffin, {\it {$ \mathcal{N} $
  = 1 superconformal blocks for general scalar operators}},  {\em JHEP} {\bf
  08} (2014) 049, [\href{http://arxiv.org/abs/1404.5300}{{\tt
  arXiv:1404.5300}}].

\bibitem{Bobev:2015jxa}
N.~Bobev, S.~El-Showk, D.~Mazac, and M.~F. Paulos, {\it {Bootstrapping SCFTs
  with Four Supercharges}},  {\em JHEP} {\bf 08} (2015) 142,
  [\href{http://arxiv.org/abs/1503.02081}{{\tt arXiv:1503.02081}}].

\bibitem{Kutasov:1995te}
D.~Kutasov, {\it {Orbifolds and solitons}},  {\em Phys. Lett.} {\bf B383}
  (1996) 48--53, [\href{http://arxiv.org/abs/hep-th/9512145}{{\tt
  hep-th/9512145}}].

\bibitem{Giveon:1999px}
A.~Giveon and D.~Kutasov, {\it {Little string theory in a double scaling
  limit}},  {\em JHEP} {\bf 10} (1999) 034,
  [\href{http://arxiv.org/abs/hep-th/9909110}{{\tt hep-th/9909110}}].

\bibitem{Teschner:1995yf}
J.~Teschner, {\it {On the Liouville three point function}},  {\em Phys. Lett.}
  {\bf B363} (1995) 65--70, [\href{http://arxiv.org/abs/hep-th/9507109}{{\tt
  hep-th/9507109}}].

\bibitem{1901}
E.~W. Barnes, {\it The theory of the double gamma function},  {\em
  Philosophical Transactions of the Royal Society of London. Series A,
  Containing Papers of a Mathematical or Physical Character} {\bf 196} (1901)
  pp. 265--387.

\bibitem{Maldacena:2001km}
J.~M. Maldacena and H.~Ooguri, {\it {Strings in AdS(3) and the SL(2,R) WZW
  model. Part 3. Correlation functions}},  {\em Phys. Rev.} {\bf D65} (2002)
  106006, [\href{http://arxiv.org/abs/hep-th/0111180}{{\tt hep-th/0111180}}].

\bibitem{Eguchi:1988af}
T.~Eguchi and A.~Taormina, {\it {On the Unitary Representations of $N=2$ and
  $N=4$ Superconformal Algebras}},  {\em Phys. Lett.} {\bf B210} (1988) 125.

\bibitem{Berkovits:1994vy}
N.~Berkovits and C.~Vafa, {\it {N=4 topological strings}},  {\em Nucl. Phys.}
  {\bf B433} (1995) 123--180, [\href{http://arxiv.org/abs/hep-th/9407190}{{\tt
  hep-th/9407190}}].

\bibitem{Antoniadis:2006mr}
I.~Antoniadis, S.~Hohenegger, and K.~S. Narain, {\it {N=4 Topological
  Amplitudes and String Effective Action}},  {\em Nucl. Phys.} {\bf B771}
  (2007) 40--92, [\href{http://arxiv.org/abs/hep-th/0610258}{{\tt
  hep-th/0610258}}].

\bibitem{Dijkgraaf:1987vp}
R.~Dijkgraaf, E.~P. Verlinde, and H.~L. Verlinde, {\it {C = 1 Conformal Field
  Theories on Riemann Surfaces}},  {\em Commun. Math. Phys.} {\bf 115} (1988)
  649--690.

\bibitem{Dixon:1986qv}
L.~J. Dixon, D.~Friedan, E.~J. Martinec, and S.~H. Shenker, {\it {The Conformal
  Field Theory of Orbifolds}},  {\em Nucl. Phys.} {\bf B282} (1987) 13--73.

\bibitem{Gluck:2005wr}
D.~Gluck, Y.~Oz, and T.~Sakai, {\it {N = 2 strings on orbifolds}},  {\em JHEP}
  {\bf 08} (2005) 008, [\href{http://arxiv.org/abs/hep-th/0503043}{{\tt
  hep-th/0503043}}].

\bibitem{Aharony:2003vk}
O.~Aharony, B.~Fiol, D.~Kutasov, and D.~A. Sahakyan, {\it {Little string theory
  and heterotic / type II duality}},  {\em Nucl. Phys.} {\bf B679} (2004)
  3--65, [\href{http://arxiv.org/abs/hep-th/0310197}{{\tt hep-th/0310197}}].

\bibitem{Aharony:2004xn}
O.~Aharony, A.~Giveon, and D.~Kutasov, {\it {LSZ in LST}},  {\em Nucl. Phys.}
  {\bf B691} (2004) 3--78, [\href{http://arxiv.org/abs/hep-th/0404016}{{\tt
  hep-th/0404016}}].

\bibitem{Kos:2014ab}
F.~Kos, D.~Poland, and D.~Simmons-Duffin, {\it Bootstrapping mixed correlators
  in the 3d ising model},  \href{http://arxiv.org/abs/1406.4858}{{\tt
  arXiv:1406.4858}}.

\bibitem{Simmons-Duffin:2015qma}
D.~Simmons-Duffin, {\it {A Semidefinite Program Solver for the Conformal
  Bootstrap}},  {\em JHEP} {\bf 06} (2015) 174,
  [\href{http://arxiv.org/abs/1502.02033}{{\tt arXiv:1502.02033}}].

\bibitem{Pappadopulo:2012jk}
D.~Pappadopulo, S.~Rychkov, J.~Espin, and R.~Rattazzi, {\it {OPE Convergence in
  Conformal Field Theory}},  {\em Phys. Rev.} {\bf D86} (2012) 105043,
  [\href{http://arxiv.org/abs/1208.6449}{{\tt arXiv:1208.6449}}].

\bibitem{Caracciolo:2014aa}
F.~Caracciolo, A.~C. Echeverri, B.~von Harling, and M.~Serone, {\it Bounds on
  ope coefficients in 4d conformal field theories},
  \href{http://arxiv.org/abs/1406.7845}{{\tt arXiv:1406.7845}}.

\bibitem{Adams:2006aa}
A.~Adams, N.~Arkani-Hamed, S.~Dubovsky, A.~Nicolis, and R.~Rattazzi, {\it
  Causality, analyticity and an ir obstruction to uv completion},  {\em JHEP}
  {\bf 0610} (2006) 014, [\href{http://arxiv.org/abs/hep-th/0602178}{{\tt
  hep-th/0602178}}].

\bibitem{Caracciolo:2009bx}
F.~Caracciolo and V.~S. Rychkov, {\it {Rigorous Limits on the Interaction
  Strength in Quantum Field Theory}},  {\em Phys. Rev.} {\bf D81} (2010)
  085037, [\href{http://arxiv.org/abs/0912.2726}{{\tt arXiv:0912.2726}}].

\bibitem{Hogervorst:2013sma}
M.~Hogervorst and S.~Rychkov, {\it {Radial Coordinates for Conformal Blocks}},
  {\em Phys. Rev.} {\bf D87} (2013) 106004,
  [\href{http://arxiv.org/abs/1303.1111}{{\tt arXiv:1303.1111}}].

\bibitem{Kim:2015oca}
H.~Kim, P.~Kravchuk, and H.~Ooguri, {\it {Reflections on Conformal Spectra}},
  \href{http://arxiv.org/abs/1510.08772}{{\tt arXiv:1510.08772}}.

\bibitem{Anselmi:1993sm}
D.~Anselmi, M.~Billo, P.~Fre, L.~Girardello, and A.~Zaffaroni, {\it {ALE
  manifolds and conformal field theories}},  {\em Int. J. Mod. Phys.} {\bf A9}
  (1994) 3007--3058, [\href{http://arxiv.org/abs/hep-th/9304135}{{\tt
  hep-th/9304135}}].

\bibitem{Douglas:aa}
M.~R. Douglas, R.~L. Karp, S.~Lukic, and R.~Reinbacher, {\it Numerical
  calabi-yau metrics},  \href{http://arxiv.org/abs/hep-th/0612075}{{\tt
  hep-th/0612075}}.

\bibitem{Headrick:2009jz}
M.~Headrick and A.~Nassar, {\it {Energy functionals for Calabi-Yau metrics}},
  {\em Adv. Theor. Math. Phys.} {\bf 17} (2013) 867--902,
  [\href{http://arxiv.org/abs/0908.2635}{{\tt arXiv:0908.2635}}].

\bibitem{Aspinwall:1996mn}
P.~S. Aspinwall, {\it {K3 surfaces and string duality}},  in {\em {Fields,
  strings and duality. Proceedings, Summer School, Theoretical Advanced Study
  Institute in Elementary Particle Physics, TASI'96, Boulder, USA, June 2-28,
  1996}}, pp.~421--540, 1996.
\newblock \href{http://arxiv.org/abs/hep-th/9611137}{{\tt hep-th/9611137}}.

\bibitem{Cheng1975}
S.-Y. Cheng, {\it Eigenfunctions and eigenvalues of laplacian},  {\em
  Proceedings of Symposia in Pure Mathematics} {\bf 27} 185--193.

\bibitem{Li1980}
P.~Li and S.-T. Yau, {\it Estimates of eigenvalues of a compact riemannian
  manifold},  {\em Proceedings of Symposia in Pure Mathematics} {\bf 36}
  205--239.

\bibitem{Hori:2001ax}
K.~Hori and A.~Kapustin, {\it {Duality of the fermionic 2-D black hole and N=2
  liouville theory as mirror symmetry}},  {\em JHEP} {\bf 08} (2001) 045,
  [\href{http://arxiv.org/abs/hep-th/0104202}{{\tt hep-th/0104202}}].

\bibitem{Bergman:1999kq}
O.~Bergman and M.~R. Gaberdiel, {\it {NonBPS states in heterotic type IIA
  duality}},  {\em JHEP} {\bf 03} (1999) 013,
  [\href{http://arxiv.org/abs/hep-th/9901014}{{\tt hep-th/9901014}}].

\end{thebibliography}\endgroup
 \end{document}